%% file: CTA_GC_limits-NewMaster.tex
\documentclass[a4paper,11pt]{article}
\usepackage{jcappub} 

\usepackage{graphicx}
\usepackage{amssymb,amsmath, eufrak}
\usepackage{subfiles}
\usepackage{hyperref}
\usepackage{footmisc}
\usepackage{booktabs}
\usepackage[T1]{fontenc}
\newcommand{\norm}[1]{\left\lVert#1\right\rVert}

\usepackage{lineno}
\usepackage{color}
\usepackage{bm}
\usepackage{subfig}
\usepackage{hyphenat}
\usepackage{seqsplit}

\newcommand{\be}{\begin{equation}}
\newcommand{\ee}{\end{equation}}
\newcommand{\bea}{\begin{eqnarray}}
\newcommand{\eea}{\end{eqnarray}}
\newcommand{\ds}{{\sf DarkSUSY}}
\newcommand{\sword}{{\sf swordfish}}

\title{Sensitivity of the Cherenkov Telescope Array to a dark matter signal from the Galactic centre}

\input{author-list-jcap.tex}

\abstract{%
We provide an updated assessment of the power of the Cherenkov Telescope Array (CTA) to search 
for thermally produced dark matter at the TeV scale, via the associated gamma-ray signal 
from pair-annihilating dark matter particles in the region around the Galactic centre.
We find that CTA will open a new window of discovery potential, significantly extending the range
of robustly testable models given a standard cuspy profile of the dark matter density distribution. 
Importantly, even for a cored profile, the projected sensitivity of CTA will be sufficient to probe 
various well-motivated models of thermally produced dark matter at the TeV scale.
This is due to CTA’s unprecedented sensitivity, angular and energy resolutions, and the planned 
observational strategy. The survey of the inner Galaxy will cover a much larger region than 
corresponding previous observational campaigns with imaging atmospheric Cherenkov telescopes.
CTA will map with unprecedented precision the large-scale diffuse emission in high-energy gamma rays, 
constituting a background for dark matter searches for which we adopt state-of-the-art models based on 
current data. Throughout our analysis, we use  up-to-date 
event reconstruction Monte Carlo tools developed by the CTA consortium, and pay special attention 
to quantifying the level of instrumental systematic uncertainties, as well as background template systematic
errors, 
required to probe thermally produced dark matter at these energies.
}

\begin{document}
\maketitle

\section{Introduction}
\label{sec:Intro}

There is compelling evidence that a cold, non-baryonic dark matter (DM) component dominates 
the matter content of the Universe at cosmological scales, contributing a fraction of about 
27\% to its total energy density~\cite{Aghanim:2018eyx}.  
The underlying nature of this DM component 
is still unknown, a century after its existence was first conjectured~\cite{Bertone:2016nfn}, 
but many hypothetical elementary particles may provide viable 
solutions~\cite{Jungman:1995df,Bertone:2004pz,Feng:2010gw}. In particular, 
even if increasingly pressured by lack of experimental evidence~\cite{Bertone:2018xtm},  
weakly interacting massive particles (WIMPs) remain one of the best-motivated candidates~\cite{Arcadi:2017kky}.
Such particles with masses and couplings at the electroweak scale would be a compelling solution to the 
DM puzzle not only because their existence could point to a way of addressing the naturalness problems 
in the standard model of particle physics (SM), 
but also because it would allow us to understand the 
presently measured DM abundance as a result of the standard thermal history of the Universe~\cite{Lee:1977ua}. 

Searches for signals of DM annihilation in astrophysical objects are sensitive to the same physical 
process as the one that, in these models, took place in the early Universe. In fact, the WIMP paradigm 
makes a clear prediction for the annihilation cross-section of interest, and hence the expected signal strength. 
Among the various possible ways of detecting such indirect detection signals, 
gamma rays are particularly promising (for a review, see Ref.~\cite{Bringmann:2012ez}). 
By far, the largest signal is expected from the region around the Galactic 
centre (GC), which is both one of the closest DM targets and features the highest DM density in our Galaxy~\cite{Bergstrom:1997fj}.

The Cherenkov Telescope Array (CTA) \cite{CTAweb} is now 
close to entering the production phase and will be the world's most sensitive gamma-ray telescope 
for a window of photon energies stretching more than three orders of magnitude,
from a few tens of GeV to above 300 TeV,
with an angular resolution that is better than any existing instrument observing at frequencies higher
than the X-ray band.
One of CTA's main science drivers is the search for a signal from annihilating DM~\cite{Acharya:2017ttl}.  
Previous estimates indicate that CTA observations of the GC have a good chance of being sensitive to 
the `thermal' annihilation cross 
section~\cite{Doro:2012xx,Silverwood:2014yza,Pierre:2014tra,Carr:2015hta,Lefranc:2015pza,Rinchiuso:2020skh}, 
i.e.~the WIMP annihilation strength required to explain 
the observed DM abundance in the simplest models from particle physics. For TeV-scale DM, notably, 
CTA might well turn out to be the {\it only} planned or existing instrument with 
this property (another potential contender being AMS-02~\cite{AMSweb} via charged cosmic 
rays, CRs~\cite{Bringmann:2006im,Cuoco:2017iax}); 
it thus provides an important tool to search for WIMPs that is highly complementary to direct 
DM searches in underground laboratories or WIMP searches at 
colliders~\cite{Bergstrom:2010gh,Cahill-Rowley:2013dpa,Arrenberg:2013rzp,Balazs:2017hxh}.

Given the imminent start of the telescope construction, and the strong science case outlined above, 
it is timely to move 
beyond existing analyses and provide more realistic sensitivity estimates to a DM signal from GC observations 
with CTA that fully take into account the current best estimates for the expected 
telescope characteristics as well as recent developments in understanding the (expected) Galactic diffuse 
emission (GDE) components in that region. In fact, CTA is expected to measure some of the 
GDE components with an unprecedented angular resolution at TeV energies 
-- which is a prominent science goal in itself.
Here we report on such an updated 
analysis and explore the most promising strategies to define signal regions and data analysis methods, 
using state-of-the-art models for astrophysical and instrumental backgrounds. 
In particular, a major motivation of this work is to study in detail the applicability of a full template fitting 
approach in the analysis of imaging atmospheric Cherenkov telescopes (IACT) data, 
a field which has traditionally mostly relied on separate 
`ON' and `OFF' regions to extract the DM signal and background, respectively (though first studies 
indicate the advantages of moving beyond that method~\cite{Silverwood:2014yza}).

We present our results for standard assumptions concerning the DM density profile, 
both in terms of expected sensitivities to the annihilation cross-section for the most commonly adopted 
annihilation channels and in the form of tabulated likelihoods that can be applied to almost arbitrary 
annihilation spectra, and we discuss in detail how these assumptions affect our conclusions. 
We further demonstrate that the DM sensitivity is, indeed, quite significantly affected by the expected GDE, 
which makes realistic modelling of this component mandatory. 
As expected for an instrument with a large collection area and excellent event 
statistics such as CTA~\cite{Bringmann:2012ez}, finally, we confirm that the sensitivity is, in large parts of the 
parameter space, limited by systematic rather than statistical uncertainties. We, therefore, put special emphasis on 
discussing this aspect, both regarding the overall uncertainty and bin-to-bin correlations
in sky positions and energy. This allows us to quantify the maximal level of systematic uncertainty that is 
required to reach the 
thermal cross-section for a given DM mass and annihilation channel (assuming a standard DM density profile). 

This article treats a range of topics, from DM to conventional astrophysics and 
instrumental properties. While we make an effort to cover all relevant aspects, which makes a 
certain overall length unavoidable, we deliberately organised the article in a way that allows the reader to 
directly skip to the (mostly self-contained) sections of interest without the need to read all preceding parts.
We start by briefly introducing high-energy gamma-ray astronomy
and the CTA observatory in Section~\ref{sec:CTA}, along with its planned observational strategy for the GC
region.
We then describe in more detail the expected DM signal and various astrophysical and instrumental 
background sources in the GC region (in Section~\ref{sec:GCtemplates}), as well as the data 
analysis techniques adopted in our analysis 
(in Section~\ref{sec:data}).  We present our findings concerning the projected sensitivity of 
CTA to a DM signal in Section~\ref{sec:results}; as this is a concise summary of our main results, 
many readers would typically directly want to jump there. In Section~\ref{sec:discussion} we discuss in 
more detail how these results depend on the 
adopted analysis strategy and the treatment of the astrophysical emission components, before concluding in 
Section~\ref{sec:conc}. 
In an extended appendix, we collect supplemental material to further support the discussion section, 
as well as more technical background information about the analysis pipeline. 
In Appendix \ref{app:Phase1}, in particular,  we show projected sensitivities 
for the reduced telescope configuration that will be implemented in the initial construction phase.

\section{Gamma-ray astronomy and CTA}
\label{sec:CTA}

\subsection{Telescope design and historical context}
\label{sec:CTA_design}

From a historic point of view, gamma-ray astronomy  started with satellite-based  studies of high energy ($> 50$ MeV) 
photon emission. In particular, the OSO-3 satellite~\cite{4323497}, launched in 1967, was the first to detect the 
Galactic centre and Galactic plane~\cite{1968ApJ...153L.203C}, as well as the presence of an isotropic extragalactic 
gamma-ray background~\cite{1972ApJ...177..341K}. In the following decades, satellite missions like 
SAS-2 (1972)~\cite{1975ApJ...198..163F}, COS-B (1975)~\cite{1975SSI.....1..245B}, EGRET 
(1991)~\cite{Hartman:1999fc} and the most recent representatives AGILE~\cite{Tavani:2008sp} and 
Fermi-LAT~\cite{Atwood:2009ez} have widely increased our knowledge of the gamma-ray sky. 
Nonetheless, due to their limited size, satellites can typically only 
cover energies below the TeV range. 

The very high-energy sky can be observed with ground-based gamma-ray telescopes. This approach was 
pioneered by the Whipple telescope~\cite{1989ApJ...342..379W} in the late 1980s, demonstrating the promise of 
the {\it imaging atmospheric Cherenkov light} technique 
which rests on imaging short flashes of Cherenkov radiation produced by cascades of relativistic charged 
particles in the atmosphere, originating from very high energy gamma rays or charged cosmic rays striking the top of 
the atmosphere (see, e.g.,~Ref.~\cite{DiSciascio:2019lse}  and references therein).\footnote{%
Current-generation IACTs are complemented by water-Cherenkov telescopes, where large water tanks provide the 
medium for Cherenkov light detection of secondary charged air shower particles that have reached the Earth's 
surface. Starting its operation in 2000, the Milagro telescope~\cite{Atkins_2004, Abdo_2008, Abdo_2009}
was able to detect gamma rays in the energy range from 100 GeV to about 100 TeV based on this technique.  
It  was succeeded by the ARGO YBT observatory~\cite{2013ApJ...779...27B} in 2007 and the HAWC 
observatory~\cite{Abeysekara:2013tza} in 2015. The next-generation water-Cherenkov 
telescope LHAASO~\cite{Bai:2019khm, Guo:2020ooo} is expected to be completed in 2021. 
}
The technique was further developed with the current set of modern IACTs, demonstrating a leap in sensitivity by 
increasing the telescope multiplicity~\cite{Bernlohr:2008kv,Becherini:2011pb}. 
In comparison, day-long observations at  sub-TeV energies with 
modern IACTs like H.E.S.S.~\cite{hess},
MAGIC~\cite{MAGIC} 
and VERITAS~\cite{VERITAS} 
very roughly result in similar sensitivities as 
year-long satellite observations. 
IACTs have mapped the very high energy gamma-ray sky, resulting in a catalogue of about two hundred TeV sources~\cite{tevcat2}, including active galactic nuclei (AGNs) as the most numerous extragalactic source class and tens of Galactic sources, most importantly supernova remnants (SNRs) and pulsar wind nebulae (PWNe).  IACT data was also used to set competitive limits on the annihilation of TeV-scale DM candidates, falling just short of reaching the theoretically motivated `thermal' cross-section value (see e.g.~\cite{Rinchiuso:2019etv,Rinchiuso:2019rrh,Zitzer:2017xlo,Doro:2017dqn}). 


CTA is the next-generation IACT gamma-ray 
observatory~\cite{Acharya:2017ttl, CTAweb}.
When completed it will 
comprise two arrays, a southern one being located at the European Southern Observatory (ESO) 
site in Chile, Atacama desert, and a northern 
one being located at the Roque de los Muchachos Observatory (ORM) site in La Palma, Canary Islands.
This combination will make CTA the first ground-based gamma-ray telescope with the capability to 
observe a large sky fraction.
The CTA design concept foresees three types of telescopes with different sizes: {\it i)} LSTs 
(Large-Sized Telescopes, 23\,m in diameter) 
 that are needed to detect the relatively small amount of Cherenkov photons from gamma rays 
 in the 20 -- 150\,GeV range, {\it ii)} MSTs (Medium-Sized Telescopes, 11.5\,m)  that aim 
to observe energies between 150\,GeV and around 5\,TeV, and {\it iii)}  a large number of
SSTs (Small-Sized Telescope, 4\,m), spread out over several square kilometers to detect the 
most energetic, but very rare gamma rays. The `baseline' goal, which we base our 
sensitivity forecast on, is to deploy 4 LSTs at each of the sites, 25 (15) MSTs in the Southern (Northern) 
hemisphere, and 70 SSTs at the southern site 
(see Appendix \ref{app:Phase1} for the effects of a slimmed-down, initial configuration).
As a result of this setup, CTA is believed to be large and sensitive 
enough to bridge the characteristic differences between current IACTs and satellite-borne gamma-ray 
telescopes, spanning a range of observable energies from 10s of GeV up to above 300 TeV. 
The large field of view cameras will also put CTA in a unique position to perform {\it surveys} of  
extended sky regions. Those currently planned include the GC survey and extended GC survey described in 
Section \ref{sec:GC_obs_strategies}, but also an extensive survey of the Galactic plane as well as an 
extragalactic survey covering a quarter of the Northern sky~\cite{Acharya:2017ttl}.

\subsection{Observational strategy of the Galactic centre}
\label{sec:GC_obs_strategies}

\begin{figure}
\centering\includegraphics[width=0.47\linewidth]{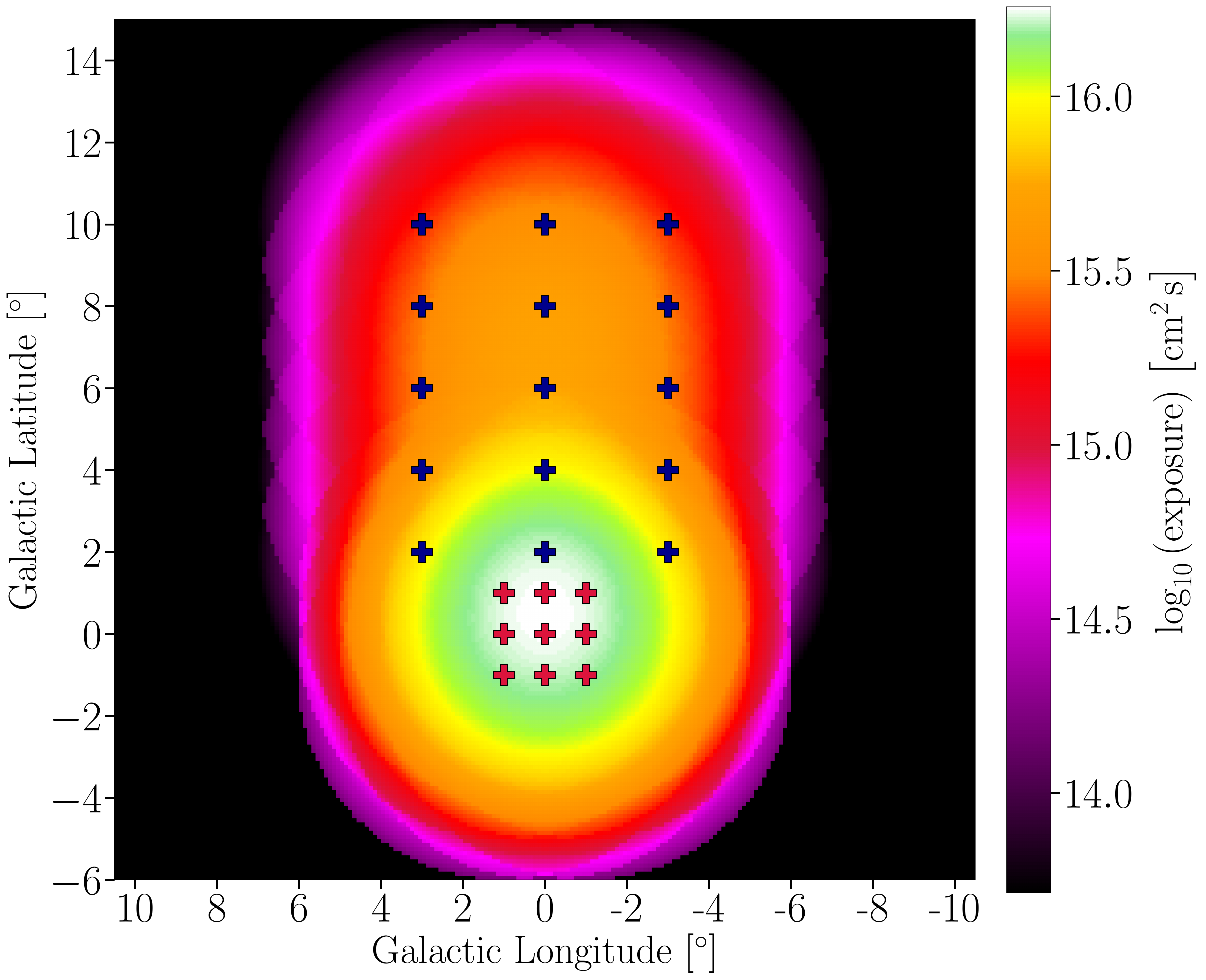}
\hspace*{0.5cm}
\centering\includegraphics[width=0.48\linewidth]{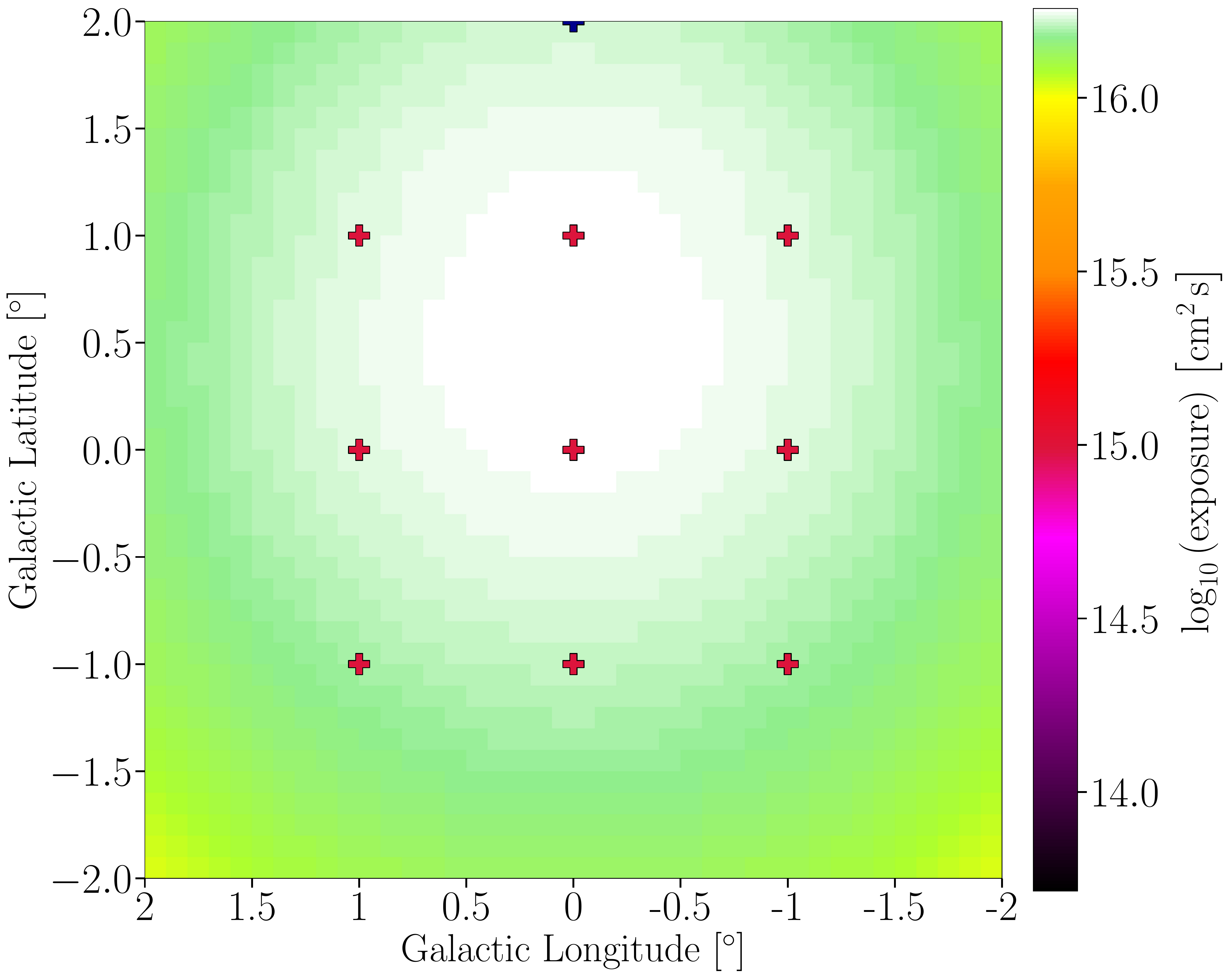}
\caption{%
The left panel shows the exposure map for CTA's Galactic centre (GC) and extended GC surveys, 
at an energy of 1\,TeV.  The right panel shows a zoom into the GC survey region. The nine 
pointing positions of the GC survey mode are marked with red crosses. The observation time for each of these
pointing positions will be 
approximately $60\,\textrm{h}$.  The 15 pointing positions of the extended survey north of the 
Galactic plane are marked with blue crosses. The observation time for each position will be  $20\,\textrm{h}$ in this case. 
\label{fig:observations_schematic}}
\end{figure}

Given its importance for both DM and astrophysical studies, a significant fraction of the 
currently planned CTA observing time for key science projects
is dedicated to a detailed exploration of the GC region~\cite{Acharya:2017ttl}. 
Surveys of extended portions of the sky will be adopted as
an observational strategy that, in scope and ambition, will surpass previous 
gamma-ray observations with IACTs. Below we detail surveys which (in part) overlap with  the Galactic center region:
\begin{itemize}
    \item[\textsc{(i)}] \textbf{Galactic centre Survey}: This is the survey strategy that CTA will follow
    according to current plans. It consists of nine individual pointing positions centred on 
    ($l = \pm1^{\circ}, 0^{\circ}$, $b = \pm1^{\circ}, 0^{\circ}$) 
    in Galactic coordinates (red crosses in Fig.~\ref{fig:observations_schematic}),
    and we evenly distribute the total observation time of  $t_{\rm{obs}} = 525\;\textrm{h}$ among them. 
    We stress that all three telescope types are part of this observation mode -- but with varying 
    actual sensitivity within the field of view (FoV), which is implicitly accounted for through the instrument 
    response functions (IRFs). 
    {\it This is the default pointing strategy as well as the default survey `region of interest' (ROI) we adopt in 
    this article.} 

    \item[\textsc{(ii)}] \textbf{Extended survey}: An observation supplementing the GC 
    survey is to scan over a region above the Galactic plane from $b = 2^{\circ}$ to $b=10^{\circ}$, 
    and $l = -3^{\circ}$ to  $l = +3^{\circ}$, with 15 additional pointing positions  
    centred on ($l = \pm3^{\circ}, 0^{\circ}$, $b = 2^{\circ}, 4^{\circ}, 6^{\circ}, 8^{\circ}, 10^{\circ}$) in Galactic 
    coordinates (blue crosses in Fig.~\ref{fig:observations_schematic}). 
    Each of those pointing directions is observed for $20\,\textrm{h}$ so that the total observation 
    time of the extended 
    survey amounts to  $t_{\rm{obs}} = 300\,\textrm{h}$ (adding to the combined 525\,h of the GC survey). 
    Due to the large region covered, this observation strategy can increase the sensitivity for
    DM distributions that are more cored around the GC (as discussed in more detail in Section \ref{sec:cores}). 
\end{itemize} 
 The planned Galactic {\it plane} survey will also overlap with the GC region. However, since it has a significantly 
 smaller exposure than the two surveys described above, we will not use  it in our  work. 
To evaluate the expected number of events for a given sky model, the CTA consortium has 
produced IRFs for the planned array configurations. 
These are based on Monte-Carlo simulations of the Cherenkov light that is generated in the 
interaction of gamma rays with the Earth's atmosphere and the subsequent measurement of this 
light by CTA telescopes, followed by event reconstruction and classification. The IRFs provide 
information on effective area, point spread function and energy dispersion as a function of energy 
and offset angle for various telescope pointing zenith angles~\cite{CTA_performance}.
In this work, we use the {\it publicly available} \texttt{prod3b-v1} IRF library, and in particular the IRF file 
\texttt{South\_z20\_average\_50h} which is optimised -- by defining background reduction cuts with 
respect to an equivalent of 50\,h of simulated Monte Carlo air showers -- for the detection of 
a point-like source at $20^{\circ}$ zenith angle (note that the GC is mostly visible from the southern site). 
Finally, for the smaller number of telescopes planned for the initial construction configuration investigated in 
Appendix \ref{app:Phase1}, we use a separate set of IRFs as described there.
A versatile tool to predict the number of expected counts, given a set of IRFs, 
is the public code \href{http://cta.irap.omp.eu/ctools/about.html}{ctools}~\cite{Knodlseder:2016nnv}
that we will make extensive use of in our analysis.


\section{Emission components for Galactic centre observations}
\label{sec:GCtemplates}

Here we discuss the various emission components that may appear in the ROI around the GC,
starting with the DM contribution and then moving on to the expected conventional astrophysical emission and 
the contribution from misidentified CRs. We conclude  with an explicit comparison of 
the emission templates that we use
(for a quick overview, see Figs.~\ref{fig:template_collection} and \ref{fig:flux-allcomponents-2}).

\subsection{Dark matter contribution}
\label{sec:DM_models}

As already stressed in the introduction, CTA provides 
unique opportunities to test the WIMP paradigm at the TeV scale.
Heavy DM candidates falling into this mass range include
the (still) most popular lightest neutralino~\cite{Jungman:1995df}, but also candidates appearing in models with 
extra dimensions~\cite{Servant:2002aq, Cembranos:2003mr},
or models involving `portals' between the standard  model and a dark 
sector~\cite{Batell:2017rol,Bandyopadhyay:2018qcv,Athron:2018hpc,Siqueira:2019wdg}, just to name a few. 
The spectral distribution of the expected gamma rays contains valuable information about the 
underlying theory. It ranges from relatively soft and featureless spectra to, as in the case of Kaluza-Klein 
DM~\cite{Bergstrom:2004cy}, rather hard spectra with an abrupt cut-off at energies corresponding to the 
DM mass; sometimes also pronounced spectral endpoint features are present~\cite{Bringmann:2007nk}, 
which have been worked out to high accuracy for theoretically well-motivated candidates like the 
Wino~\cite{Rinchiuso:2018ajn,Baumgart:2018yed,Beneke:2018ssm,Beneke:2019vhz} and 
Higgsino~\cite{Aparicio:2016qqb,Kowalska:2018toh,Beneke:2019gtg,Rinchiuso:2020skh}. 
Further aspects making the TeV scale particularly interesting
from a theoretical perspective include the large flux enhancements that are possible due to the
Sommerfeld effect~\cite{Hisano:2004ds} and the fact that this scale is close to the unitarity
limit for thermally produced DM~\cite{Griest:1989wd,Smirnov:2019ngs}.

In general, the {\it prompt emission component} of the differential gamma-ray flux, per unit energy and solid angle, 
that is expected from annihilating 
DM particles $\chi$ with a density profile $\rho_\chi(\mathbf{r})$ is given by 
(see e.g.~Ref.~\cite{Bringmann:2012ez})
\be
\label{DMflux}
  \frac{d\Phi_{\gamma}}{d\Omega\, dE_\gamma} (E_\gamma,\psi) = 
  \frac{1}{4\pi} \int_\mathrm{l.o.s}
  d\ell(\psi) \rho_\chi^2(\mathbf{r}) 
  \left({\frac{\langle\sigma v\rangle_\mathrm{ann}}{2S_\chi m_{\chi}^2} \sum_f
  B_f\frac{dN_\gamma^{f}}{dE_\gamma}}\right) \,,
\ee
where the integration is performed along the line of sight (l.o.s.) in the observing 
direction ($\psi$). Particle physics parameters that enter here -- contained in the parenthesis --
are the average velocity-weighted
annihilation cross-section $\langle\sigma v\rangle_\mathrm{ann}$\footnote{%
Here, the average is performed with respect to Galactic velocities {\it today}. The WIMP relic density,
in contrast, depends on $\langle \sigma v\rangle$ averaged over the DM velocities in the early 
universe~\cite{Gondolo:1990dk}. The numerical value for this latter quantity that is needed to match the 
cosmologically observed DM abundance is often referred to as the {\it `thermal' cross section}. 
While this is also the generically expected numerical value for $\langle\sigma v\rangle_\mathrm{ann}$ entering in Eq.~(\ref{DMflux}) for models with velocity-independent $\sigma v$, there are many particle
physics examples where the annihilation rate today can be larger than in the early universe, in 
particular in the presence of 
resonances~\cite{Griest:1990kh,Baer:2003bp,Kakizaki:2005en,Ibe:2008ye,Arina:2014fna} 
or the already mentioned Sommerfeld effect.
}, 
the DM mass $m_\chi$, a symmetry factor 
that is $S_\chi=1$ ($S_\chi=2$) if the DM particle is (not) its own antiparticle,
the annihilation branching ratio $B_f$ into channel $f$ and  the number $N_\gamma^{f}$ of photons 
per annihilation. 
If the annihilation rate (and spectrum) is sufficiently independent of the small Galactic DM
velocities $v(\mathbf{r})$, as for the simplest DM models,
the factor in parenthesis can be pulled outside the line-of-sight and angular integrals.\footnote{%
In practice, we will assume that  the DM velocities are sufficiently small that rest-frame spectra 
can be used for $dN_\gamma^{f}/dE_\gamma$, thus neglecting the small boost. 
}
Spatial and spectral information contained in the signal 
then factorise, and hence are uncorrelated, such that the flux from a given angular region $\Delta\Omega$
becomes simply proportional to what is conventionally defined as the `$J$-factor',
\be
\label{Jfactor}
J\equiv\int_{\Delta\Omega} d\Omega\int\!\!d\ell\,\rho_\chi^2\,.
\ee
For simplicity, and in order to make our limits directly comparable to corresponding limits in the literature,
we will in the following assume that all of the astrophysically observed DM consists of a single type of
self-conjugate particles $\chi$ (i.e.~$S_\chi=1$); if only a fraction $f_\chi$ of the total DM component annihilates, 
all reported limits weaken by a factor of $f_\chi^2$.

\subsubsection*{Spatial distribution} \label{sec:DMdistribution}
Calculating the $J$-factor to sufficient precision requires a good knowledge of the DM distribution.
The average {\it local} DM density at the Sun's distance from the GC, 
which we take as the canonical $r_{\odot}=8.5\;\textrm{kpc}$ (though recent precision
measurements rather indicate a value closer to 
$r_{\odot}=8.2\;\textrm{kpc}$~\cite{2019A&A...625L..10G,Abuter:2020dou}), 
can be determined relatively well by observations. Here we follow the common practice of using 
$\rho_{\odot}=0.4$\,GeV/cm$^3$, noting that the uncertainty associated with this value is typically quoted 
to be a factor of less than about 2~\cite{Catena:2009mf, McMillan:2011wd,Benito:2019ngh,Karukes:2019jxv}. 
The DM content in the inner kpc of the Milky Way (MW),  
in contrast, is almost 
unconstrained {\it observationally} because the baryonic component largely dominates the
gravitational potential in that region~\cite{Iocco:2011jz,Gammaldi:2016uhg,Benito:2019ngh,Karukes:2019jxv}
(for an early discussion, see also Ref.~\cite{Evans:2003sc}).

Numerical $N$-body {\it simulations} of collision-less cold DM clustering -- {\it not} including baryonic feedback -- 
have consistently demonstrated, on the other hand, 
that DM halos should develop a universal density profile during cosmological structure formation,
following the gravitational collapse of initially small density 
perturbations~\cite{Navarro:1995iw,Navarro:1996gj,Zavala:2019gpq}. Largely independent of the virial mass,
in particular, such WIMP DM halos today should be `cuspy', with the logarithmic slope at small 
galactocentric distances $r$ being roughly $d(\log\rho)/d(\log r)\approx-1$.  
Recent simulations rather tend to favour an Einasto 
profile~\cite{1965TrAlm...5...87E}, 
\be
\label{einasto}
\rho_{{\rm Einasto}}\!\left(r\right)=\rho_{s}\exp\!\left(-\frac{2}{\alpha}\left[\left(\frac{r}{r_{s}}\right)^{\alpha}-1\right]\right){\rm ,}
\ee
which is slightly shallower in the central-most parts of the halo than the form originally suggested by 
Navarro, Frenk and White (NFW)~\cite{Navarro:1995iw,Navarro:1996gj}. In our analysis we will adopt 
benchmark values of $\alpha=0.17$ and $r_{s}=20$ kpc (and hence $\rho_s=0.081$\,GeV/cm$^3$). 
This is both compatible with the most recent observations~\cite{Karukes:2019jwa}  and inside the 
expected range of these parameters for simulated halos with the mass of the MW~\cite{Diemand:2008in}. 
 
More realistic simulations of MW-like halos necessarily have to include a baryonic component. 
Baryons can radiate away energy and angular momentum, leading to the formation of disks and
much more concentrated densities in the central halo region. The correspondingly larger gravitational 
potential will then also affect the DM component, leading to a significant steepening
of the DM profile (and hence an increase of the $J$-factor) if this process happens 
adiabatically~ \cite{Blumenthal:1985qy,Gnedin:2004cx,Gustafsson:2006gr}.
On the other hand, feedback from star formation and supernovae, in particular if happening on
short time-scales and hence not adiabatically, leads to the formation of central {\it cores} of roughly 
constant density in the DM profile~\cite{DiCintio:2013qxa}.
The current state of the art in simulations suggests that the latter mechanism can often be decisive 
in smaller galaxies  (with masses $\lesssim 10^{12}\,M_{\odot}$), while in larger galaxies 
(like the MW, or more massive), the former effect often dominates -- i.e.~baryons tend to contract 
rather than dilute the central DM distribution (for a recent discussion specifically applying to the 
MW, using Gaia DR2 data, see Ref.~\cite{Cautun:2019eaf}).
Even though there has been significant
progress in including baryonic effects in hydrodynamical simulations of structure 
formation~\cite{Cautun:2019eaf, Vogelsberger:2019ynw,Sawala:2015cdf,Tollet:2015gqa,Pillepich:2017jle,Springel:2017tpz}, it should be noted that
resolving the scales at which the relevant astrophysical processes happen is still far from achievable. 
This means that these simulations need to rely on phenomenological prescriptions, rather than
prescriptions directly based on first principles, which makes it challenging to assess whether the DM halo of 
a galaxy with the specific properties of the MW -- also taking into account its position in the local 
group -- should be expected to develop a sizeable core or not. Still, it seems unrealistic
to obtain core sizes much larger than about 1\,kpc (see e.g.~the comparison of different simulation results
in Ref.~\cite{Gammaldi:2016uhg}), even though this might be consistent from a purely observational point 
of view~\cite{Bovy:2013raa}.

In light of this discussion we will consider a second, purely phenomenologically motivated benchmark 
profile with core sizes of $r_{c}=0.5$\,kpc and 1\,kpc. For this {\it cored Einasto} profile we adopt
\be
\label{einasto_core}
\rho_{{\rm coredEinasto}}\!\left(r\right)=
\begin{cases}
\rho_{{\rm Einasto}}\!\left(r_{c}\right) & \textrm{if}\;r\leq r_{c}\\
\rho_{{\rm Einasto}}\!\left(r\vphantom{r_{c}}\right) & \textrm{if}\;r>r_{c}
\end{cases}\,,
\ee
using the same Einasto parameters as for the benchmark described after Eq.~(\ref{einasto}).
This choice is motivated by the attempt to bracket the expected sensitivity of CTA to a 
WIMP annihilation signal, thus roughly serving as a `worst-case scenario'. 
We stress however that our selection of benchmark DM profiles is based on theoretical expectations 
for {\it cold and collisionless} DM, like 
WIMPs, rather than on the arguably even larger uncertainty inferred from observations alone.

\begin{figure}[t!]
\centering
\begin{minipage}{0.499\textwidth}
\includegraphics[width=\linewidth]{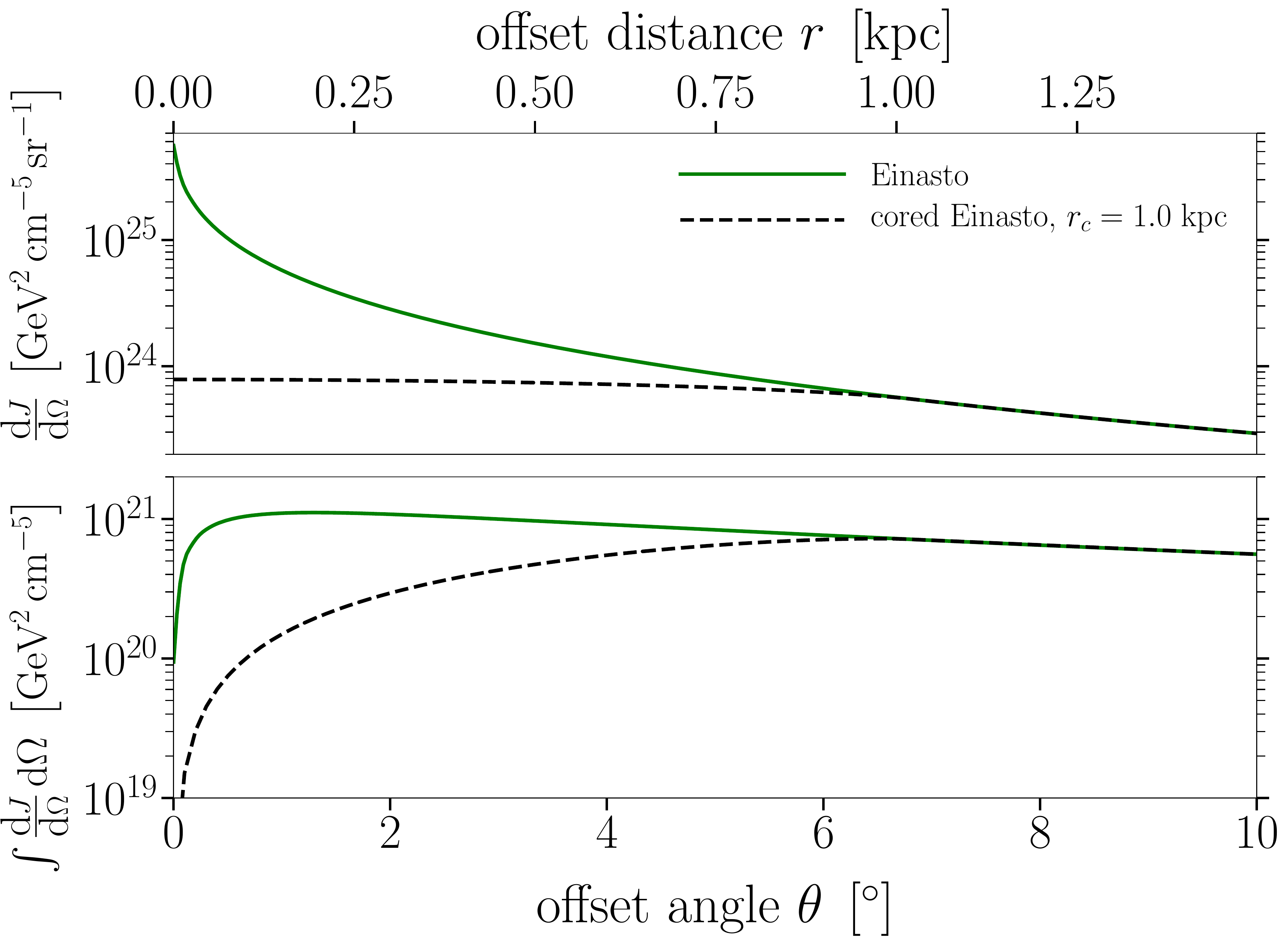}\\[0.1cm]
\end{minipage}
\begin{minipage}{0.482\textwidth}
\includegraphics[width=\linewidth]{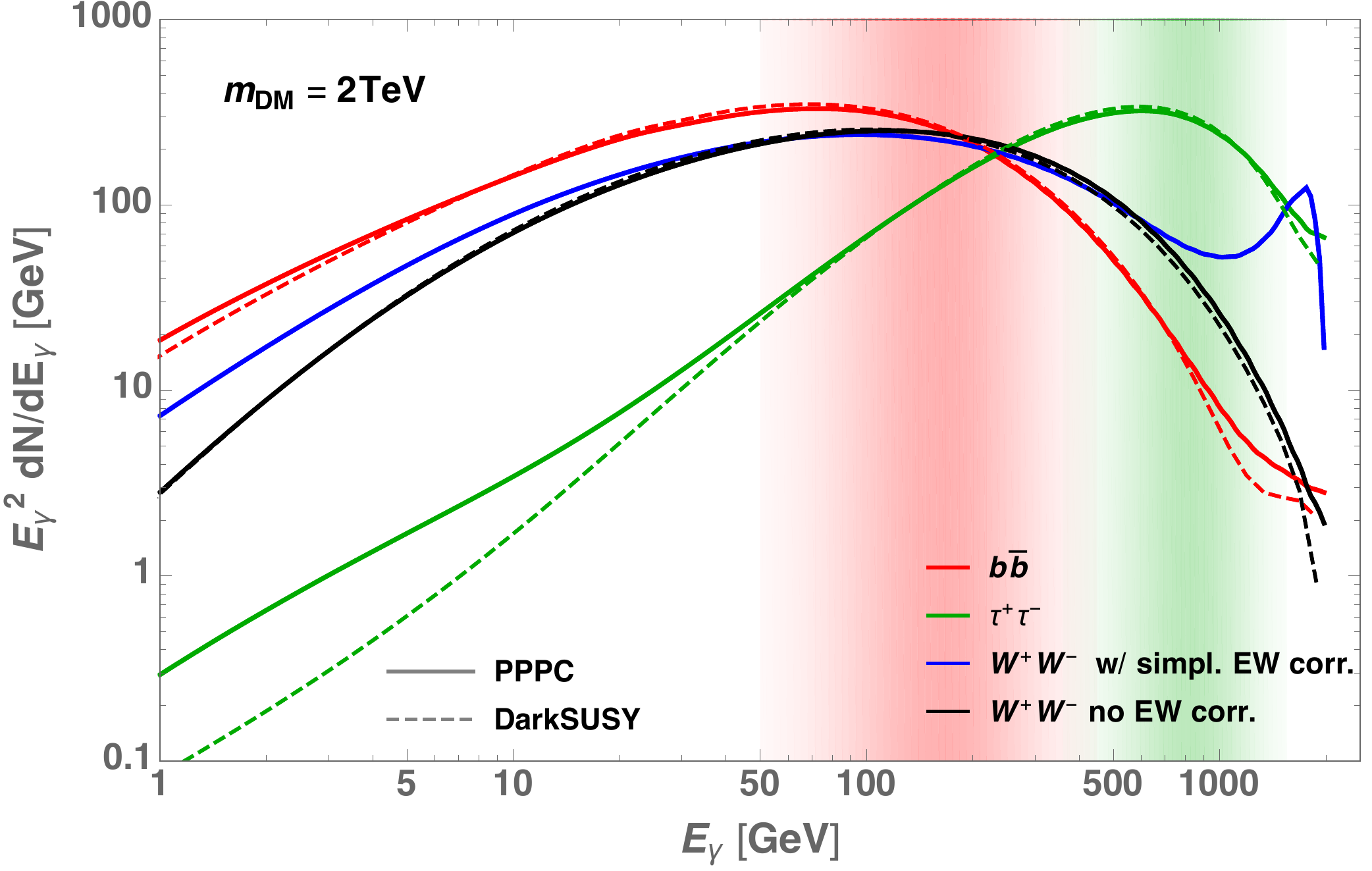}
\end{minipage}
\caption{{\it Left:} Summary of $J-$factor profiles considered in this study, with solid (dashed) lines
corresponding to the case of an Einasto (cored Einasto) profile for the DM density. The top panel 
shows the differential 
$J$-factor for an observation pointing an angle $\theta$ away from the GC (translating to a physical offset 
distance $r=r_\odot \tan\theta$), while the bottom panel shows the  $J$-factor integrated over annuli of 
width $0.1^\circ$ as a function of the average annulus radius.
{\it Right:} Summary of benchmark DM spectra adopted in this analysis for various final states, indicated 
by solid lines with different colours. Dashed lines show the corresponding spectra obtained with an 
alternative event generator. 
The shaded regions illustrate where the DM signal would be most important compared to a background that 
falls like a power law, $E^{-n_\gamma}$, with a fiducial value of $n_\gamma=2.7$ for the spectral index. 
For better visibility, these regions are only indicated for $b\bar b$ (red) and $\tau^+\tau^-$ (green) final states.
 \label{fig:DM_benchmarks}}
\end{figure}

In the left panel of Fig.~\ref{fig:DM_benchmarks} we show the resulting radial and angular profile for our 
benchmark DM distributions, both in terms of the differential $J$-factor and the integrated $J$-factor for annuli around the 
GC with a width of $0.1^\circ$ (corresponding to the resolution of the morphological analysis that we will adopt). 
These $J$-factors have been calculated with {\sf \href{http://www.darksusy.org/}{DarkSUSY}}~\cite{Bringmann:2018lay}, 
and independently cross-checked with standard SciPy~\cite{2020SciPy-NMeth} integration routines. 
For our analysis, we use instead
{\sf \href{https://clumpy.gitlab.io/CLUMPY/}{CLUMPY}}~\cite{bonnivard2016clumpy, charbonnier2012clumpy, Hutten:2018aix} 
to generate HEALPix~\cite{Gorski_2005}-based $J$-factor sky maps of the inner region of the MW. 
Extracting the $J$-factors from these maps, we find (at least) percent-level 
agreement with what is plotted in Fig.~\ref{fig:DM_benchmarks} for annuli centred at $\theta\gtrsim1^\circ$;  
at smaller scales, on the other hand, the values extracted from the HEALPix maps are systematically smaller, at the level 
of $\mathcal{O}$(10\,\%). Pre-empting the general discussion of our results in Section \ref{sec:discussion}, we conclude 
that this discrepancy is a highly sub-dominant source of the overall uncertainty in our final sensitivity estimates,
not the least because we do not include the central 0.3$^\circ$ in our analysis (thus masking the emission 
of Sgr A$^*$, see below) and because the constraining power for a DM signal typically 
originates from a significantly larger sky region than from the innermost $\sim1^\circ$ (as detailed in Appendix \ref{app:InfoFlux}). 

\subsubsection*{Spectral distribution}
\label{sec:DM_spectra}

The dominant source of prompt gamma-ray emission from DM 
is expected to stem from the tree-level annihilation of WIMP(-like)  particles into pairs of leptons, quarks, 
Higgs or weak gauge bosons. The primary annihilation products for non-leptonic 
channels then hadronise and decay, producing secondary photons mainly through the eventual decay of
neutral pions. The resulting photon spectra 
${dN_\gamma^{f}}/{dE_\gamma}$ for a given annihilation channel $f$ can be estimated with event 
generators like {\sf Pythia}~\cite{Sjostrand:2007gs} or {\sf Herwig}~\cite{Corcella:2000bw}. 
Owing to the large multiplicity of pions produced in the event
chains, these spectra are typically of a rather universal form, lacking pronounced features apart
from a soft fall-off towards the kinematical limit $E_\chi=m_\chi$~(see, e.g., Ref.~\cite{Bringmann:2012ez}).
For leptonic final states, in contrast, the production of pions is kinematically impossible 
(or, for $\tau^+\tau^-$, strongly suppressed). The result is a  harder gamma-ray spectrum, 
from final state radiation in lepton decays, with a sharper cutoff at $E_\chi=m_\chi$.

The spectrum from a given two-body annihilation channel is in principle uniquely defined apart 
from intrinsic uncertainties originating from how different event generators implement the 
hadronisation and decay chains~\cite{Cembranos:2013cfa} . The dependence on the DM 
model enters the calculations explicitly when radiative corrections are taken into account, 
which lead to three- (or more) body final states (for a detailed discussion, see Ref.~\cite{Bringmann:2017sko}).
In particular, it is well known that an additional 
photon in the final state can both significantly enhance the annihilation rate and lead to very characteristic
spectral features around the kinematic endpoint at 
$E_\gamma=m_\chi$~\cite{Bergstrom:1989jr,Flores:1989ru,Bringmann:2007nk}, while final state gluons only 
slightly change the photon spectrum expected from
quark final states~\cite{Bringmann:2015cpa}. The effect of an additional electroweak gauge or 
Higgs boson in the final state has also been investigated in 
detail,  again showing a large model-dependence for the resulting particle 
yields~\cite{Kachelriess:2009zy,Ciafaloni:2011sa, Bell:2011if, Garny:2011cj, Garny:2011ii, Bringmann:2017sko}. 
When including {\it electroweak corrections} here, we will do so in a form that is sometimes referred to
as `model-independent'~\cite{Ciafaloni:2010ti} (as implemented in  the `Poor Particle Physicist Cookbook',
\href{http://www.marcocirelli.net/PPPC4DMID.html}{PPPC}~\cite{Cirelli:2010xx}). Specifically, the 
underlying assumption is that the contribution from weakly interacting bosons radiated from the initial 
DM states and virtual internal propagators can be neglected. This is, for example, 
satisfied in contact-type interactions of electroweak singlet DM; for Majorana DM like 
the supersymmetric neutralino, on the other hand, the resulting photon spectra can 
differ substantially~\cite{Bringmann:2017sko}.
It should also be stressed that all radiative corrections mentioned so far only concern leading order effects,
and that there has recently been significant progress in including higher-order effects by consistently treating
leading logarithms~\cite{Beneke:2018ssm,Baumgart:2018yed,Beneke:2019gtg}. 
While these effects start to change the photon 
spectra appreciably for DM masses above the TeV scale, we will not take them into account here. 

In the right panel of Fig.~\ref{fig:DM_benchmarks} we plot the photon spectra from {\sf PPPC} for selected 
benchmark annihilation channels (solid lines). For the case of $W^+W^-$ the effect of the above-described 
implementation of electroweak corrections is largest; we therefore also indicate the spectrum {\it without} these 
corrections (which can be thought of as a very rough means of bracketing the model-dependence of such 
corrections). For comparison, we furthermore show spectra without electroweak corrections obtained 
from \ds\ (dashed lines). 
We note that for soft backgrounds that fall like $d\Phi/dE_\gamma\propto E^{-n_\gamma}$, 
with $n_\gamma\gtrsim2$, 
the dominant contribution of these spectra to DM {\it limits} derives from the energy range where 
$E^{n_\gamma}_\gamma {dN_\gamma}/{dE_\gamma}$ 
(rather than ${dN_\gamma}/{dE_\gamma}$) peaks;
we indicate this by shaded areas for typical soft ($b\bar b$) and hard ($\tau^+\tau^-$) channels,
respectively. This demonstrates that, in the energy range relevant for our analysis, 
uncertainties due to different event generators, as well as how exactly {\it simplified} 
radiative corrections (in the above sense) are implemented, should not significantly affect DM limits 
-- with the notable exception of the second peak that is expected~\cite{Bergstrom:2005ss} near the 
kinematical endpoint for $W^+W^-$ final states.

We finally remark that, in general, DM does not only produce prompt emission of gamma rays as 
described by Eq.~(\ref{DMflux}). For typical annihilation channels, in particular, the 
same processes that
produce prompt gamma rays also produce high-energy leptons, which leads to an {\it inverse Compton} (IC)
component in gamma rays from the upscattering of cosmic microwave background (CMB) 
photons, thermal dust and starlight~\cite{Regis:2008ij,Cirelli:2010xx}. 
For hadronic channels both the fraction and the distribution of energy that goes to electrons is comparable to 
that going into photons,  
but since the upscattered IC photons have on average significantly lower energy than the promptly produced 
gamma rays, the latter are much more easily discriminated against typical backgrounds 
(that fall with energy faster than the signal). 
Multi-TeV DM models annihilating to leptonic channels, on the other hand, 
and to some extent also $W^+W^-$ final states, produce sufficiently hard electrons to lead to a potentially 
distinguishable contribution of IC photons above the CTA threshold 
(see, e.g., Refs.~\cite{Cirelli:2009vg,Belikov:2009cx,Abazajian:2010zb}).
Morphologically, however, the IC signal is much more diffuse (because TeV electrons propagate 
several hundred pc before being stopped~\cite{Delahaye:2010ji}), and hence more difficult to model and 
detect against backgrounds. In fact, predicting the exact IC morphology requires detailed knowledge of 
both starlight distribution and electron propagation near the GC, thus introducing significant modelling 
uncertainty. Here we will therefore not include this component, but note that once there is evidence for a 
prompt DM emission signal, the detection of the associated IC component would 
provide a compelling cross-check of its nature.

\subsection{Conventional astrophysics}
\label{sec:conv_astrop}

Radio and X-ray data reveal the GC to be a very active region with non-thermal emitters such as 
star clusters, radio filaments and Sagittarius A* (see, e.g.,~Refs.~\cite{Genzel:2010zy,Morris:1996th}).  
In gamma rays at energies below $100$\,GeV the GC region is not significantly 
brighter than the rest of the Galactic plane~\cite{Ackermann:2012pya} 
-- despite the high number of confirmed energetic sources and the fact that this region contains about 
10\% of the total Galactic molecular gas content \cite{Morris:1996th,2017arXiv170505332M,2017IAUS..322..164B}. 
At TeV energies, on the other hand,  
IACT~measurements have shown that the GC region clearly stands out, as described below~\cite{Aharonian:2006au,Abdalla:2017xja,Archer:2016ein,Abramowski:2016mir}.
The astrophysical gamma-ray emission consists mainly of {\it i)} interstellar emission (IE) produced as secondary 
emission from CRs interacting with the interstellar medium 
(gas and dust, ISM) and fields (in particular the interstellar radiation field, ISRF), 
{\it ii)} localised gamma-ray {sources}\footnote{%
{Throughout this work we use the term `sources' rather loosely, implying (catalogue) objects that can 
be either {\it point-like}  or  {\it extended}.  Even though IE is also a {source} of gamma 
rays, strictly  speaking, we thus mostly use the term here to distinguish (intrinsically) {\it localised} from 
{\it diffuse} emission.  }
} 
 (both individually resolved, and a cumulative emission from a sub-threshold population)
and {\it iii)} possibly the emission from the base of the Fermi bubbles, all described in more detail below. 
Note that all these components (except for individually resolved sources) are extended and 
together also sometimes collectively referred to as Galactic Diffuse Emission (GDE).

\subsubsection*{Interstellar emission}
\label{subsec:intro_GDE}

The IE extends along the Galactic plane and is the brightest emission component in the Fermi-LAT data \cite{Ackermann:2012pya}.
At high energies, the most relevant processes are {\it i)} CR interactions with the ISM gas, 
producing gamma rays predominantly through the neutral pion channel,
and {\it ii)} IC scattering, in which CR 
electrons up-scatter ISRF and CMB photons to gamma-ray energies. These two components have distinct 
morphologies: the first, so-called `hadronic emission' tracks that of the gas, while the morphology of IC 
emission is determined by the distribution of CR leptons and the ISRF. Theoretical modelling of the 
IE emission depends on a significant number of parameters related to the injection spectra of 
CRs, their spatial  distribution in the Galaxy, diffusion properties but also properties of the interstellar 
medium and ISRF. This implies a high level of modelling uncertainty, adding to significant degeneracies 
between some of the involved parameters (for a review see Ref.~\cite{Ackermann:2012pya}).        

Despite its relative brightness, the spatial extension of the IE has made it notoriously difficult to be detected 
with IACTs  (partially motivating efforts to determine TeV diffuse emission using the
decade-long exposure of  Fermi-LAT~\cite{Neronov:2019ncc}). 
Notable progress has been made by H.E.S.S.~through the detection of a bright diffuse gamma-ray emission
from the so-called Galactic Ridge, originating from dense molecular clouds in the 
central 200 pc~\cite{Aharonian:2006au,Abdalla:2017xja,Archer:2016ein}, as well as that of cumulative emission 
from the Galactic plane {(providing latitude and longitude profiles, but not spectral 
information~\cite{Abramowski:2014vox}).  Advances in determining the large-scale diffuse emission along the 
Galactic plane have also been made by water-Cherenkov telescopes at higher energies, first with 
MILAGRO~\cite{Abdo:2008if}  and more recently with HAWC~\cite{Zhou:2017lgv}, 
though mostly at longitudes $\gtrsim30^{\circ}$ due to the geographical location of these instruments.
CTA is expected to spatially resolve large-scale emission at TeV energies with unprecedented precision 
and angular resolution. 
We will  test a set of specific IE models (IEMs) to gauge the impact of the 
associated modelling uncertainties on the sensitivity to a DM signal.
These models are chosen to sample a representative range of 
realistic theoretical possibilities, given the current state of knowledge:

\begin{itemize}
\item Gaggero {\it et al.}~\cite{Gaggero:2017jts} studied diffuse models that can  simultaneously explain 
H.E.S.S.\ data in the region of the Galactic Ridge, and Fermi-LAT data in the surrounding region, 
at lower energies. They discuss two possibilities to reconcile these measurements:
 
\begin{itemize}

\item The {\bf \emph{Base}} model rests on often-adopted, simplifying assumptions concerning CR diffusion, 
and in particular assumes a constant diffusion coefficient across the Galaxy. The large scale diffuse emission 
measured by Fermi-LAT and the emission from the Galactic Ridge (H.E.S.S.) then must have a different origin, 
with the latter postulated to originate from a so-far unknown source related, e.g., to 
transient emission from the GC. 
In our analysis, we thus simply add the H.E.S.S.~Ridge~template~\cite{Aharonian:2006au} to the original
form of the  Base model (see also Appendix \ref{app:Ridge}). 
The large-scale emission predicted in this model is very soft, nominally already in some tension with 
the Fermi-LAT data~\cite{Gaggero:2017jts} (which would be alleviated in the presence of yet another 
emission component like, e.g., unresolved sources). 

\item The {\bf \emph{Gamma}} model relaxes the assumption of a spatially constant CR diffusion coefficient, 
allowing instead for a radial dependence where diffusion is more efficient closer to the GC. This implies 
harder and brighter gamma-ray emission in the innermost Galactic regions, explaining 
simultaneously the bright Galactic Ridge in the very centre and the large scale diffuse emission measured by 
Fermi-LAT.  This model predicts a relatively bright emission also outside the ridge. 

\end{itemize}

\item  The {\bf\emph{Pass-8 Fermi IEM}} was derived based on the detailed analysis of {8} years {\it Pass 8} Fermi-LAT data.\footnote{%
Namely gll$\_$iem$\_$v07 model, available at 
\url{https://fermi.gsfc.nasa.gov/ssc/data/access/lat/BackgroundModels.html}.  For  details see 
\url{https://fermi.gsfc.nasa.gov/ssc/data/analysis/software/aux/4fgl/Galactic_Diffuse_Emission_Model_for_the_4FGL_Catalog_Analysis.pdf}} 
Special care was 
taken for the model to describe the high-energy ($\geq 50$ GeV) spectrum, and it is, therefore,
more reliable regarding high energy extrapolation than previous versions. It uses different gas maps than the 
Gaggero {\it et al.} models and is tuned to the LAT data over the entire sky (as opposed to the 
{\it  Base} model, which is more heavily based on theoretical expectations).\footnote{%
Let us stress here that even though both the {\it Pass-8 Fermi IEM} and the {\it Gamma} model result from  
a fit to data, only the procedure for determining the former includes a template for sub-threshold point 
sources. The {\it Fermi IEM} should thus indeed exclusively describe IE, while the {\it Gamma} 
model may implicitly include a contribution from sub-threshold sources. Note that the \emph{Base} and 
\emph{Gamma} IEMs are based on HI gas maps  with a resolution of 0.5 deg, while the \emph{Pass8} 
model uses improved HI maps with $16'$ resolution.}
\end{itemize}

\subsubsection*{Resolved and sub-threshold sources}
\label{subsec:intro_point_sources}

The Galactic plane survey of the H.E.S.S.~collaboration
has discovered six TeV-bright objects in the GC region~\cite{H.E.S.S.:2018zkf}, that were later studied also with MAGIC and VERITAS. In particular, G0.9+0.1, 
HESS J1745-290 and 
HESS J1741-302 are best fit as point sources, while HESS J1745-303 and HESS J1746-308 are extended 
sources (with an extension of $0.2^{\circ}$ and $0.15^{\circ}$, respectively\cite{Abdalla:2017xja}); 
G0.9+0.1 is identified as a composite supernova remnant hosting a PWN in its core (see also the 
VERITAS analysis~\cite{Archer:2016ein}) while HESS J1745-290 
coincides with the position of Sagittarius (Sgr) A$^{\ast}$ (c.f.~further characterisations by 
MAGIC~\cite{Ahnen:2016crz} and VERITAS~\cite{Archer:2016ein}), 
the supermassive black hole at the centre of the MW. 
The sixth source, HESS J1746-285, is very close to two TeV-emitting sources detected by 
VERITAS~\cite{Archer:2016ein} and MAGIC~\cite{Ahnen:2016crz}. However, as reported by 
H.E.S.S.~\cite{Abdalla:2017xja}, this source is possibly the combination of a part of the Galactic 
ridge and a yet unknown emitter. Hence, we do not consider this source in our analysis. For 
the remaining five sources we adopt circular masks centred on the respective 
source position (taken from Ref.~\cite{Abdalla:2017xja}), with an 
energy-independent radius of $0.3^{\circ}$ for point sources and $0.6^{\circ}$ for extended sources. 
Our masking scheme is indicated in Fig.~\ref{fig:observations_schematic}.

The Galactic centre region presumably also houses many sources that are too faint to be detected 
with the current generation of IACTs, as well as a component of even fainter sources, below the CTA 
detection threshold, that will contribute to the diffuse emission. For example, while the Fermi-LAT catalogue 
of hard sources (2FHL\cite{Ackermann:2015uya}) lists no sources in our ROI,  
4FGL~\cite{Fermi-LAT:2019yla} lists 16 identified sources, three of which are 
tagged as candidate TeV emitters listed in the online TeV source catalogue TeVCAT~\cite{tevcat2}. Since  
the CTA source detection threshold is still unknown especially in crowded regions like  the Galactic centre, 
we will use a single template for {\it all} but  the brightest sources detected by current IACTs.

Given that there are only a few TeV sources currently known in the region, predicting the contribution of faint 
sources comes with considerable uncertainty. Within the context of the CTA Galactic plane survey, 
significant consortium effort has recently gone into modelling the population properties of the most numerous 
Galactic TeV sources (PWNe, SNRs and binaries)~\cite{GPSPreliminary}. 
We use the gamma-ray templates derived in that work (applying a $\mu$Crab lower flux threshold, while for the higher flux cut-off we used the detection  threshold from the H.E.S.S. Galactic plane survey, \cite{2018A&A...612E...1F}, namely, for the GC region, a flux of 5 mCrab at energies $> 400$ GeV) and refer for all details to that upcoming publication.
We will discuss the impact of that template on our analysis in Sec \ref{sec:PS_FB}.

\subsubsection*{Fermi bubbles}
\label{sec:FB}

The gamma-ray emission from the Fermi bubbles (FB) has been studied extensively since their discovery in 
2010~\cite{Su:2010qj,Fermi-LAT:2014sfa}. Even though the FB outshine the IE at high latitudes 
due to their hard spectrum, 
their shape close to the Galactic plane is challenging to distinguish from the bright IE. Here we will rely 
on a recent analysis~\cite{TheFermi-LAT:2017vmf} determining the morphology and spectrum 
at the base (i.e.~the low-latitude part) of the FB, and use these spatial and spectral templates to gauge 
the potential impact of the FB on the search for DM signals. A projection of CTA's sensitivity to the base
of the FB based on the same spatial and spectral template has been derived in Ref.~\cite{Yang:2018bfb}.
We stress however that the exact shape and spectra of the FB close to the GC are highly uncertain
-- though a re-examination of the FB base using nine years of LAT data~\cite{Herold:2019pei} 
confirmed the previously reported hard power law without indications of a cutoff up to energies of 1\,TeV. 

\subsection{Residual cosmic-ray background}
\label{sec:CR_bckg}

CR events misidentified as gamma rays make up the highest portion of detected events, 
outshone only by the brightest sources. The core of the issue is that the CR proton (electron) 
fluxes are $10^4$ ($10^2$) times higher than the diffuse flux of gamma rays expected from the Galaxy 
(at $\sim100$ GeV).  While hadron-induced showers can be distinguished from electromagnetic showers 
based on their shape, with an (energy-dependent) background rejection rate better than $10^{-2}$,
CR electrons present an essentially irreducible background 
(preliminary studies indicate that some rejection may be possible~\cite{electrons1}, but not on short 
time scales). Besides, while the spectrum of CR protons and electrons is well measured 
below a few TeV \cite{Aguilar:2014fea,Aguilar:2014mma,Adriani:2018ktz,Ambrosi:2017wek}, significant uncertainties 
about the number of events passing all analysis cuts remain, making the exact spectrum and normalisation
of this {\it  intrinsically isotropic} component challenging to model
(the {\it measured} distribution of events within CTA's field of view, in contrast, is determined by the 
instrument response to cosmic-ray background, which is {\it not} isotropic -- especially at high energies).
On top of this, the atmosphere itself, acting as an effective calorimeter, introduces additional uncertainties. 

We will see, however, that uncertainties in isotropic parts of the background components mostly affect our analysis by 
changing the signal-to-noise ratio, which in fact turns out to be a subdominant effect. A bigger impact on 
the  DM sensitivity results from varying (in time and space) unresolved backgrounds, for example 
small-scale anisotropies in an otherwise largely isotropic emission.
These could originate, for example, by the presence of aerosols in the atmosphere which can also introduce a strong bias in energy reconstruction 
and deteriorate the energy resolution (even though this is to some degree addressed by dedicated studies of atmospheric conditions by 
CTA monitoring instruments).

In our analysis, the modelling of the misidentified CR component relies on extensive Monte Carlo simulations 
of CR showers and their subsequent event reconstruction, allowing us to obtain the expected number of CR 
misidentified events for a given set of IRFs.
(This is in contrast to the more conventional `ON/OFF' technique\footnote{%
In this work we use the term `ON/OFF' in a sense often seen in the DM context, referring to 
the existence of `ON'-signal and `OFF'-background measurement regions. In the wider IACT 
community, in contrast, the term sometimes refers to an observation mode where the ON region is
at the centre of the FoV, while the OFF region is not  taken during the same observation period
-- to distinguish it from 
{\it wobble} mode observations, where ON and OFF regions are both chosen off centre and measured 
during the same observation period~\cite{1994APh.....2..137F}.
\label{foot:onoff}
} 
 which does not rely on MC
simulations and makes it instead possible to adjust the CR background model directly to the data; see, 
e.g., Ref.~\cite{Knodlseder:2019coy, Mohrmann:2019hfq}.)
The underlying IRFs do not include small-scale anisotropies (which is an issue shared with the ON/OFF
technique), which might be present in the 
real data due to, e.g., uneven atmospheric conditions.
Because the corresponding systematic uncertainties  have not yet been studied in 
detail, we will include them in a parametric way (as described in Section~\ref{sec:data}).


\begin{figure}[t!]
\centering\includegraphics[width=0.9\linewidth]{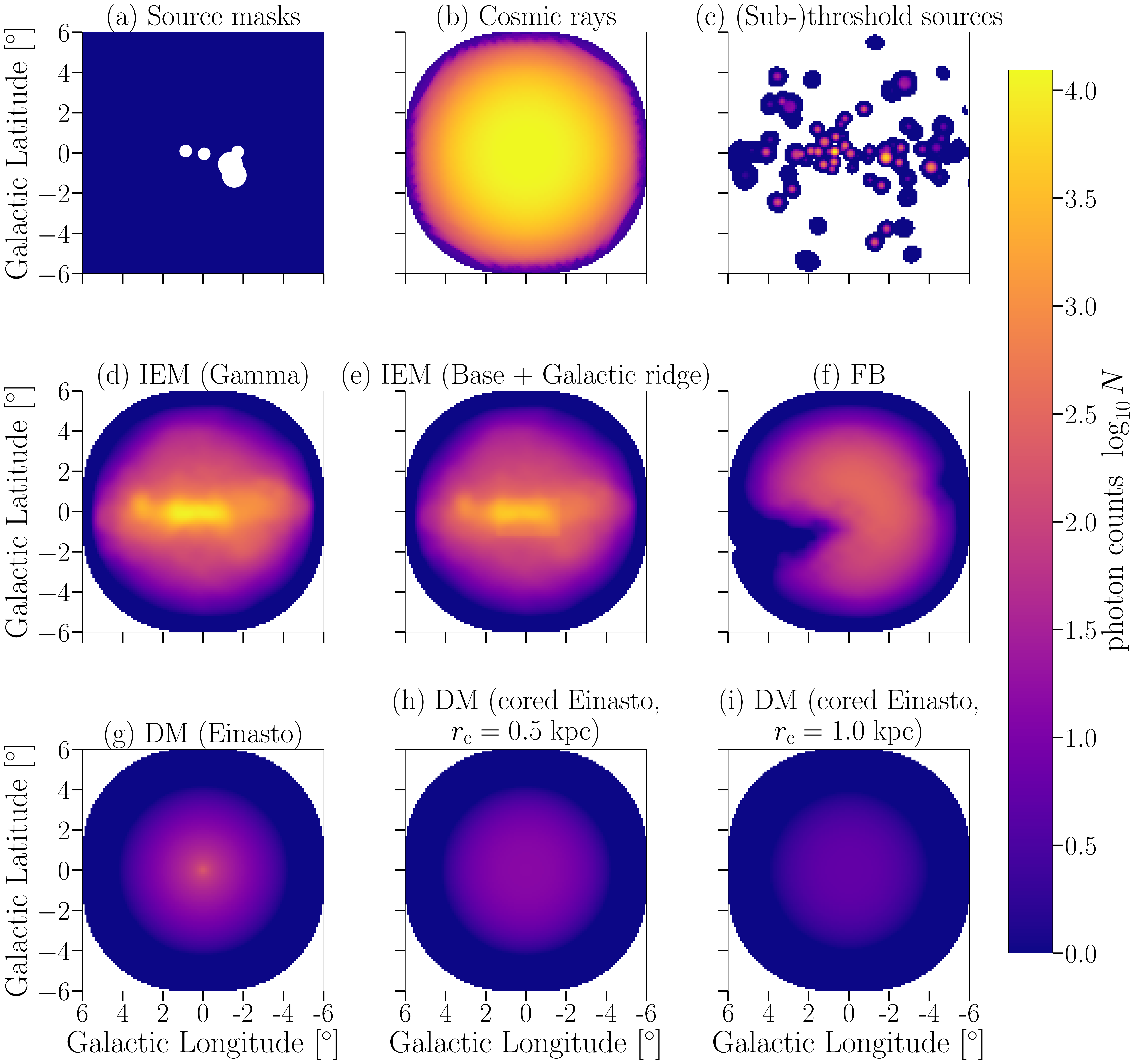}
\caption{Background and signal templates computed by {\sf ctools} for the GC survey observation, 
showing the expected photon counts in the energy range from 100 to 500 GeV. The
(logarithmic) colour code indicates the number of expected counts $N$ per $0.1^\circ\times 0.1^\circ$ pixel.
See text for a description of each of the individual templates shown here.
 \label{fig:template_collection}}
\end{figure}

\subsection{Emission templates and caveats}
\label{sec:caveats}

To summarise our discussion of emission models, we compare in Fig.~\ref{fig:template_collection} the 
total count maps in the 100 -- 500\,GeV range that result from our benchmark emission templates
(as generated by {\sf ctools}, for the GC survey mode described in Section ~\ref{sec:GC_obs_strategies}). 
From top left to bottom right, these correspond to:

\begin{itemize}
\item  residual CR background events, generated from \texttt{prod3b-v1} IRFs (Section \ref{sec:GC_obs_strategies})
\item interstellar emission, as  predicted in the {\it Gamma} and the {\it Base} model 
(Section \ref{subsec:intro_GDE})
\item a realisation of sub-threshold sources (Section \ref{subsec:intro_GDE})
\item the Fermi bubbles (Section \ref{sec:FB})
\item the DM emission template (Section \ref{sec:DM_models}) for the Einasto  
profile with and without a constant density core, as indicated.
For definiteness we choose here $m_\chi=2$\,TeV for the DM mass, and an annihilation cross-section 
$\langle\sigma v\rangle= 3\times10^{-26}\;\textrm{cm}^3\textrm{s}^{-1}$ to $b\bar b$ final states.
\end{itemize}
In Fig.~\ref{fig:flux-allcomponents-2} we show the energy dependence of the various
components, by plotting the total number of expected counts, during the entire Galactic centre 
survey observation, per energy bin. We first note a relatively sharp increase in the number of counts
for {\it all} components at energies $\gtrsim 60$\,GeV; the origin of this feature is a corresponding increase in 
the effective area of the array, as we pass above the MST energy threshold.
When it comes to the comparison of the various physical components, furthermore, Figs.~\ref{fig:template_collection} 
and \ref{fig:flux-allcomponents-2}  call for a number of pertinent comments:
\begin{enumerate}
\item CR contamination clearly dominates all other emission components. 
The CR electron flux up to TeV energies has been well-measured by a number of instruments, including 
AMS-02~\cite{Aguilar:2014fea}, the Calorimetric Electron Telescope (CALET)~\cite{Adriani:2018ktz}, and 
the  Dark Matter Particle Explorer (DAMPE)~\cite{Ambrosi:2017wek}.  In the figure we show the 
DAMPE spectrum to guide the eye as to the level of expected background. CR electron fluxes
constitute, as discussed in Section \ref{sec:CR_bckg}, 
an essentially irreducible background to gamma-ray searches. Given the importance of electrons
up to the TeV energy range, it thus will be particularly hard to further improve the CR rejection
efficiency at these energies.

\item The {\it Gamma} and {\it Base} IEMs are based on the same target gas and  ISRF maps, but on 
different assumptions concerning CR diffusion. This results both in different spectra and in different 
morphologies, with the {\it Gamma} model being significantly brighter in the central regions. 
For comparison,  {\it Pass 8 Fermi} IEM (only shown in Fig.~\ref{fig:flux-allcomponents-2}, not in 
Fig.~\ref{fig:template_collection}) 
features a flux very similar in spectrum and normalisation to that of the {\it Gamma} model, however it is based 
on different target gas and  ISRF maps, as well as on  different assumptions about 
CR diffusion  (the morphologies of templates based on different IEMs are compared in more detail in  
Appendix \ref{app:GDEmorphology}).

\item Unresolved sources and FB are among the most uncertain emission components,
and a mis-modelling of their morphology could potentially mimic, at least partially, the DM template.
This is aggravated by the fact that the fluxes of these components are at least 
comparable to that from the annihilation of thermally produced DM.
Potentially, this could thus have a significant impact on DM searches,  causing fake signal detections or
artificially strong limits. Note that a separate study of sub-threshold sources for CTA is still ongoing and 
we hence only use the specific realisation of such a population shown in Fig.~\ref{fig:template_collection}c;
the eventual analysis of real data, beyond the scope of this work,  will have to be based on an 
average over many sub-threshold source realisations. 
We return to these issues in Section \ref{sec:discussion}. 
\end{enumerate}

\begin{figure}[t!]
\centering
\includegraphics[width=0.85\linewidth]{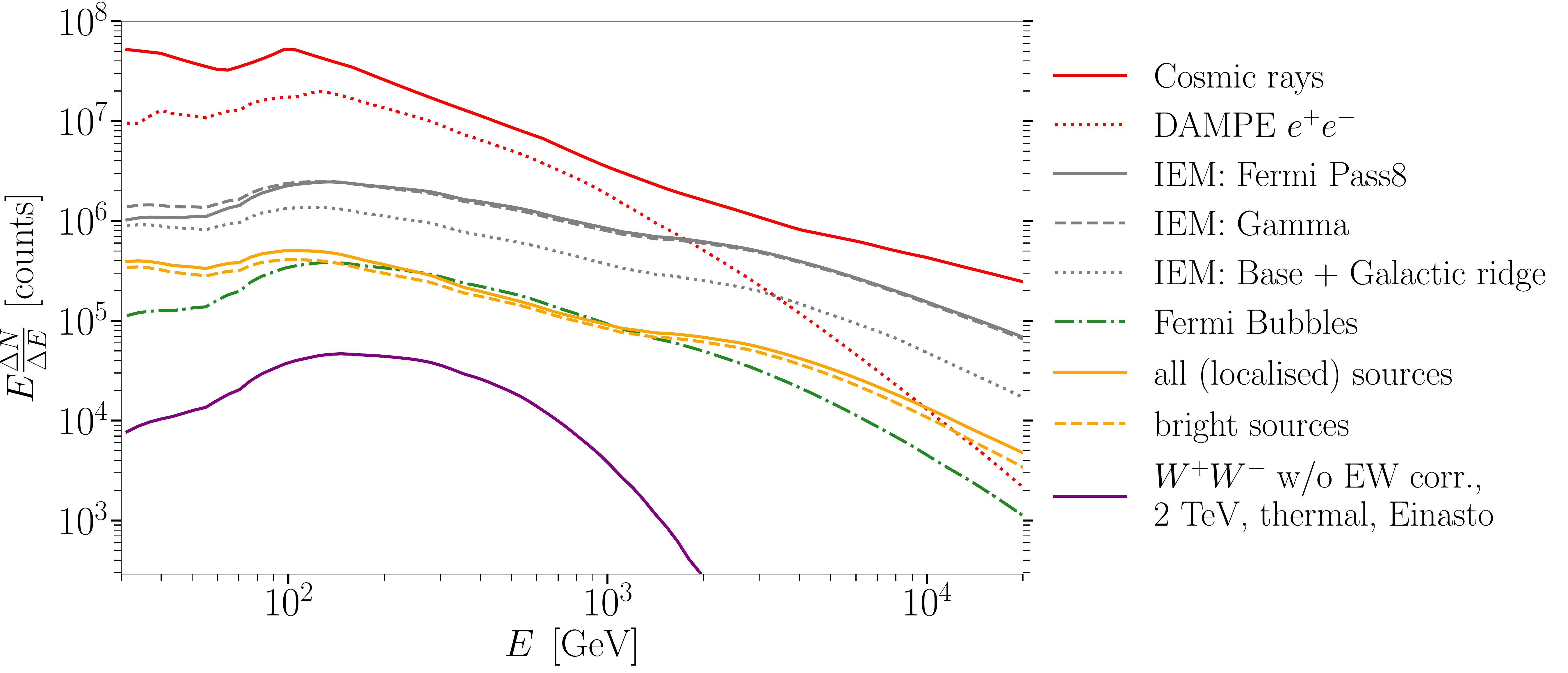}
\caption{Differential counts (per bin) expected from the GC survey, multiplied by the bin energy $E$,
for the emission components considered in this work -- including the total CR background (from electrons
and hadrons, solid red line), 
three alternative IE models, localised source components (both bright individual sources and unresolved
sources), Fermi bubbles and the DM spectrum (assuming $m_\chi=2$\,TeV, 
$\langle\sigma v\rangle= 3\times10^{-26}\;\textrm{cm}^3\textrm{s}^{-1}$ and dominant annihilation to 
$W^+W^-$). 
For a very rough translation of these counts to fluxes, one needs to divide by the CTA
effective area
($9\cdot10^5$\,m$^2$ at 1\,TeV~\cite{CTA_performance}),
the angular size of our analysis ROI (0.037\,sr, not including direction cuts) and the total observation time of 525\,h.
\label{fig:flux-allcomponents-2}}
\end{figure}

We conclude this section by mentioning another aspect of astrophysical modelling that may
appear as a relevant issue once the analysis chain is confronted with real data, but which would be 
premature to include in the 
present modelling of emission components given the current lack of knowledge and robust data. 
The IC component of the interstellar emission, in particular,
is more difficult to model since it does not, unlike hadronic emission, correlate with gas 
maps. Besides, the IEMs used here (and more generally in the majority of the relevant literature) 
assume {\it steady-state} solutions for CR propagation, based on smoothly distributed source populations.  
That assumption is expected to fail at energies $\gtrsim 100$\,GeV because of the small energy loss time of 
electrons,  implying that the morphology of the IC emission changes significantly and becomes sensitive 
to the CR electron injection history~\cite{Porter:2019wih}. In particular, electrons are contained closer to 
the sources, which in turn introduces a significant granularity in the IC templates and lowers the 
strength of large-scale IC emission by up to about $30\%$~\cite{Porter:2019wih}. While the latter 
effect would facilitate the detection of a DM signal,  the former (i.e.~the difficulty to model overlapping 
`point-like' emission sources) could present a non-negligible challenge for future  DM searches at 
these energies. We leave a more detailed study of these aspects for future works, but note that the 
difference between {\it Base} and {\it Gamma} models should capture (in part) the 
impact of the latter effect on the DM sensitivity.


\section{Data Analysis}\label{sec:methodology}
\label{sec:data}

The traditional way to constrain DM annihilation with IACTs is the so-called {ON/OFF} approach
(e.g.~\cite{Lefranc:2015pza,Pierre:2014tra,Silverwood:2014yza,Abdallah:2016ygi,Acharya:2017ttl}), 
which rests on the definition of two spatially separate, different kinds of ROIs (often within the same FoV):
in the `ON' region the  signal is expected to be the strongest while 
in the `OFF' region it is expected to be subdominant. Under the hypothesis that we know how the background scales between OFF and ON regions 
(solid angle/acceptance effects are routinely corrected for; see, for instance, the $\kappa$ 
factors in Eq.~(\ref{eq:ONOFFlikelihood})),
it is, in principle, possible to {\it measure} the 
background under the same observational conditions.
Such an approach is complementary to {template-based}
morphological analyses, more typical in the context of 
satellite-borne instruments (e.g.~\cite{Charles:2016pgz}), 
where different emission components are described by templates that are fitted to binned data.
While the template analysis offers the possibility to incorporate spatially varying backgrounds,
there can be a remaining systematic uncertainty 
related to the exact form of the adopted templates
(for attempts to address these limitations see e.g.~SkyFACT~\cite{Storm:2017arh}).
Possible reasons for not using the template approach in most past IACT analyses include 
{\it i)} their relatively small 
FoV 
{\it ii)} the residual CRs being the only 
background component, assumed to be effectively the same in the ON and OFF regions at the 
energies of interest here; 
and {\it iii)} the complexity of robustly modelling this background.

Only more recently it was realised that template fitting may be a powerful technique for the analysis of IACT 
data~\cite{Knodlseder:2019coy, Mohrmann:2019hfq} (see also Ref.~\cite{Lefranc:2015pza} for a 'hybrid' approach).
To fully exploit the power of CTA with its larger FoV, higher background  rejection and 
higher flux sensitivity compared to previous experiments, and to achieve a corresponding increase 
in DM sensitivity, the background needs to be modelled in higher detail and with more components 
than required for current instruments.
So far, astrophysical modelling was not done in a very detailed way and CR uncertainties were 
mostly treated in a simplified manner~\cite{Silverwood:2014yza,Balazs:2017hxh}. 
One of the main motivations of this work is 
to study the applicability of the template fitting 
approach in detail (later, in Appendix \ref{app:onoff},  we will also directly confront this method with the 
traditional ON/OFF approach).

\paragraph{Template Analysis} 
\label{sec:morph_analysis_pipeline}

We employ a binned likelihood based on Poisson statistics
$\mathcal{L\!}\left(\left.\bm{\mu}\right|\bm{n}\right)=\prod_{i,j} e^{-\mu_{ij}} {\mu_{ij}^{n_{ij}}}/{\left(n_{ij}\right)!}$,
where $\bm{\mu}=\{\mu_{ij}\}$ denotes the model prediction and $\bm{n}=\{n_{ij}\}$ the (mock) data counts, for bins 
in energy (indicated by an index $i$) and angular position on the sky (indicated by an index $j$). 
The model is given by a set of background templates as shown on Fig .~\ref{fig:template_collection}, 
$\{\mu_{ij}^X\}$, a signal template
for the DM component, $\mu_{ij}^\chi$, and normalisation parameters $\bm{A}$ for the relative weight of these
templates:
\begin{equation}
\mu_{ij}(A^{\chi},A^X_{i})=A^{\chi}\mu_{ij}^{\chi}+\sum_X A^X_{i}\mu_{ij}^X\,.\label{eq:model_counts}
\end{equation}
For any given signal template -- defined by the adopted DM density profile and annihilation spectrum --
we thus introduce a global normalisation parameter $A^{\chi}$ that is directly proportional to the annihilation 
strength $\langle\sigma v\rangle$ that we want to constrain, c.f.~Eq.~(\ref{DMflux}).  For the background 
components $X$ -- CRs, IE, Fermi bubbles and unresolved sources, depending on the 
analysis benchmark -- we instead adopt 
normalisation parameters $\{A^{\rm X}_i\}$ that may vary in each energy bin, where $A^{\rm X}_i\equiv1$ 
corresponds
to the (expected) default normalisation of the templates as summarised in Section \ref{sec:caveats}.
This ansatz accounts in an effective way for uncertainties in the spectral properties
 of the templates, thereby rendering the resulting DM limits more conservative. 
 It should be stressed that {\it by construction} this method thus relies more on the morphological than on the 
 spectral information in the templates, which is partially motivated by the excellent angular resolution
 of CTA. We will discuss this point in more detail below when explicitly introducing
  systematic uncertainties.
 Mock data, finally,  are prepared for each of the background components $X$ by drawing
 the number of photon counts in a given bin, $n_{ij}^{X}$, from a Poisson distribution with mean $\mu_{ij}^{X}$. 
 Summing these contributions then gives the total number
of counts per bin, $n_{ij}=\sum_X n_{ij}^X$.

We generate all count maps using {\sf ctools}.\footnote{%
We use {\tt ctmodel} to obtain 3D data cubes with the mean photon counts of each emission template, 
and {\tt ctobssim} to produce an event list (both in the form of .fits files) containing MC realisations of the data. 
}
As our benchmark binning scheme we choose -- unless explicitly stated otherwise
(see also Section \ref{sec:bm_sys} for a discussion) -- square spatial bins of width $0.1^{\circ}$,
roughly corresponding to the typical PSF,
and 55 spectral bins in the range from 30 GeV to 100 TeV chosen such that their width is given by 
the energy resolution at the central bin  energy, at the two standard deviations ($2\sigma$) level.\footnote{%
For our standard IRFs, this corresponds to a bin width of $\Delta E/E=0.52$ for the 
lowest energy bin, decreasing to $\Delta E/E=0.12$ at $E\sim4$ TeV, before increasing again to
$\Delta E/E=0.17$ at the high-energy end.
}
We restrict our analysis to circular FoV regions with a radius of 5$^\circ$ around the respective
pointing direction of the array (c.f.~Fig.~\ref{fig:observations_schematic}).

To derive an upper bound on the DM normalisation $A^{\chi}$, for a fixed DM template $\bm{\mu}^\chi$ and a 
given data set $\bm{n}$, we define the test statistic
\begin{equation}
\label{eq:TS_stat_reach}
\textrm{TS}\!\left(A_{\chi}\right)=\min_{\{A_i^X\}}\left(-2\ln\!\left[\frac{\mathcal{L}\!\left(\left.\bm{\mu}(A^\chi,A_i^X) \right|\bm{n}\right)}{\mathcal{L}\!\left(\left.\bm{\hat{\mu}}\right|\bm{n}\right)}\right]\right)\,,
\end{equation}
where $\bm{\hat{\mu}}\equiv\mu(A^\chi=\hat A^\chi, A^X_i=\hat A^X_i)$ denotes the model counts in 
Eq.~(\ref{eq:model_counts}) for the best-fit values of {\it all} normalisation parameters (i.e.~both for DM and 
background components) obtained by maximising the likelihood. 
This test statistic is distributed 
according to a $\chi^{2}$-distribution with one degree of freedom~\cite{Wilks:1938dza}, 
so a (one-sided) upper limit on $A^{\chi}<A^{\chi}_\mathrm{max}$ at 
$95\%$ ($99\%$) Confidence Level (C.L.) corresponds to a TS value of 2.71 (5.41).

It is straightforward to extract the {\it mean expected limit}, $\langle A^{\chi}_\mathrm{max}\rangle$, and its 
variance, 
$\sigma^2=\langle A^{\chi}_\mathrm{max}-\langle A^{\chi}_\mathrm{max}\rangle\rangle^2$, 
by compiling Monte Carlo realisations of mock data sets, and then take limits for each of those 
according to the above prescription. As this is computationally rather intensive, however, we will 
instead typically utilise a single `representative' set of data, the so-called \emph{Asimov data set}, 
$\bm{n}_A$: for a Poissonian process, this corresponds to the expected number of counts 
per bin one would obtain with an infinitely large sample of individual Poisson realisations of a 
given background or signal model, i.e.~$\bm{n}_A=\mu(A^\chi=0,A^X_{ij}=1)$~\cite{2011EPJC...71.1554C}. 
In principle, this approach can also be used to estimate the variance of the expected upper limits. However, 
we checked that in its simplest implementation~\cite{2011EPJC...71.1554C} this does not lead to a reliable
estimate once systematic uncertainties (to be discussed below) are taken into account; whenever we present
`sidebands' to expected limits, these are thus based on full Monte Carlo calculations.

\paragraph{Treatment  of Systematic uncertainties}
\label{sec:systematics}

For a future experiment, instrumental systematic uncertainties are by nature hard to quantify. However, 
we can still estimate the possible effects in a general manner by introducing uncertainties that are 
correlated among the data bins (as is typical for instrumental systematic errors).
Similarly, correlated systematic errors can also account for additional systematic uncertainties in the IEM 
templates that are not already captured in the template analysis.
Such correlated uncertainties may partially degrade morphological differences of the 
background/signal templates and, hence, weaken their constraining power over
the signal component.

Correlated Gaussian uncertainties (with zero mean) are fully defined in terms of their covariance matrix, $K$. 
For our purposes, this may encompass 
\begin{itemize}
\item[\textsc{(i)}] spatial bin -- spatial bin correlations, 
\item[\textsc{(ii)}] energy bin -- energy bin correlations and/or 
\item[\textsc{(iii)}] spatial bin -- energy bin correlations.
\end{itemize}
As described below, we will only consider the first two types of correlations. 
To apply the covariance matrix description of systematic errors, we follow 
the approach outlined in Refs.~\cite{Edwards:2017mnf, Edwards:2017kqw}, and implemented in the 
publicly available Python package \sword~\cite{www_swordfish}.
In particular, we change the construction of the model prediction in Eq.~(\ref{eq:model_counts}) 
(but not that of the data $\bm{n}$) in the following way: Instead of varying the background templates 
by normalisation parameters $A_i^{\rm X}$ per energy bin to account for background fluctuations, 
we set these normalisation parameters to unity and explicitly introduce Gaussian `background perturbations' 
$\Delta B$ -- related, e.g., to uncertainties of the reconstruction of events -- 
for each individual template bin $k$,
\begin{equation}
\left(\mu_K\right)_k\equiv \sum_X \mu_{k}^X+\Delta B_k + A^{\chi}\mu_{k}^{\chi}\label{eq:model_counts_cov}\,.
\end{equation}
Here, the sum runs over the model templates $X$ to be examined, the index $k$ comprises both spatial and energy bins, i.e.~$k\in[1,\mathcal{N}]$ with 
$\mathcal{N}$ being the product of the number of spatial pixels and the number of energy bins.  
In principle, the different templates can give rise to different background perturbations, 
i.e.~$\Delta B_k = \sum_X\Delta B_k^X$. 
Including the Gaussian prior on the background variations $\Delta B_k$ in the likelihood function 
(and neglecting a constant determinant) then yields
\begin{equation}
\mathcal{L\!}\left(\left.\bm{\mu}\right|\bm{n}\right)=\prod_{k=1}^{\mathcal{N}}\frac{\mu_k^{n_k}}{\left(n_k\right)!}e^{-\mu_k}\times \exp\!\left[-\frac{1}{2}\Delta B_k\sum_{l=1}^{\mathcal{N}}\left(K^{-1}\right)_{kl}\Delta B_l\right]\,,
\end{equation}
where $K_{ij} \equiv \langle \Delta B_i \Delta B_j \rangle$ is the covariance matrix 
(and we assume $\langle \Delta B_i \rangle = 0$).
Profiling over the nuisance parameters $\Delta B_i$, this reduces to a log-likelihood function that only depends on the 
signal normalisation $A^\chi$ (again omitting terms that are constant in the model parameters): 
\begin{equation}
-2\ln{\mathcal{L}\!\left(\left.\bm{\mu_K}\right|\bm{n}\right)}= 
\min_{\bm{\Delta B}}\left\{\sum_{k=1}^{\mathcal{N}}\left[\vphantom{\int}n_k\ln{\left(\mu_K\right)_k} - \left(\mu_K\right)_k\right] 
- \frac{1}{2}\sum_{k,l=1}^{\mathcal{N}}\left[\vphantom{\int}\Delta B_k \left(K^{-1}\right)_{kl} \Delta B_l\right]\right\}\rm{.}
 \label{eq:loglikelihood_cov}
\end{equation}
For systematic uncertainties that are {\it uncorrelated} between the background templates $X$, which 
is the case we consider here, we have $\langle \Delta B_i^X \Delta B_j^Y \rangle = 0$ for $X\neq Y$. The last term 
in the above equation can then be written as $\sum_{k,l}\Delta B_k \left(K^{-1}\right)_{kl} \Delta B_l=
\sum_X\sum_{k,l}\Delta B_k^X \left(K^{-1}\right)^X_{kl} \Delta B^X_l$, where $K=\sum_X K^X$ 
is now understood to be the {\it total} correlation matrix.

Upper limits on the DM signal are derived by constructing a test statistic in full 
analogy to Eq.~(\ref{eq:TS_stat_reach}), \emph{mutatis mutandis}. 
Concerning the concrete construction of covariance matrices, the simplest way to parameterise spatial 
correlations is by an
 $\mathcal{N}_{\rm S}\times\mathcal{N}_{\rm S}$ matrix $K_{\rm S}$, with
\begin{equation}
\left(K_{\rm S}\right)_{jj'} = \sigma_{{\rm S}}^2 \exp\!{\left(-\frac{1}{2}\frac{\norm{\vec{r}_j - \vec{r}_{j'}}^2}{\ell_{\rm S}^2}\right)}\rm{,}
\end{equation}
where $\mathcal{N}_{\rm S}$ refers to the number of spatial bins in the ROI, 
$\sigma_{{\rm S}}$ denotes the magnitude of the spatial systematic uncertainty,  $\ell_{\rm S}$ the 
spatial correlation length, $\vec{r}_j$ is the central position of the $j-$th spatial template bins 
in degrees of Galactic longitude and latitude, and we use the norm on the unit sphere for the distance 
between two spatial bins.
$\sigma_{{\rm S}}$ and $\ell_{\rm S}$ may in general depend on the position in the template but,  for 
simplicity, we assume them to be constant here. By analogy, energy correlations can be parameterised by an
 $\mathcal{N}_{\rm E}\times\mathcal{N}_{\rm E}$ matrix 
\begin{equation}
\left(K_{\rm E}\right)_{ii'} = \sigma_{{\rm E}}^2 \exp\!{\left[-\frac{1}{2}\left(\frac{\log_{10}\!{\left(E_i/E_{i'}\right)}}{\ell_{\rm E}}\right)^2\right]}\,,
\end{equation}
where $\mathcal{N}_{\rm E}$ refers to the number of energy bins, $\sigma_{{\rm E}}$ denotes the 
magnitude of the spectral systematic uncertainty, $\ell_{\rm E}$ the energy correlation length (in dex, i.e.~per decade)  
and $E_i$ is the central value of the $i-$th energy bin.
In general, the covariance matrix is then given by the tensor product $K=K_{\rm E}\otimes K_{\rm S}$.
In our analysis, however, we will restrict ourselves to considering correlations of type \textsc{(i)} and \textsc{(ii)} 
from the aforementioned list, which can be understood as particular instances of the most general case.
They can be constructed as follows:
\begin{itemize}
\item[Type \textsc{(i)}] $K_S$ describes correlations among the spatial template bins. To exclude any 
further energy correlation between different energy bins, one has to assume an infinitesimally small 
energy correlation length $\ell_E$. Thus, $K_E$ should be diagonal, i.e.~each energy bin is 
exclusively correlated to itself. In other words, the full covariance matrix is the tensor product of the 
identity matrix in energy space and $K_S$, $K=\mathbb{I}\otimes K_S$. 
\item[Type \textsc{(ii)}] Spectral correlations among a template's energy bins are described by $K_E$. 
In this case, however, one cannot assume an infinitesimally small spatial correlation 
length to describe the full matrix $K$: 
otherwise $K_E\otimes\mathbb{I}$ would predict a correlation of every spatial bin with its own copy in 
different energy bins, allowing the spatial bins to vary independently of each other and thereby erase
 the morphological information one wants to preserve.
Instead, one needs to assume an infinitely large spatial correlation length such 
that all spatial bins are varied as an ensemble, i.e.~$K_S$ must be chosen as a dense matrix where every 
element is equal to 1.
\end{itemize}
As a {\it default} assumption, we will adopt a 1\% overall normalisation error (corresponding
to one of the design requirements of CTA~\cite{Acharya:2017ttl}), $\sigma_S=0.01$, 
and a spatial correlation length of $\ell_S=0.1^\circ$ (roughly motivated by the typical size of the PSF). 
We also do not explicitly assume any energy correlations in the default analysis pipeline, 
as these turn out to affect our analysis much less. 
All these choices will be explicitly revisited and discussed in Section~\ref{sec:bm_sys}.


\paragraph{ON/OFF analysis}

For comparison with the more traditional approach, we also perform a likelihood analysis with the same 
energy binning as in the template approach,  but effectively using spatial bins with a ring morphology
(implemented as multiple ON regions). 
Here we do not model the background components as above (because the background is by definition 
determined in the OFF region), so the total joint-likelihood function $\mathcal{L}$ is
 a function of only two parameters, namely the DM mass $m_\chi$ and the velocity-weighted annihilation 
 cross section $\langle\sigma v\rangle$.  Our construction of ON and OFF regions near the GC closely follows that by 
H.E.S.S.~\cite{Abdallah:2016ygi, PhysRevLett.120.201101}, adapted to the planned GC survey of CTA. 
We provide further analysis details in Appendix \ref{app:onoff}, and discuss how this 
approach compares to the results from our baseline analysis strategy based on template fitting 
-- with particular emphasis on the fact that CTA is also expected to pick up astrophysical `signal' 
components that most likely are different in the two ROIs.


\section{Projected dark matter sensitivity}
\label{sec:results}

In this section we present the main results of our analysis, namely the sensitivity of CTA to a DM signal,
focussing exclusively on the following benchmark settings:
\begin{itemize}
\item GC survey observation strategy, masking bright sources as indicated in 
Fig.~\ref{fig:observations_schematic}. 
\item Asimov mock data set based on CR background and IE {\it Gamma} model templates. 
\item  Template fitting analysis based on $0.1^{\circ}\times0.1^{\circ}$ 
spatial bins and 55 energy bins between 30 GeV and 100 TeV (and a width corresponding to the 
energy resolution at the $2\sigma$ level). 
Our default treatment of systematic uncertainties implements a 1\% overall normalisation error
and a spatial correlation length of $0.1^\circ$ (but no energy correlations).
\end{itemize}
 In the subsequent Section \ref{sec:discussion}, we will discuss how our results are affected by modifying the 
 benchmark assumptions listed above. 

\subsection{Expected dark matter limits}

\begin{figure}[t!]
\centering\includegraphics[width=0.49\linewidth]{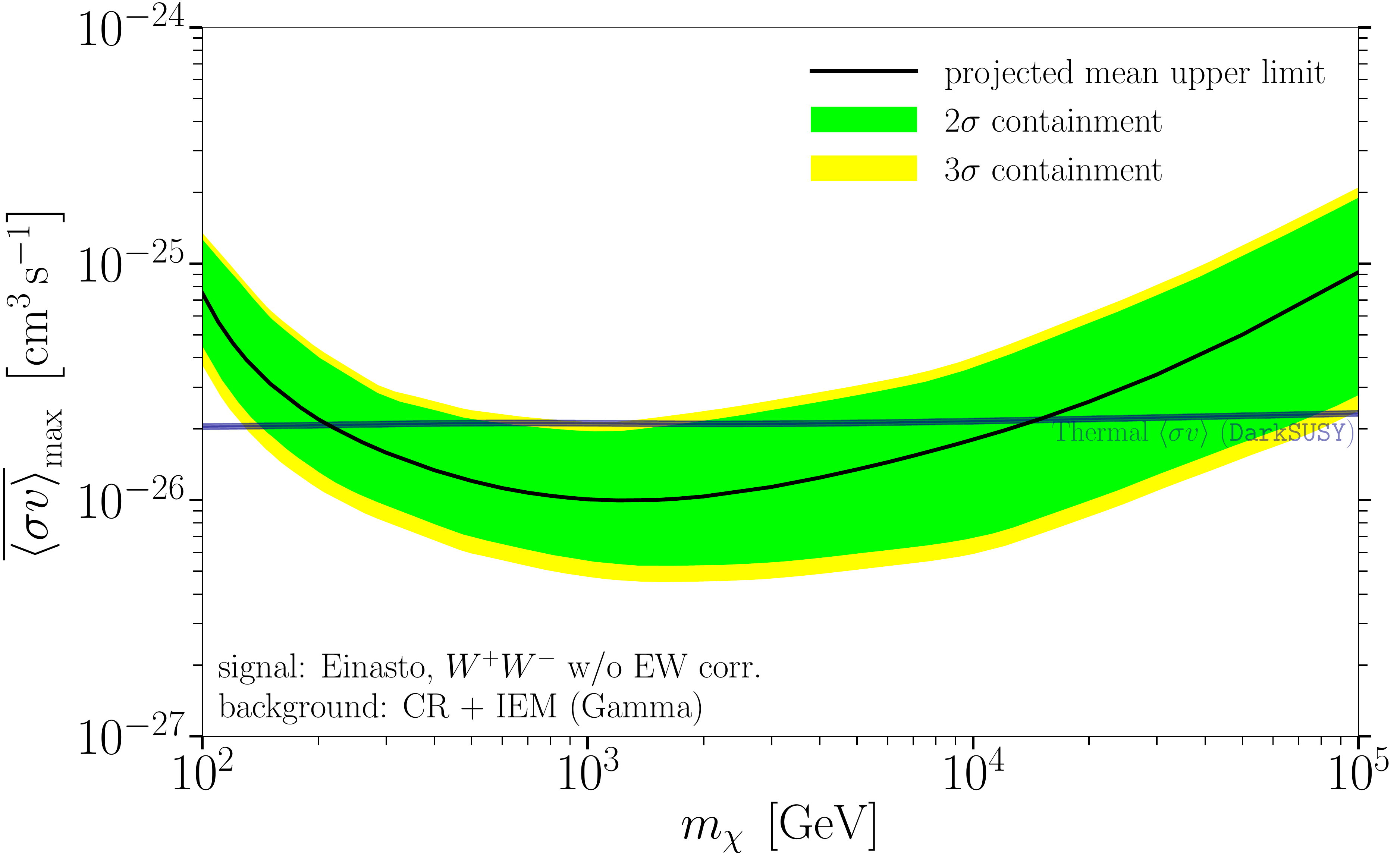}
\includegraphics[width=0.49\linewidth]{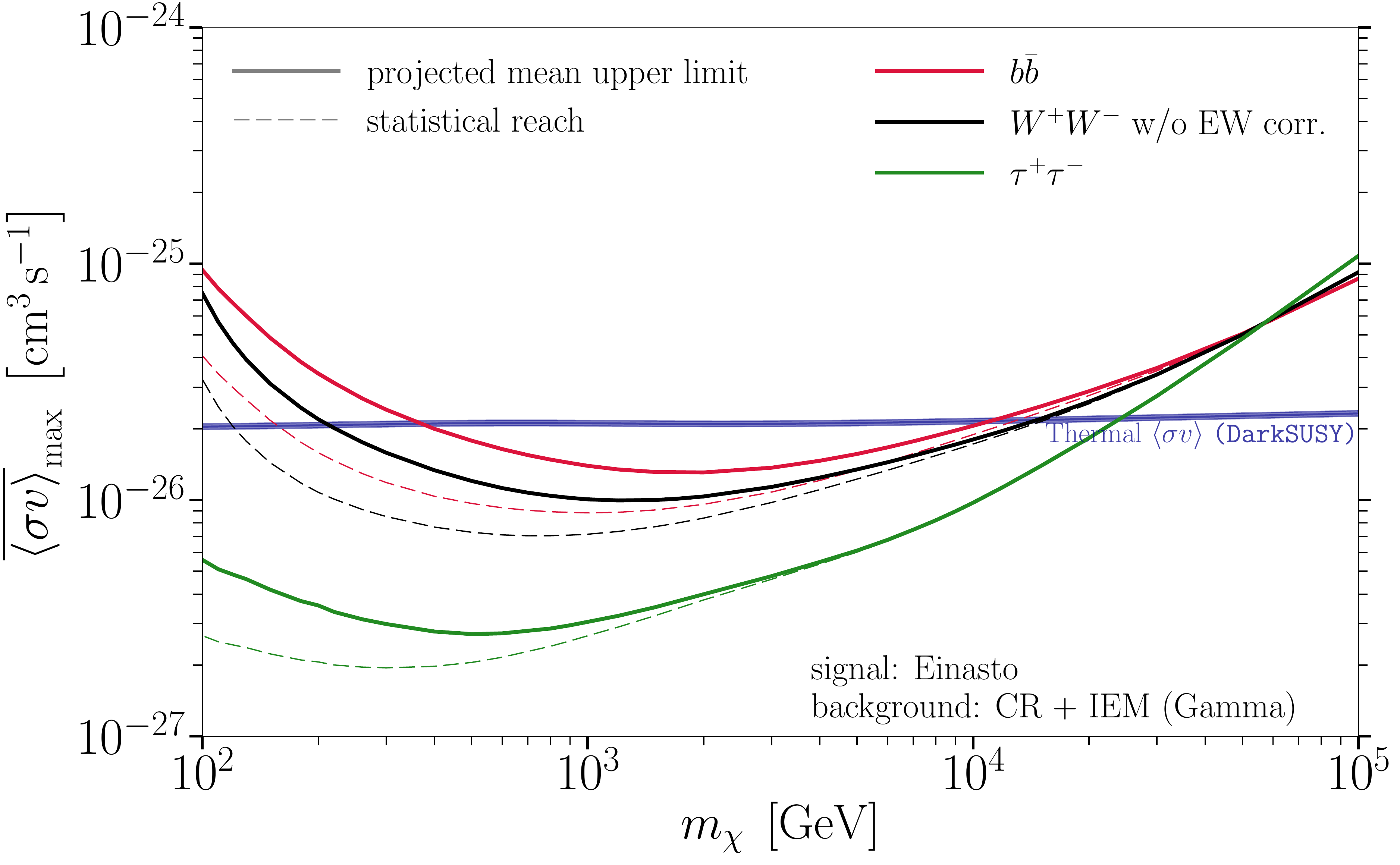}
\caption{Sensitivity of CTA to a DM annihilation signal, at 95\%\,C.L., based on our 
benchmark treatment of the expected instrumental systematic uncertainty.  
Following common practice, this is presented
in terms of projected mean upper limits on the average velocity-weighted annihilation cross section, as  a
function of the DM mass $m_\chi$. Solid lines show the sensitivity based on our benchmark settings, while dashed 
lines show the reach assuming no systematic uncertainty in the spatial templates.
We also indicate the `thermal' cross-section that for the simplest 
DM models leads to a relic density within the $3\sigma$ range of the DM abundance observed by
Planck~\cite{Aghanim:2018eyx, PLA}.
\emph{Left panel:} Sensitivity to DM annihilation into $W^+W^-$ final states (black), without electroweak 
corrections (see Section \ref{sec:DM_spectra} for a discussion). 
The green (yellow) band indicates the $2\sigma$ ($3\sigma$) 
scatter of the projected limits (based on Monte Carlo realisations).
\emph{Right panel:} DM annihilation into $\bar bb$ (red), $W^+W^-$ (black) and $\tau^+ \tau^-$ (green), 
respectively. 
}. \label{fig:bb_MCband}
\end{figure}

The most often considered `pure' annihilation channels for heavy DM candidates are those resulting 
from $\bar bb$, $W^+W^-$ and $\tau^+\tau-$ final states (in the order of increasingly harder spectra). 
In Fig.~\ref{fig:bb_MCband} we show the expected limits for DM models where annihilation into
these final states dominates, for a DM template based on the Einasto DM profile given in Eq.~(\ref{einasto}).
For comparison, we also indicate the cross-section needed to thermally produce DM in the early universe in order 
to match the cosmologically observed DM abundance. Specifically, we use \ds\ to calculate this cross-section,
following the treatment of Ref.~\cite{Gondolo:1990dk} under the assumption of self-conjugate DM particles 
annihilating with a velocity-independent $\sigma v$. We thereby improve upon similar recent 
results~\cite{Steigman:2012nb,Bringmann:2018lay} 
by using an updated temperature dependence of the number of relativistic degrees of freedom during and after 
freeze-out~\cite{Drees:2015exa} and  the newest Planck data for the observed value of 
$\Omega_\chi h^2=0.120$~\cite{Aghanim:2018eyx, PLA}.

We see from Fig.~\ref{fig:bb_MCband} that CTA indeed has the potential to test the `thermal' annihilation 
cross-section for a wide range of DM masses, in particular for the slightly harder gamma-ray spectrum 
that results from $W^+W^-$ final states. As pointed out in the introduction, this makes CTA perhaps the 
most promising instrument to test the WIMP paradigm for DM masses at the TeV scale, 
providing indeed one of its major science cases. Let us stress that we confirm this expectation 
{\it after} including our benchmark treatment of systematic uncertainties -- which we consider realistic 
given the obvious limitation that our analysis describes an instrument 
yet to be built (see Section \ref{sec:bm_sys} for a discussion). 
For comparison, we also indicate the mean projected limits that {\it would} result if only statistical errors were 
included in the analysis.\footnote{%
More precisely, these limits follow from the template analysis  detailed in 
Section \ref{sec:morph_analysis_pipeline}, without adding a correlation matrix to describe instrumental 
systematic errors. As discussed there, allowing for independent normalisations of the spatial templates, per 
energy bin, already is an effective way of including systematic uncertainties in the {\it spectral} templates.
}
As expected, limits are not affected in the statistics-limited case of the low photon counts
in models with large DM masses (as well as the background components at these high energies). 
For DM masses significantly below 10\,TeV, on the other hand, 
the limits clearly become dominated by systematic uncertainties rather than by statistical errors because,
for a given annihilation cross-section, both the background and the 
signal fluxes are much higher.

\subsection{Generalised flux sensitivities per energy bin}
\label{sec:limit_general}

Actual spectra for a given DM model rarely coincide exactly with those of the `pure' channels 
discussed above. 
While the use of `pure' channel limits is standard practice, limits as presented in Fig.~\ref{fig:bb_MCband}
thus have reduced practical applicability.
In Fig.~\ref{fig:flux_sensitivity_DMGDE} we therefore provide limits to the (spatial) DM template 
in a different way, more independent of the spectral model. Concretely, we show energy-flux 
sensitivities obtained by applying the likelihood function defined in 
Eq.~(\ref{eq:loglikelihood_cov}) {\it per energy bin}. Here we assumed for simplicity that the flux 
is described by a power law $\textrm{d}\Phi/\textrm{d}E \propto E^{-2}$ rather than following one of the 
explicit annihilation spectra considered above. 
As expected (and checked explicitly) the spectral form has only a minor effect on the result because 
the per bin contribution to the total likelihood is mostly affected by the photon count inside that energy bin 
(provided the bins are, as in our case, chosen sufficiently small \cite{Ackermann:2015zua}).
This makes this result more universal, motivating us to also indicate 
the change in the full likelihood $\mathcal{L}_i$ per energy 
bin (and to make it available in tabulated form~\cite{bringmann_torsten_2020_4057987}). 
To a reasonable approximation, this can be used to constrain the signal normalisation, at 95\%\,C.L., 
of an almost arbitrary smooth DM spectrum $dN_\gamma/dE_\gamma$, where $N_\gamma$ is the 
number of photons {\it per annihilation} process. Concretely, this corresponds to requiring
\be
\label{eq:sens_est}
\sum_i \Delta\ln \mathcal{L}_i \left[C_\chi E_i \, dN/dE_\gamma\right]<2.71\,,
\ee
where the sum runs over all energy bins, with central energy $E_i$,
and the correct flux normalisation is ensured by using 
\be
\label{eq:sens_est_norm}
C_\chi\simeq \left(2.5\cdot10^{21}\;\textrm{GeV}^2/\right. \left.\textrm{cm}^5\right)
\times  \langle\sigma v\rangle_\mathrm{ann}\, S_\chi^{-1} m_\chi^{-2}\,,
\ee 
c.f.~Eq.~(\ref{DMflux}).\footnote{%
The normalisation obviously depends on the chosen profile and, {\it for cuspy profiles},
scales roughly with the $J$-factor. For more cored profiles, see the discussion in Section~\ref{sec:cores}.
} 
For DM spectra varying more strongly with energy than $E^{-2}$, integrating over the energy
inside each bin, rather than using the mean number of photons in each bin as in Eq.~(\ref{eq:sens_est}),
would provide a slightly more accurate estimate (while highly localised spectral features, such as monochromatic gamma-ray lines~\cite{Bergstrom:1997fj}, would warrant a different analysis strategy that leads to significantly 
better limits than indicated in Fig.~\ref{fig:flux_sensitivity_DMGDE}~\cite{Ackermann:2015lka}).
Let us stress that we provide here the tabulated binned likelihoods $\mathcal{L}_i$ only  for convenience,
to allow for quick and simple estimates of sensitivities to DM models not covered in our analysis;
all our results are based on the full procedure detailed in Section \ref{sec:data}
rather than on the `short-cut' defined by Eq.~(\ref{eq:sens_est}).

\begin{figure}[t]
\centering\includegraphics[width=0.71\textwidth]{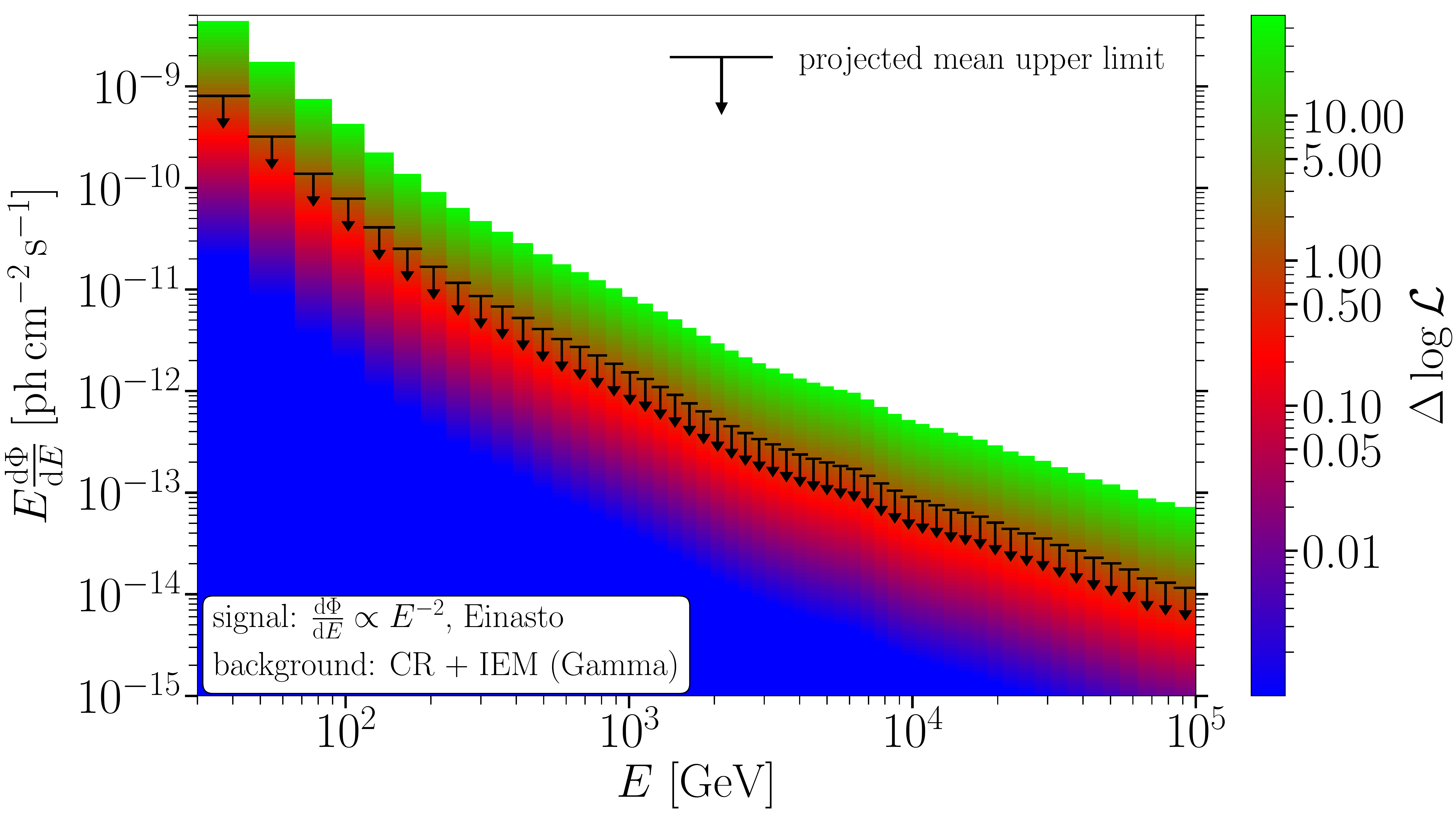}
\caption{Differential flux (times $E$) as a function of source energy $E$. The (logarithmic) colour scale shows the 
change in the total likelihood due to the contribution from a DM signal in the respective energy bin,
assuming a DM-induced flux locally scaling as $\textrm{d}\Phi/\textrm{d}E \propto E^{-2}$.
To guide the eye, the black symbols indicate the value of $\Delta \log\mathcal{L}$ that corresponds to a $2\sigma$ upper 
limit on such a DM signal, from a single bin. Limits on DM models with arbitrary spectra, taking into account all energy bins, 
can be approximated as in Eq.~(\ref{eq:sens_est}). The full likelihood table is available for download at 
zenodo~\cite{bringmann_torsten_2020_4057987}.
\label{fig:flux_sensitivity_DMGDE}}
\end{figure}

\subsection{Extended dark matter cores}
\label{sec:cores}

Let us now address the impact of the assumed DM density profile. 
We re-iterate from the discussion in Section \ref{sec:DMdistribution} that the Galactic DM
distribution within the inner few kpc (and even more so for $r\ll1$\,kpc) is rather  uncertain and 
not very well observationally constrained -- but at the same time the distribution is crucial for 
estimating the overall strength of the annihilation signal. The situation in the GC is, in general, different from 
the situation in dwarf spheroidal galaxies, where kinematic data allow us to constrain the $J$-factor 
sufficiently~\cite{Bonnivard:2015xpq} to warrant including them in the likelihood analysis, fully 
marginalising over the profile parameters~\cite{GeringerSameth:2011iw,Drlica-Wagner:2015xua,Ahnen:2016qkx}. 
One of the reasons behind this is that for dSphs we typically observe the entire DM halo and 
therefore the full  `bolometric' DM emission which -- unless considering extreme examples -- only 
depends weakly on the DM density profile~\cite{Ackermann:2011wa} (for the GC, on the other hand, 
IACTs have traditionally just observed the inner region, with its highly uncertain DM density). 
For that reason, we focus here on discussing the benchmark profiles introduced in Section
\ref{sec:DMdistribution}, intended to bracket realistic and more conservative expectations;
in Appendix \ref{app:profiles}, we complement this by a more detailed discussion based on a 
larger set of density profiles. 
Let us already mention, however, that due to the extent of the GC survey 
(reaching up to 15 degrees, Fig. \ref{fig:observations_schematic}),  CTA will actually observe the 
entire inner 1\,kpc region for which the DM density is most uncertain, which will in fact  
significantly reduce the standard uncertainties in predicting the sensitivity to a DM signal from the GC.

For the conservative case we take a core of constant DM density, as in Eq.~(\ref{einasto_core}), 
that reaches out to about 1\,kpc (recall that 
even larger cores may be compatible with observational data, but that it is very challenging 
to produce those in hydrodynamical numerical simulation with standard {cold} DM; in fact, the expectation
is a profile {\it steeper} than the standard Einasto case~\cite{Cautun:2019eaf}). 
For such large cores, the limits are clearly expected to weaken because if a signal template is highly
degenerate with the misidentified CR background, c.f.~Fig.~\ref{fig:template_collection}, 
it almost constitutes a blind spot for morphological analyses.
The second reason why limits should weaken is that the signal strength is directly proportional to the
$J$-factor that -- due to its $\rho^2_\chi$ dependence -- benefits from a local concentration of DM. 
This effect, however, is less relevant because of the
large ROI we adopt in our analysis; as expected from Fig.~\ref{fig:DM_benchmarks}, the $J$-factor
{\it integrated} over the full ROI\footnote{%
In practice, the whole ROI does not contribute uniformly to the signal discrimination
power, and the region with the highest Signal-to-Noise Ratio (SNR) is different for cuspy and cored
profiles; see Appendix \ref{app:InfoFlux} for a more detailed discussion. 
}
 should not deviate too much between the two profiles
(we find $J_{\rm Einasto} = 7.1\cdot10^{22}\;\textrm{GeV}^2/\textrm{cm}^5$ and 
$J_{\rm core\,(1\,kpc)} = 3.9\cdot10^{22}\;\textrm{GeV}^2/\textrm{cm}^5$), respectively).
Previous studies of DM annihilation at the
GC (e.g.~Refs.~\cite{Abramowski:2011hc,Abdallah:2018qtu}), in contrast, typically used a smaller ROI and hence faced a 
much larger difference in the $J$-factors between cuspy and cored profiles.

\begin{figure}[t]
\centering\includegraphics[width=0.71\linewidth]{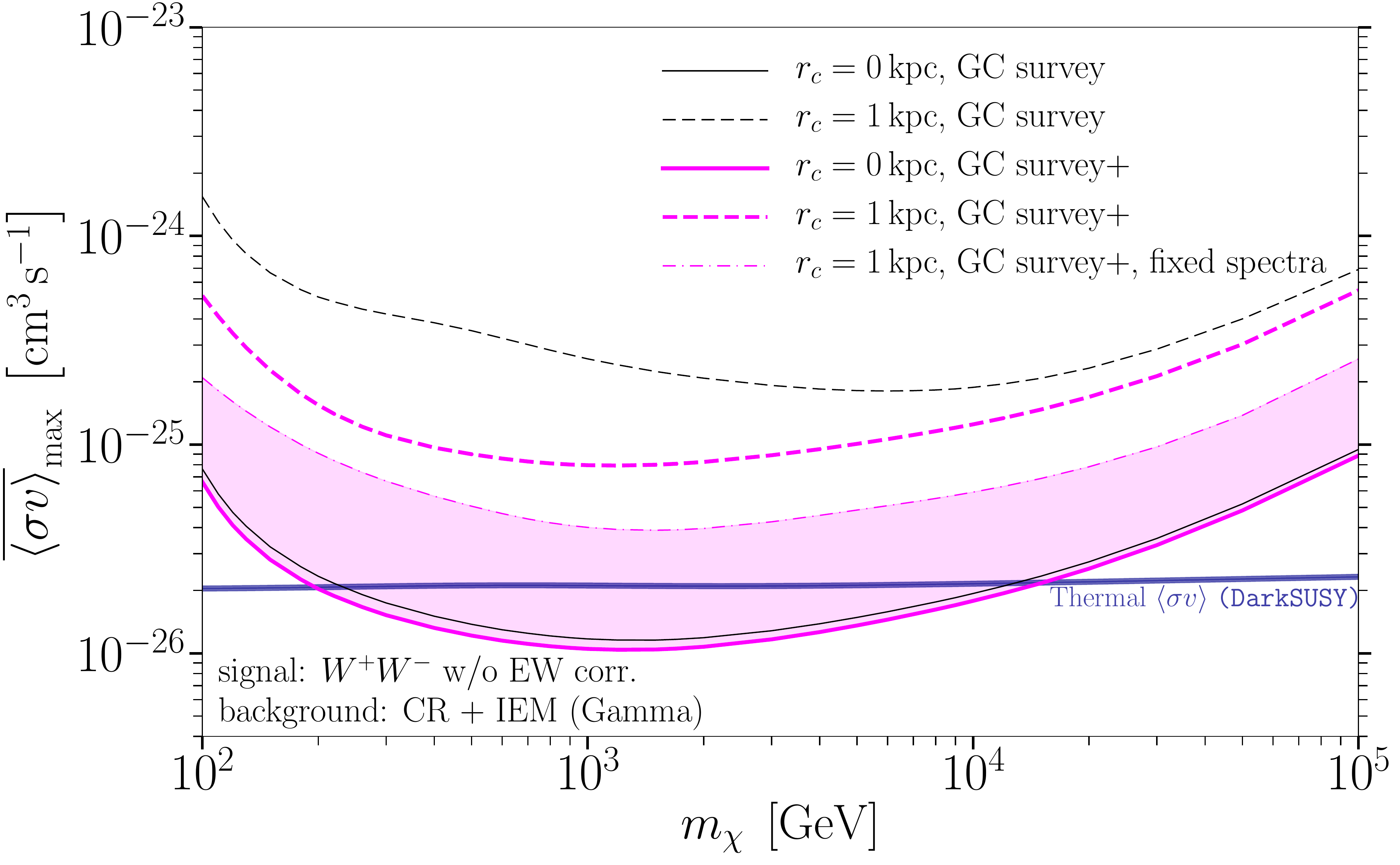}
\caption{CTA sensitivity to a DM signal, for the $W^+W^‐$ channel, comparing the case of an Einasto profile 
without core (solid) to that of an Einasto profile with a 1\,kpc core (dashed). 
Black lines show the sensitivity with the base survey only and magenta lines show the sensitivity from 
adding extended survey observations. Modelling the spectral information with greater care may lead to a 
further improvement of the sensitivity to a cored profile, as indicated by  the magenta dash-dotted line 
(see text for more details, including a discussion of the shaded area).
In order avoid excessive use of computational resources, the sensitivity predictions in this figure are 
based on only 20 (equally log-spaced) energy bins.
\label{fig:cored_templates_wwo_extended}} 
\end{figure}

In Fig.~\ref{fig:cored_templates_wwo_extended} we show how our limits for the $W^+W^-$ channel 
for the baseline Einasto profile (black solid line) worsen by about one order of magnitude when 
assuming large core sizes (black dashed line). 
The planned {\it extended GC survey} would clearly help to better distinguish even
such an extended DM signal and would hence significantly improve limits
in this case (magenta dashed line). For the Einasto profile (solid magenta line), on the other hand, the effect
is minimal; here, the template discrimination is already so good for the standard survey
that it is only the (slight) increase in observation time that implies a corresponding improvement. 
As discussed in Appendix \ref{app:profiles},  additional {\it spectral} information can help significantly in 
the presence of large cores (but not for our benchmark case of a cuspy profile). 
With the thin dash-dotted magenta line, we indicate the maximal further improvement of limits that could be 
achieved this way (mimicking the situation where the astrophysical background components
are well determined by complementary measurements; see Appendix \ref{app:profiles} for details). 
The magenta band can thus be interpreted as a rough estimate
of how much our default projected DM limits could realistically be affected if the DM density profile turns
out to be significantly less peaked than in the standard Einasto case.
For similar reasons, we refrain from providing the full binned likelihood as in 
section \ref{sec:limit_general}: unlike in the case of
cuspy profiles the normalisation given in Eq.~(\ref{eq:sens_est_norm}) now depends on an {\it effective}
$J$-factor that is itself energy-dependent (because the size of the region with highest SNR is
energy-dependent, see discussion above), which in turn makes the translation to
different density profiles and spectral shapes much less straight-forward.

We conclude that a large core in the DM distribution would indeed worsen the CTA sensitivity
to DM annihilation -- but much less severely than naively expected (or indicated by previous studies). 
This implies that, for DM masses in the TeV range, many models of thermally produced DM could 
be probed even in this highly unfavourable situation (both because of the statistical scatter in 
the expected mean limit, and because annihilation rates exceeding the `thermal' rate  by a factor 
of a few are by no means unusual, for example in the context
of simple supersymmetric models~\cite{Athron:2017qdc,Athron:2017yua}).
It is also worth stressing again that Fig.~\ref{fig:cored_templates_wwo_extended}
summarises our assessment of what could be coined a `realistic worst-case scenario'; in reality, 
the situation can also be significantly {\it better} than the benchmark case of an Einasto profile,
because of DM density spikes very close to the GC (see again Appendix \ref{app:profiles} for examples).

\section{Discussion} 
\label{sec:discussion}

In this section we turn to a discussion of our main results and how the projected sensitivities 
depend on the benchmark choices that we have adopted for our analysis.
As stressed previously, one of the biggest challenges in the template fitting approach is a 
realistic account of systematic uncertainties both in the performance of the instrument and in the modelling of 
the templates. The parameters crucial to the description of the relevant physical effects are not only the magnitude of 
the systematic uncertainty, but also the correlation lengths, both in morphology and energy. 
Correlation matrices are an adequate way to describe these effects, as detailed in 
Section \ref{sec:morph_analysis_pipeline}. We start by discussing instrumental systematic errors (related to the 
event reconstruction and hence mostly caused by misidentified CRs), which tend to dominate over 
astrophysical uncertainties (to be further discussed in section \ref{subsec:GDEbintobin}).
While the importance of instrumental systematic uncertainties is already clearly seen in the right panel of Fig.~\ref{fig:bb_MCband},  
we note that they are potentially easier to study in real data (e.g.~by defining special high-quality photon event classes).

\subsection{Instrumental systematic uncertainties} 
\label{sec:bm_sys}

In Fig.~\ref{fig:correlations} we consider the impact of changing the overall {\it spatial correlation lengths} $\ell_S$ on 
our limits. Here we keep the overall systematic error amplitude to a fixed level of 
$\sigma_S^\mathrm{instr}=1$\%, corresponding to our benchmark choice. 
Compared to our benchmark correlation length of $\ell_S=0.1^\circ$, corresponding to the typical PSF,
limits worsen by a factor of up to three at intermediate DM masses, when $\ell_S$ is comparable 
to the spatial extension of the DM signal ($\lesssim 1^\circ$ for the Einasto DM profile). When signal 
and correlation lengths are sufficiently different, on the other hand, $\ell_S\lesssim 0.1^\circ$ or $\ell_S\gtrsim 1^\circ$, 
the impact of varying the correlation length on the sensitivity is generally milder -- though the limit of very 
large correlation lengths would correspond to fixing the overall amplitude and hence result in the `statistical' limit
(with no systematic uncertainty in the spatial templates) indicated with a dashed line. 
Performing a similar exercise to explore the effect of {\it energy correlations} shows that these have a significantly weaker 
impact on the sensitivity to a DM signal, for our benchmark case of cuspy DM density profiles 
(but see Appendix \ref{app:profiles} for a discussion of how this changes in the presence of cores).

\begin{figure}[t]
\centering\includegraphics[width=0.71\linewidth]{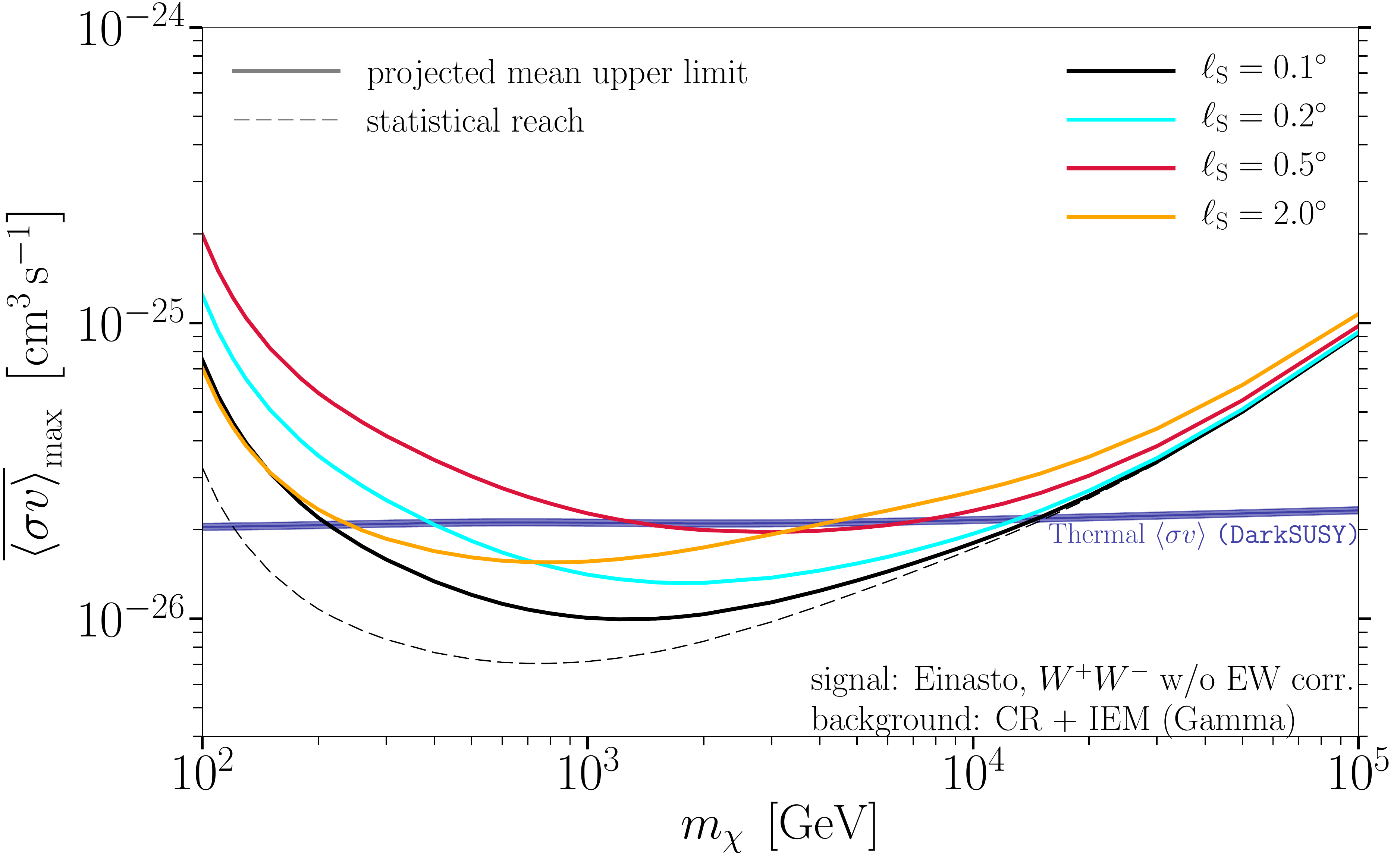}
\caption{Solid lines show how the projected CTA sensitivity to a DM annihilation signal, for the $W^+W^-$ channel, 
varies with the spatial correlation length $\ell_S$, 
when keeping the instrumental systematic error-related fluctuation amplitude to a fixed level of 
$\sigma_S^\mathrm{instr}=1$\%. For comparison, the dashed 
line show the statistical reach. \label{fig:correlations}
}
\end{figure}

While our benchmark choice to account for systematic uncertainties may appear reasonable from the point of 
view of the expected 
instrument performance, we stress that this choice is not unique and other values of both $\sigma$ and $\ell$ may 
turn out to characterise the instrument more accurately. On the other hand, our benchmark scenario is by no 
means a carefully selected singular point, either, in the sense that various combinations of parameters would lead 
to similar conclusions. In fact, we can turn the problem around and ask for the {\it required} level of systematic errors 
allowing CTA to probe the thermal cross-section for standard assumptions about the DM profile. This question is 
explored in Fig.~\ref{fig:thermal_req}. In the left panel, we fix the DM mass to $m_\chi=2$\,TeV and show the 
combinations of 
amplitude and spatial correlation length for which the thermal cross-section can be reached (green shaded area),
while in the right panel we fix the fluctuation amplitude $\sigma_S$ and vary the correlation length and DM mass.
The figure illustrates, as already seen in the discussion of Fig.~\ref{fig:correlations}, that one wants to 
avoid values of $\ell_s$ comparable in extension to the DM signal.
In particular, for the design goal of $\sigma_S=1\%$, it is indeed crucial that the spatial correlation in 
mis-reconstructed events does not significantly exceed (a few times) $\ell_S\sim0.1^\circ$. If the 
overall systematic uncertainty can be improved to be less than $1\%$, on the other hand, larger spatial 
correlations could be accepted.

\begin{figure}[t]
\centering
\includegraphics[width=0.48\linewidth]{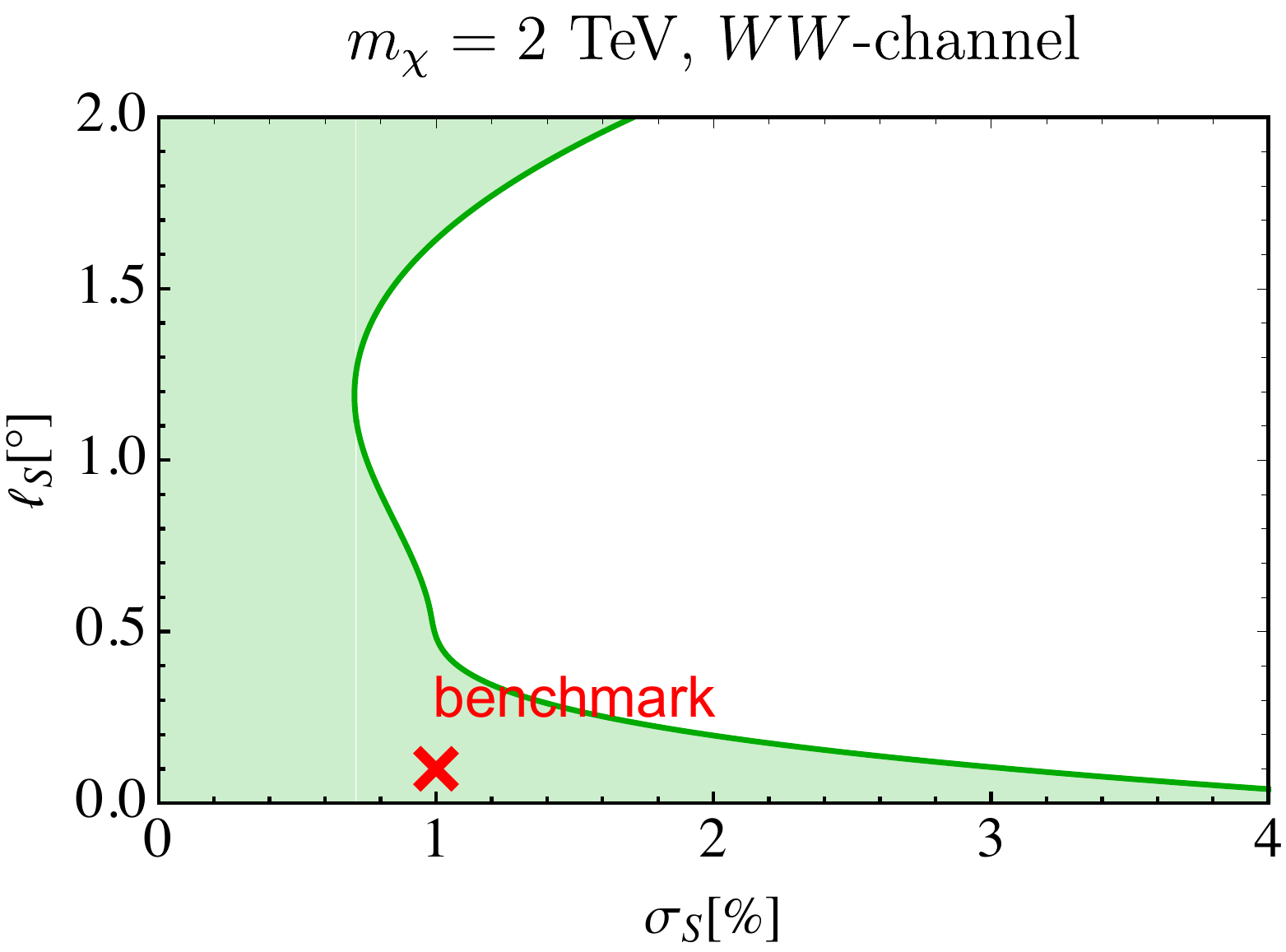}
\includegraphics[width=0.49\linewidth]{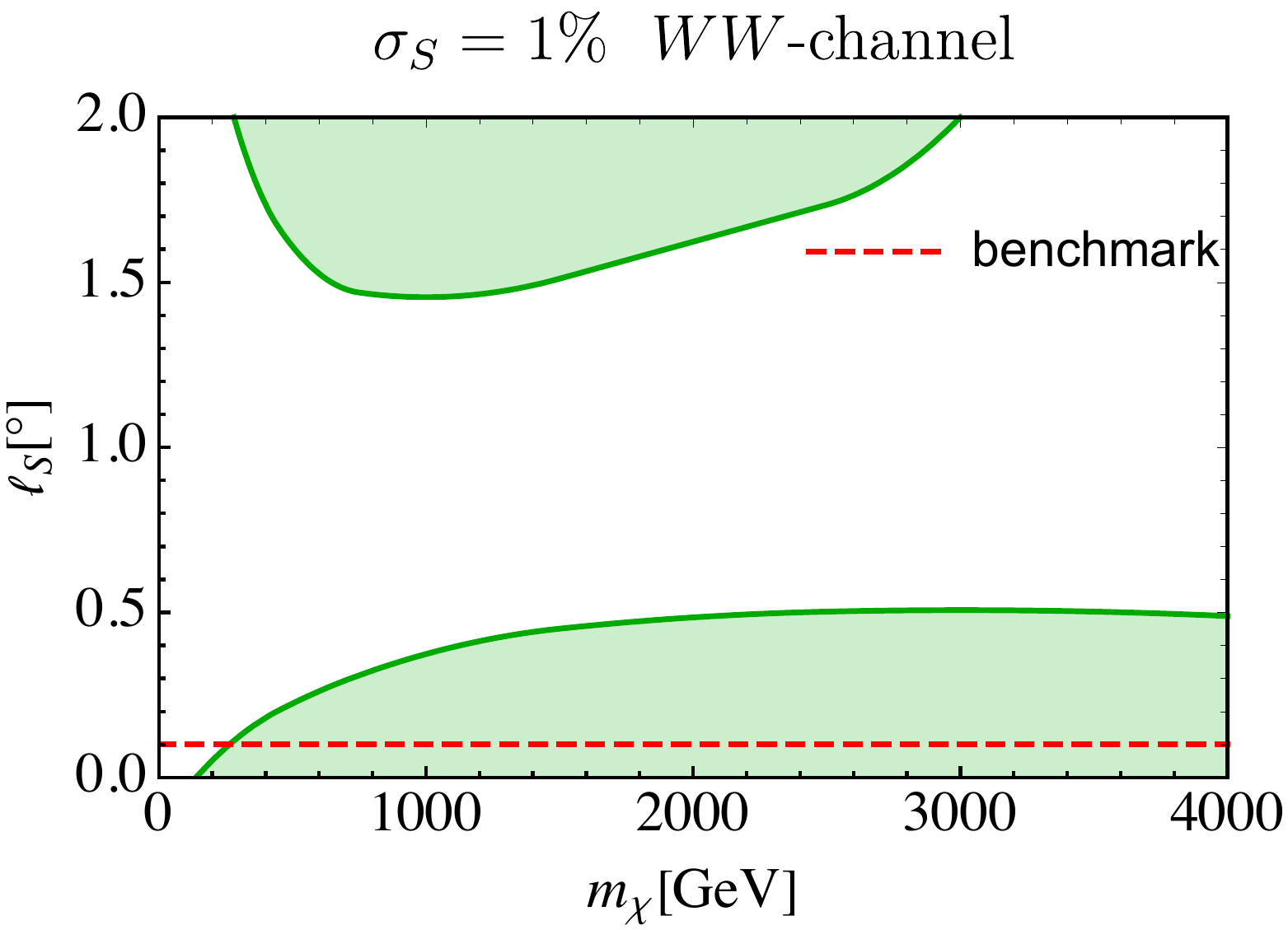}
\caption{
Contour plot (in green) showing the requirement on the instrumental systematic errors for which 
CTA is expected to reach the thermal cross section sensitivity to DM as a function of the amplitude 
and spatial correlation length of the instrumental systematic errors.
For comparison, we  mark the systematic benchmark settings used per default in this work (red cross
in the left panel and dashed red line in the right panel). 
{\it Left.} For fixed DM mass $m_\chi=2$\,TeV. 
 {\it Right.}  For a fixed overall systematic error amplitude of $\sigma^\mathrm{instr}_S=1\%$.
   \label{fig:thermal_req}
}
\end{figure}

In the last part of this section we briefly comment on the impact of  a set of internal (i.e.~not  yet publicly available) 
IRFs, based on tighter cuts for gamma/hadron separation and optimised for extended 
source detections (unlike the standard IRFs, which are optimised for point source detection). 
We checked explicitly that even with such optimised event cuts it will be hard to 
improve upon the sensitivity based on our standard IRFs. This can be 
understood by noting that, in the energy  range where  the DM sensitivity is best (few 100\,GeV -- few TeV), 
the misidentified CR background is already electron dominated,  see Fig.~\ref{fig:flux-allcomponents-2}, and 
it cannot 
be reduced significantly with current event reconstruction techniques. At the lowest and highest 
energies on the other hand, the tighter cuts can in principle 
substantially reduce the background, but only at the cost of reducing the effective area. 
We note, however, that these optimised 
IRFs still use the same event reconstruction scheme as the standard ones, so if an improved 
reconstruction algorithm becomes available in the future (benefitting from e.g.~deep learning~\cite{Shilon:2018xlp}) 
one may hope for more significant improvements.

\subsection{Uncertainties in astrophysical components}
\label{sec:astro_uncertainty}

\begin{figure}[t]
\centering\includegraphics[width=0.71\textwidth]{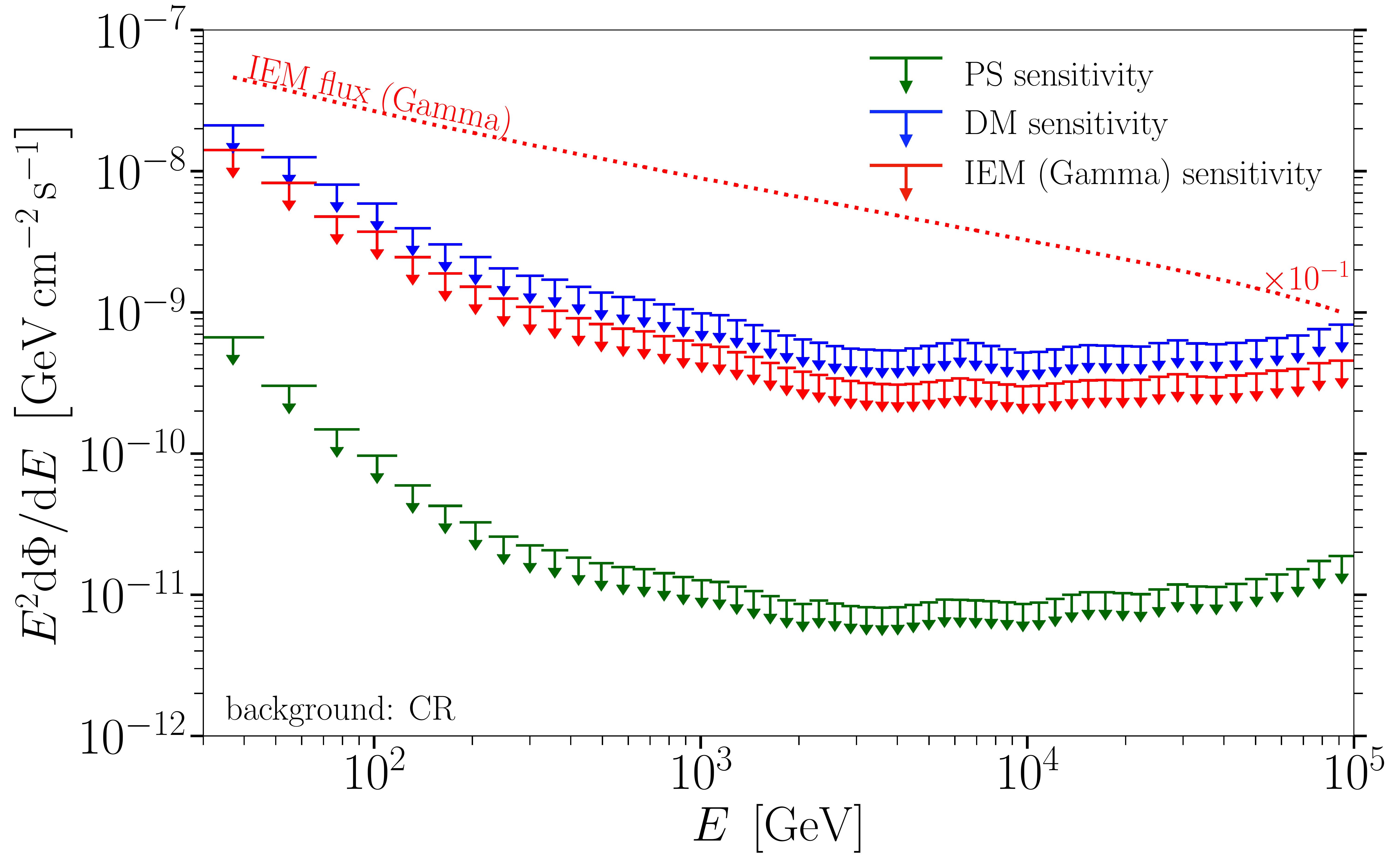}
\caption{ Differential flux sensitivity (time $E^2$) per energy bin to various gamma-ray templates in the GC region,
corresponding to  $2\sigma$ upper limits on mock data sets based on the residual CR background alone.
The point source (PS) limit shows the statistical reach, for better comparison with published results, while 
the IEM and DM sensitivities adopt our benchmark scheme for the treatment of systematic errors 
($\ell_{\textrm{S}}=0.1^{\circ}$, $\sigma_{\textrm{S}}=1\%$). 
For definiteness, we choose here a DM model with mass $m_\chi=100$\,TeV annihilating to bottom 
quarks and the IE {\it Gamma} model (the dashed line show the emission intensity of the latter).
\label{fig:flux_sensitivity_components}}
\end{figure}

For the DM sensitivity targeted at by CTA, the astrophysical backgrounds discussed in Section \ref{sec:GCtemplates} 
will have a stronger impact than for current generation instruments. In this section we aim to assess the 
associated uncertainties.
For comparison, to set the scale, let us start by directly comparing the bin-by-bin integrated energy-flux 
sensitivities for different emission templates; this is done in Fig.~\ref{fig:flux_sensitivity_components}, for mock 
data produced from the CR background alone. As expected, the sensitivity to the IE component (red) is 
significantly better than 
the flux expected in the {\it Gamma} model (dotted red line) -- implying that it is indeed very likely 
that CTA {will perform detailed measurements of this component}. In comparison, the sensitivity to the DM 
template (blue) is worse by a factor of about two, related to the fact that the morphology of the IEM 
(elongated along the Galactic plane and following that of the gas  column density)
is  less degenerate with the (much more isotropic) background of misconstructed CRs than the spherical DM signal. 
 We checked explicitly that masking the Galactic ridge by excluding the region
 ($-1.5^\circ<l<1.9^\circ,-0.5^\circ<b<0.5^\circ$) 
from the analysis does not change the relative sensitivity to the DM and IE component, which 
confirms that the difference is indeed related to the large-scale diffuse emission.
For comparison, we indicate in the same figure also the flux sensitivity to point sources (green) obtained 
with our binned likelihood approach, which agrees very well with the official CTA point-source sensitivity based  
on an ON/OFF technique.\footnote{%
For the point source sensitivity, we  also require the detection of at 
least 10 photons per energy bin, following the standard CTA procedure~\cite{CTA_performance}. 
This condition is always  satisfied for our extended emission templates. \label{ft:cta-performance}} 

\subsubsection*{IE template uncertainty}
\label{subsec:GDEbintobin}

\begin{figure}[t!]
\centering\includegraphics[width=0.49\linewidth]{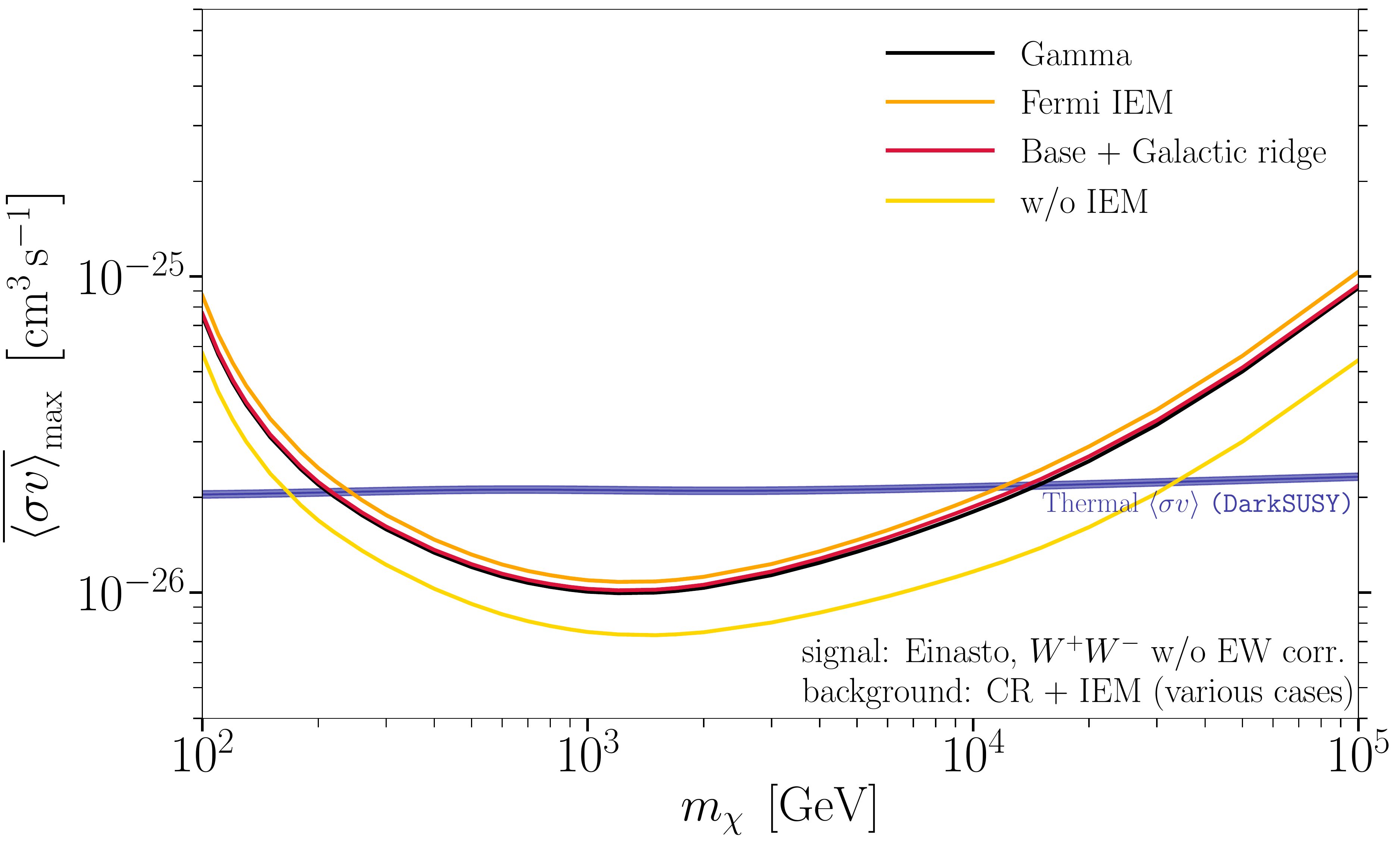}
\centering\includegraphics[width=0.49\linewidth]{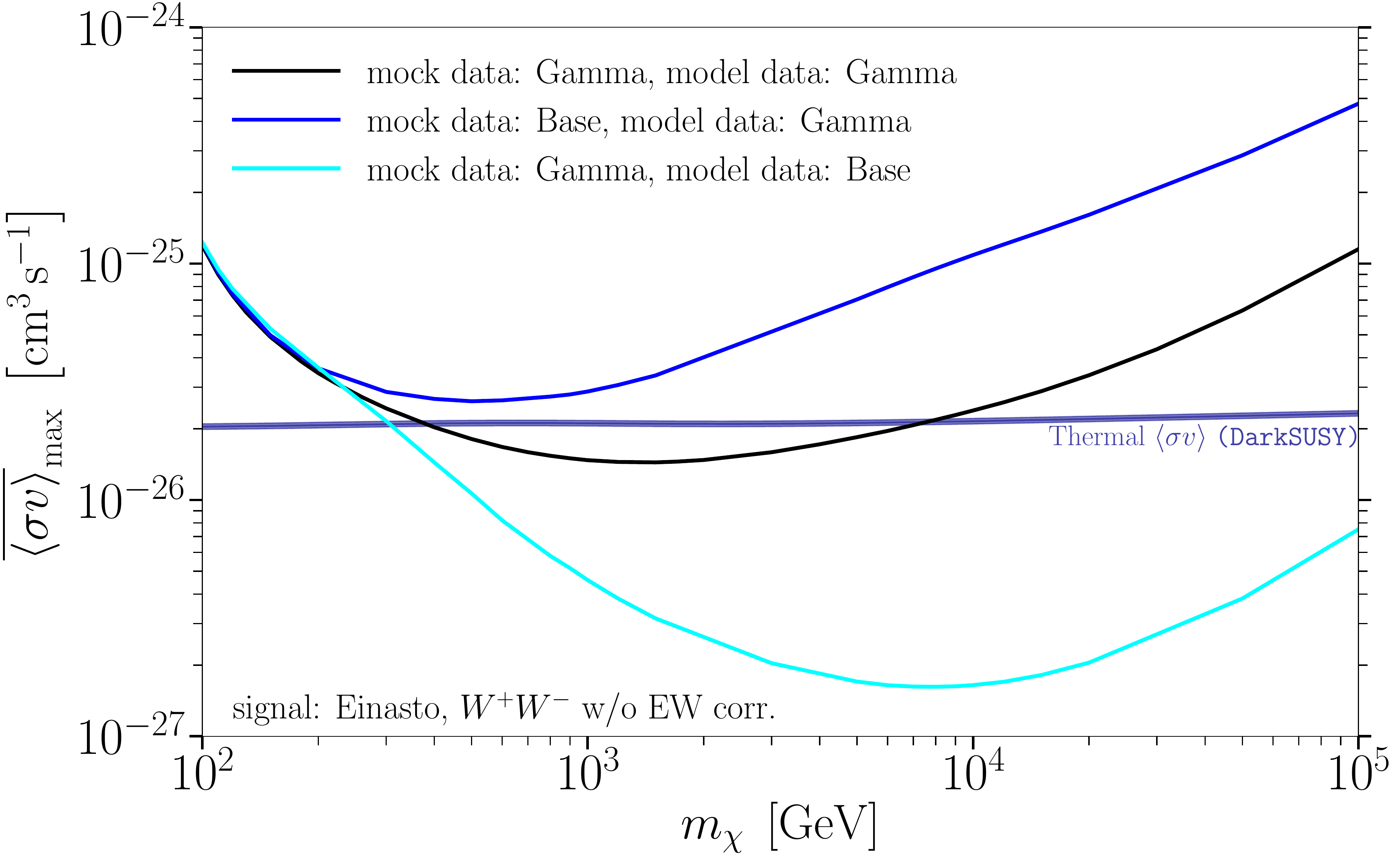}
\caption{{\it Left:} CTA sensitivity to the  DM annihilation channel for cases with no IE (yellow line) and with different  
IE models discussed in the text (note that the red line, for the Base IEM, largely overlaps with the black line,
for our benchmark Gamma IEM). 
{\it Right:} 
Illustration of the projected DM sensitivity that would result from  
purposefully choosing a wrong IE model, compared to the one used in producing the mock data 
(sensitivities after adopting a  Galactic plane mask of
$-6.0^\circ<l<6.0^\circ,-0.6^\circ<b<0.6^\circ$).
Note that the large variation of limits shown here illustrates an extreme case, 
unrealistic to be encountered in the analysis of real data (see text for a discussion).
\label{fig:sensitivity_GDE}}
\end{figure}

Let us  now discuss in more detail how the modelling of the IE component affects the DM sensitivity.
We start by exchanging our benchmark {\it Gamma model} for the alternatives mentioned in 
Section \ref{sec:morph_analysis_pipeline}. The result is shown in Fig.~\ref{fig:sensitivity_GDE} (left panel), for
the {\it Base} (red) and {\it Pass8-Fermi} (orange) models.
Adding IE thus degrades the DM sensitivity by almost a factor of two (black line). 
On the other hand, the effect of choosing IEMs more degenerate with the DM 
template (the {\it Gamma} model) or less `structured' emission models (the {\it Base} model) is marginal. 
The apparent conclusion that the impact of varying the IE emission only mildly affects the sensitivity,
 however, is clearly overly {\it optimistic} because it implicitly relies on having access to the {\it true} 
 emission model (modulo instrumental systematic uncertainties) when analysing the real data-set.

In the right panel of the same figure, we therefore relax this assumption and explore a different extreme, a 
{\it pessimistic} case, in which we purposefully chose a `wrong' IE model with respect to the one used to 
produce the mock data (here we also use a Galactic plane mask, $-6.0^\circ<l<6.0^\circ,-0.6^\circ<b<0.6^\circ$, since we do not attempt any IE modelling; masking
however has limited impact on the IEM uncertainty, given that the differences extend beyond the plane).
As illustrated, the limits could in this case {\it artificially} improve or worsen significantly more, 
by a factor of up to one order of magnitude. Let us stress
that this plot is  only meant to serve as an illustration, deliberately using two extreme IE modelling scenarios, 
neither of which is optimised for our energy range and ROI. With real data at hand, one would inspect the
residuals to judge the quality of the fit and immediately discard models that deviate so significantly from the data
as in the cases shown in this figure. In general, any residuals correlating with the galactic plane would 
strongly suggest that the IE modelling needs to be improved  (and that DM limits obtained in this case are 
not realistic). Also the understanding of the IE itself is clearly expected to improve once real data are available. 
It is worth pointing out, however, that the issue of 
mis-modelling background components, as explored in this figure, could rather easily lead to fake signal claims and 
is hence much more relevant when assessing the DM {\it discovery} potential (rather than the potential to 
constrain a signal).

\begin{figure}[t]
\centering\includegraphics[trim={0cm -0.5cm 0cm 0cm},width=0.3\linewidth]{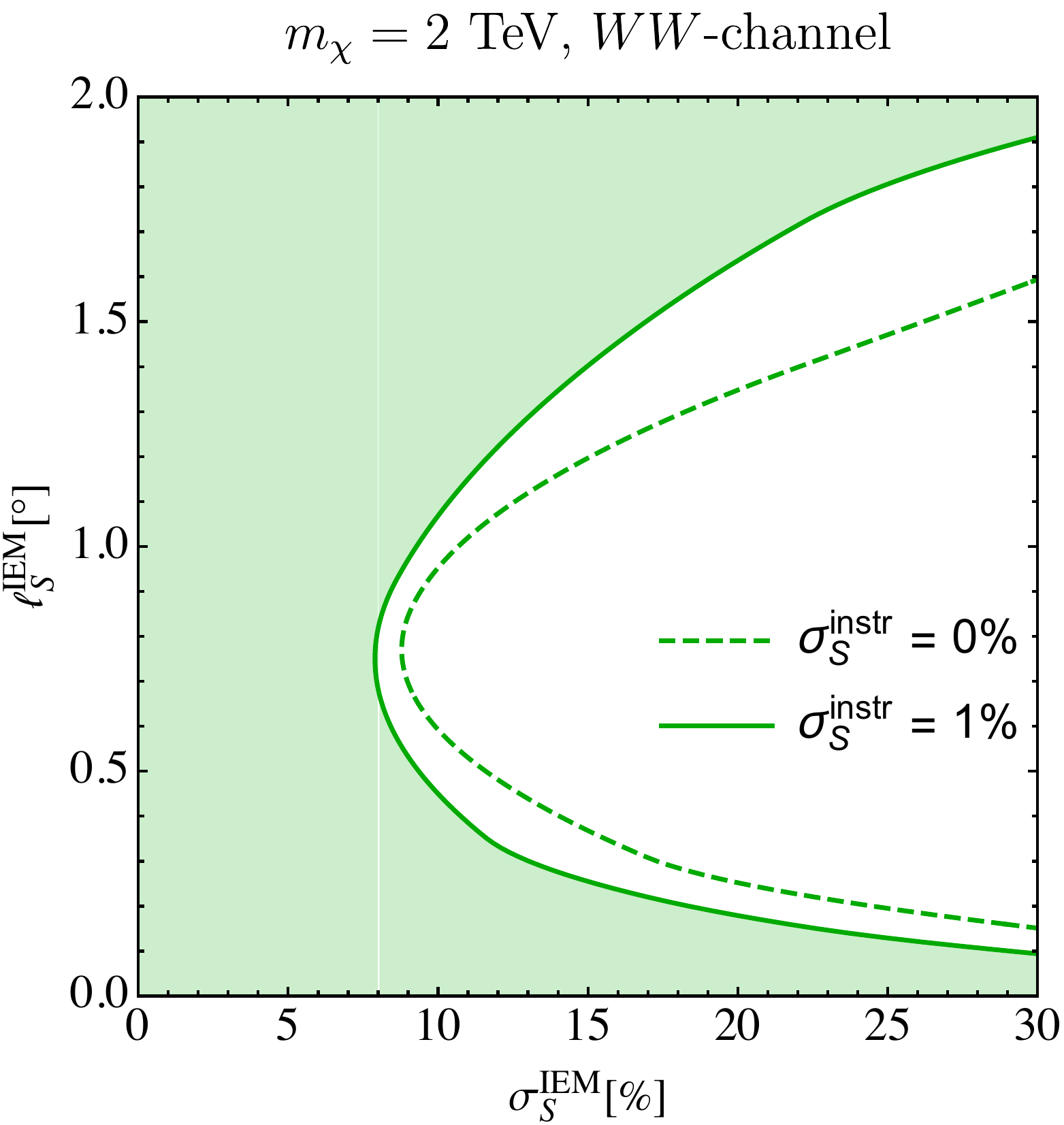}
\centering\includegraphics[trim={0cm -0.1cm 0cm 0cm},width=0.315\linewidth]{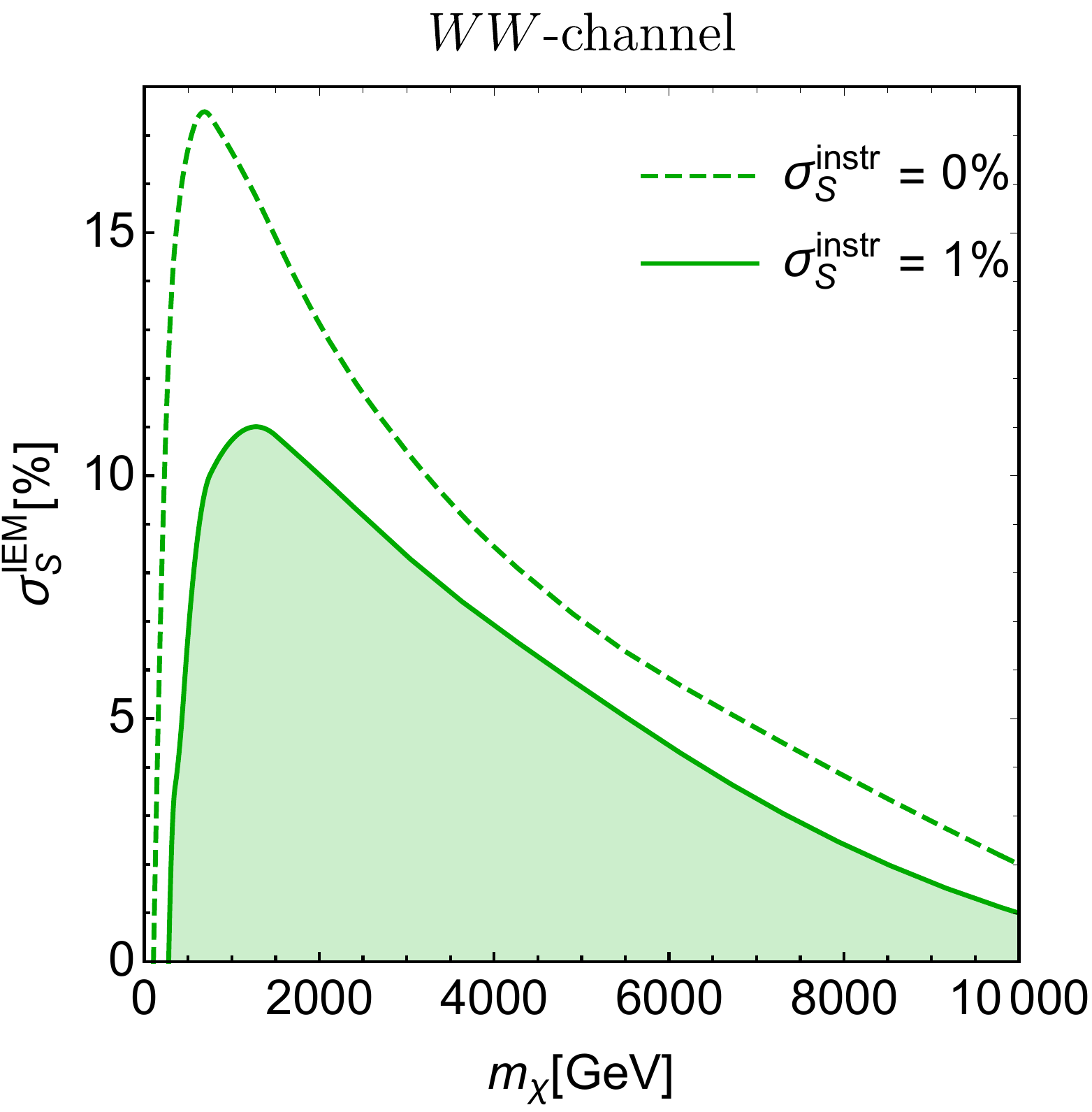}
\centering\includegraphics[width=0.325\linewidth]{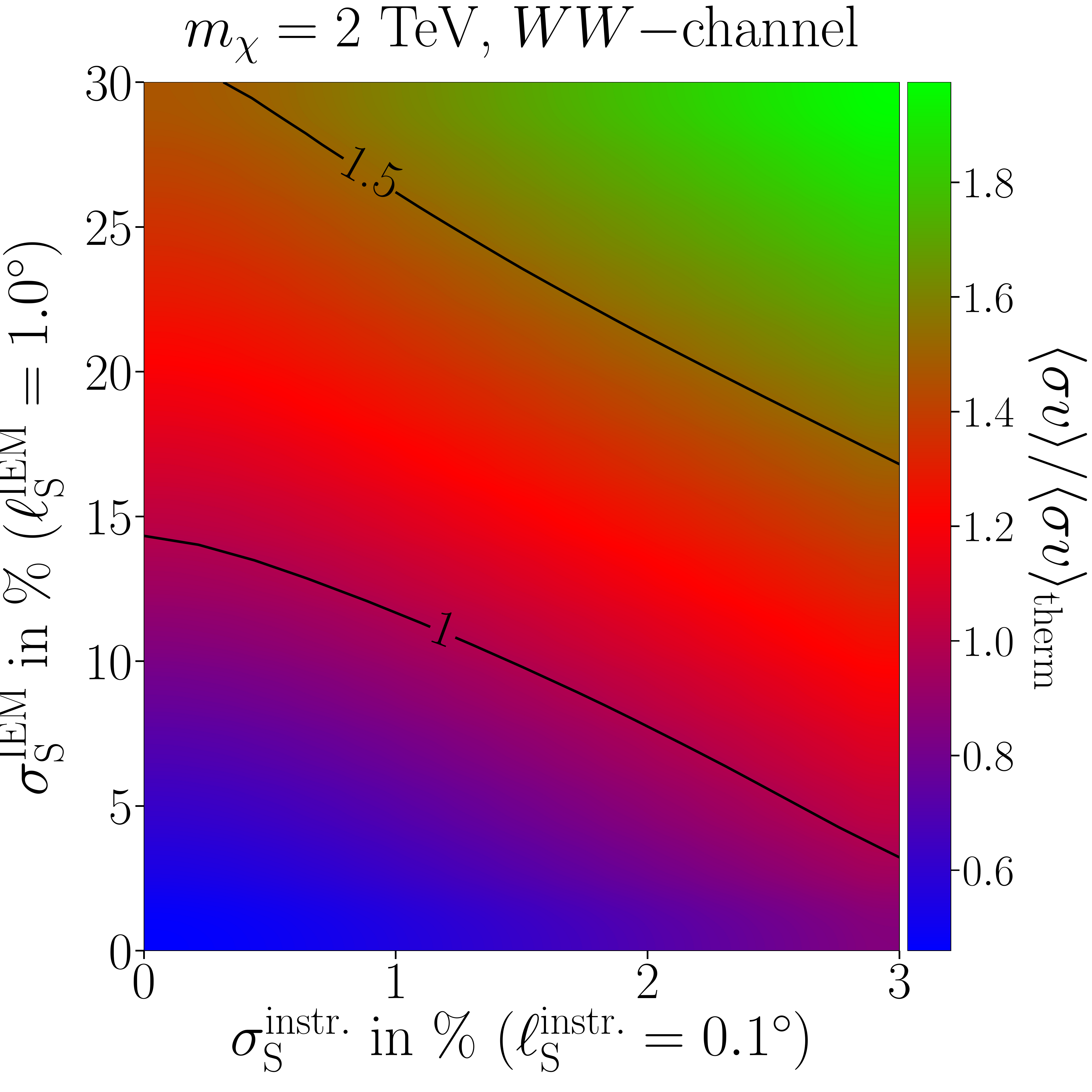}
\caption{%
{\it Left:} The maximal uncertainty in the GDE template that 
can be tolerated for CTA to reach the `thermal' cross-section sensitivity,
as a function of the amplitude 
and spatial correlation length of the IEM systematic errors, for fixed levels of 
overall instrumental systematic errors of $\sigma_S^\textrm{instr}=1$\% (solid line)
and $\sigma_S^\textrm{instr}=0$\% (dashed line).
{\it Middle:} 
The maximal uncertainty in the overall IEM systematic error for CTA to reach the `thermal' cross-section 
sensitivity, as a function of the DM mass, for $\sigma_S^\textrm{instr}=1$\% (solid line)
and $\sigma_S^\textrm{instr}=0$\% (dashed line).
{\it Right:}
{Mutual impact of instrumental and GDE model systematic uncertainties on DM limits 
(for the annihilation channel $\chi\chi\rightarrow W^+W^-$, at fixed DM $m_{\chi} = 2$ TeV and for an 
Einasto profile), fixing $\ell_{\rm S}^{\rm instr.}=0.1^\circ$ and $\ell_{\rm S}^{\rm GDE}=1.0^{\circ}$. 
The colour scheme visualises the 
upper limit on the annihilation cross-section in units of the thermal annihilation cross-section for a given 
set of fluctuation amplitudes $\left(\sigma_{\rm S}^{\rm instr.}, \sigma_{\rm S}^{\rm GDE}\right)$ in 
percent.}
\label{fig:uncertainty_GDE}}
\end{figure}

A different approach, more in line with our general treatment of systematic uncertainty, is to parameterise the {GDE} 
uncertainties via the correlation matrix. In Figure \ref{fig:uncertainty_GDE} we define the part of the parameter 
space $l_s ^{\rm GDE}$ vs.~$\sigma_s ^{\rm GDE}$  for the GDE uncertainty for which we can reach the thermal 
cross-section
(shown in green). In the left panel of that figure we focus on a fixed DM mass of 2\,TeV, in two limiting cases 
where the instrumental systematic uncertainty $\sigma_S^\textrm{instr}$ is negligible (dashed lines) or  
corresponds to our benchmark-setting (solid lines). This figure demonstrates that, realistically speaking, it will be 
very challenging to reach the thermal cross-section if the {GDE} uncertainty significantly 
exceeds $\sim$5 - 10\,\%, at least for correlation lengths comparable  to the DM signal shape 
(roughly $0.2^\circ-1^\circ$). To put this into perspective, we stress that template analyses based on Fermi-LAT
data have already successfully been used to identify new emission components at the level of $\sim 10\%$ of 
the data, for example the Galactic Center Excess and the Fermi  Bubbles~\cite{TheFermi-LAT:2017vmf}. 
One of  the main `tools' to identify new emission components or inadequate modelling is to inspect fit residuals, 
guiding model improvement in an iterative procedure. We envision that a similar approach will be possible 
also with CTA data.
In the middle panel of that figure, we show that an even better understanding of the {GDE} component
at the few-percent level
is needed for DM masses both significantly below and above around 1\,TeV.

In the above discussion, we have mostly emphasised whether the specific value of the `thermal' cross-section can 
be reached. On the other hand, even for many thermally produced DM candidates, one expects annihilation rates 
that can easily be a factor 
of a few above this value. We, therefore, present a complementary view of the above considerations by showing, 
in  the right panel of Fig.~\ref{fig:uncertainty_GDE},  the  cross-section that can be probed in the
$\sigma_S^\textrm{GDE}$ vs.~$\sigma_S^\textrm{instr}$ plane for  a given choice  of the spatial correlation 
lengths ($l_S ^\textrm{GDE}=1 ^\circ$, $l_S ^\textrm{instr}=0.1 ^\circ$). As the figure shows, CTA will be 
able to probe models with only slightly enhanced annihilation rates even in the presence of instrumental 
systematic errors exceeding the current design goal {\it and} {GDE}  uncertainties as large as $30\%$ (and more).
Let us briefly mention that the above discussion is also closely related to 
that of applying a mask, which is the traditional choice of limiting the impact of such uncertainties. 
We take a more detailed look at this in Appendix \ref{app:masking}, concluding that for a template analysis the 
benefit of masking is at best unclear given that uncertainties in the GDE component are distributed 
across the entire ROI.

\subsubsection*{Localised sources and Fermi bubbles}
\label{sec:PS_FB}

Apart from the cosmic-ray induced interstellar emission, we expect two further 
contributions in our ROI, namely those connected to the low-latitude end of the Fermi bubbles as well as 
sub- and above-threshold sources. Here we briefly investigate how uncertainties in these components affect 
our DM limits.

\begin{figure}[t]
\centering\includegraphics[width=0.71\linewidth]{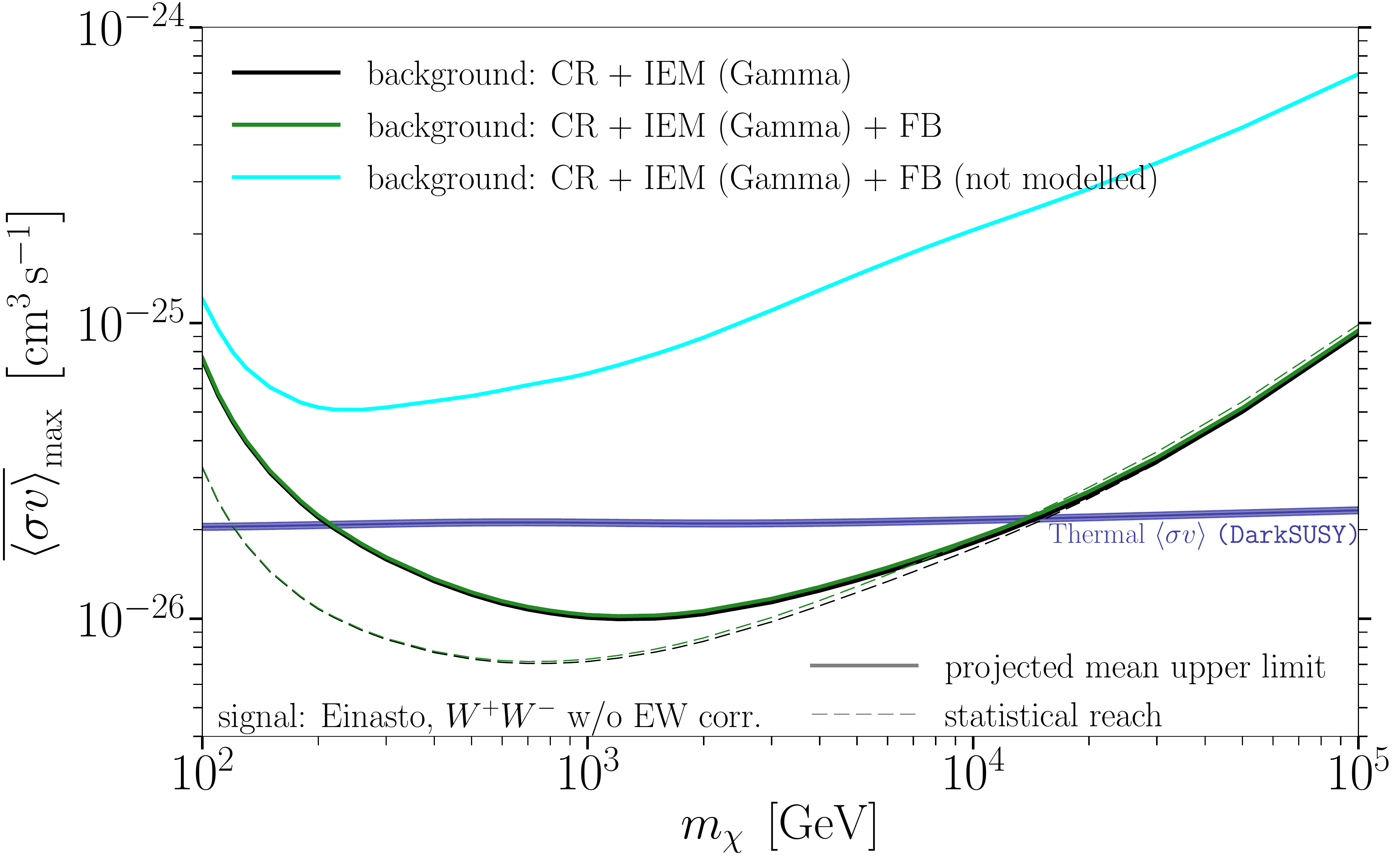}
\caption{CTA sensitivity to a DM signal for our standard analysis settings (black solid line, as in 
Fig.~\ref{fig:bb_MCband}), after adding an FB template (green solid line, largely overlapping with the black solid line)
and the result of an analysis where the FB emission is present in the mock data, but not accounted for 
in the fitting procedure (cyan solid line). Dashed lines show the `statistical' reach (neglecting systematic 
uncertainties in the spatial templates) for the former two cases.
\label{fig:sensitivity_FB}}
\end{figure}

{\bf Fermi Bubbles}: In Fig.~\ref{fig:sensitivity_FB} 
we show the impact of adding the FB template to our benchmark set-up. As anticipated, this impact is rather 
limited because there is almost no degeneracy between the symmetrical DM and the off-centre FB templates. 
We remark, however, that if the FB emission (and in particular its 
morphology) turns out to be significantly different at CTA energies compared to the extrapolation based on the 
Fermi-LAT data, the impact of this emission component might be higher. To illustrate this, we indicate in the
figure the limiting case  where the FBs are left completely unmodelled; for TeV DM, this would (artificially)
worsen the DM sensitivity  by up to an order of magnitude. Studying this effect in more detail, 
which would involve studying the systematic uncertainties related to the physics modelling of the FBs, is beyond 
the scope of this  work. 

{\bf Non-diffuse sources}: Concerning the emission from bright (resolved) localised sources, our baseline procedure is to 
mask them in order to limit their impact on the fitting procedure. While such a procedure works well 
with mock data, it was realised already with the Fermi-LAT analysis that 
masking in crowded regions poses several challenges related to the choice of a proper mask size and the  
modelling of the source emission that `leaks' outside the mask. Besides, cumulative emission from sub-threshold 
sources could result in a significant diffuse emission component that is challenging to model. A detailed treatment of these issues is also beyond the scope of this work. Here we instead limit ourselves to verifying that leakage of sources, for the current five bright sources we consider, affects the limits by roughly $30\%$ (when we mask sources with a $0.1^\circ$ mask, while the emission effectively extends out to approximately $0.5^\circ$ for lower energies). We further checked explicitly that sub-threshold sources will not affect the DM sensitivity with respect to the
benchmark scenario as long as this contribution is modelled `perfectly',\footnote{%
Recall that our 
template for sub-threshold point sources only describes one (specific) realisation of the source population. 
A dedicated study of modelling sub-threshold source populations is ongoing, 
which will allow an improved treatment of this component in follow-up work.
} 
c.f.~Fig.~\ref{fig:sensitivity_GDE} (left
panel) for a similar conclusion regarding IE models. In the most pessimistic case on the other hand, 
where such a component is present  in the data but left unmodelled, the impact on the DM sensitivity can reach 
up to a factor of a few; unlike in the case explored in Fig.~\ref{fig:sensitivity_GDE} (right panel), we note 
that this typically results in an artificial {\it strengthening} of the projected DM constraints, 
as the IE model overcompensates for the sub-threshold emission.
We note however that, similar to the case of IEM, the potential impact of mis-modelling this component 
is likely much higher in the case of the {\it detection} of a potential DM signal.

\section{Summary}
\label{sec:conc}

The CTA observatory will have excellent charged particle discrimination, an improved angular resolution, a 
wide energy coverage and a larger effective area than currently operating atmospheric Cherenkov telescopes, 
making it a promising instrument to detect large-scale gamma-ray emission. 
All these attributes of CTA are also highly relevant for DM 
searches, motivating the expectation that CTA will significantly improve upon the DM sensitivity 
of existing IACTs and possibly make a major discovery.
Indeed, probing a potential signal from annihilating DM has long been stated as one of the  primary scientific 
goals of CTA which is soon entering the construction phase.\footnote{%
The first telescope installed on a CTA site is LST‐1, which is a prototype LST at the CTA‐North site. LST‐1 
is intended to become the first of four LSTs installed at that site.
}

This article represents the most detailed assessment of the CTA sensitivity to DM 
signals at the GC so far, taking 
into account details of the planned observational strategy of that region and the latest IRF versions. 
We also use state-of-the-art modelling of the emission components in this region,  including interstellar 
emission (IE), known and unknown (subthreshold) sources and Fermi bubbles, based on  
currently  available measurements coming mainly from Fermi-LAT and H.E.S.S. data. 

From the perspective of fundamental physics, our  most important finding is the confirmation that CTA 
will  be able to reach the `thermal' cross-section for TeV-scale DM, a milestone for probing the WIMP 
paradigm at these masses, for a large range of well-motivated assumptions about the instrument's performance; 
Figs.~\ref{fig:thermal_req} and \ref{fig:uncertainty_GDE}
present a new and refined way of presenting these conclusions.
In particular, these figures quantify the level to which the instrumental systematic errors need to 
be controlled if CTA is to achieve its designated goal of testing the `thermal' annihilation cross-section of DM.
We also find that the existence of a cored DM distribution would clearly deteriorate CTA's sensitivity to a DM signal  
-- but only by a factor of a few, c.f.~Fig.~\ref{fig:cored_templates_wwo_extended},
i.e.~significantly less than expected from earlier studies adopting smaller ROIs. While the 
`thermal' cross-section would likely (just) be missed in that case, it is important to stress that there
is still a plethora of well-motivated particle physics models predicting DM candidates which would 
leave a  sizeable gamma-ray signal in CTA data even in this case (for example in the context
of simple supersymmetric models~\cite{Athron:2017qdc,Athron:2017yua}).

In arriving at the above conclusions, a major motivation of our work was also to explore the most promising 
data analysis procedures. We therefore confronted the traditional ON/OFF analysis technique with a template fitting 
analysis procedure, still not yet widely explored by IACT collaborations.  In Appendix \ref{app:onoff} we 
demonstrate that for the  set  of  Fermi-LAT inspired IE models and 
given the CTA sensitivity, this technique leads to a decidedly better performance.  
As is typical for template  fitting procedures, the main source of  uncertainty lies in the systematic errors 
originating from the event classification and/or the modelling of the emission components. We include such 
systematic errors directly in the likelihood,  in a parametric way, thereby accounting for their spatial and energy 
correlations (sections \ref{sec:bm_sys} and \ref{subsec:GDEbintobin}). Ideally, the output of such an `agnostic' 
approach to studying the impact of instrumental systematic errors 
can be used for future IRF optimisation once CTA is fully operational.

It is worth stressing that the relative impact of the various sources of uncertainties on the DM sensitivity 
was difficult to judge prior to this study. For example, CTA is expected to greatly advance the measurement of 
large-scale interstellar emission at TeV energies, thus effectively identifying an additional background
 component and thereby potentially lowering  the constraining power of CTA to a DM signal (compared to the situation
 where such a background would not be present). An important 
 result of our analysis is that the impact of this large-scale diffuse component is rather limited, 
as long as the associated modelling uncertainties are not very large 
($\sigma_S^{\rm IE}\lesssim 10\%$). While a smaller value of $\sigma_S^{\rm IE}$ presently 
appears rather optimistic given the best existing models, 
it is expected to become a realistic assumption once  
the IE models can be tuned to actual CTA measurements. 
In that case {\it instrumental} systematic uncertainties will continue to play the dominant role in constraining a 
signal. 
For larger uncertainties in the IE component on the other hand, 
the impact on the DM sensitivity can be comparable to that from purely instrumental effects.

\begin{figure}[t]
\centering\includegraphics[width=0.49\linewidth]{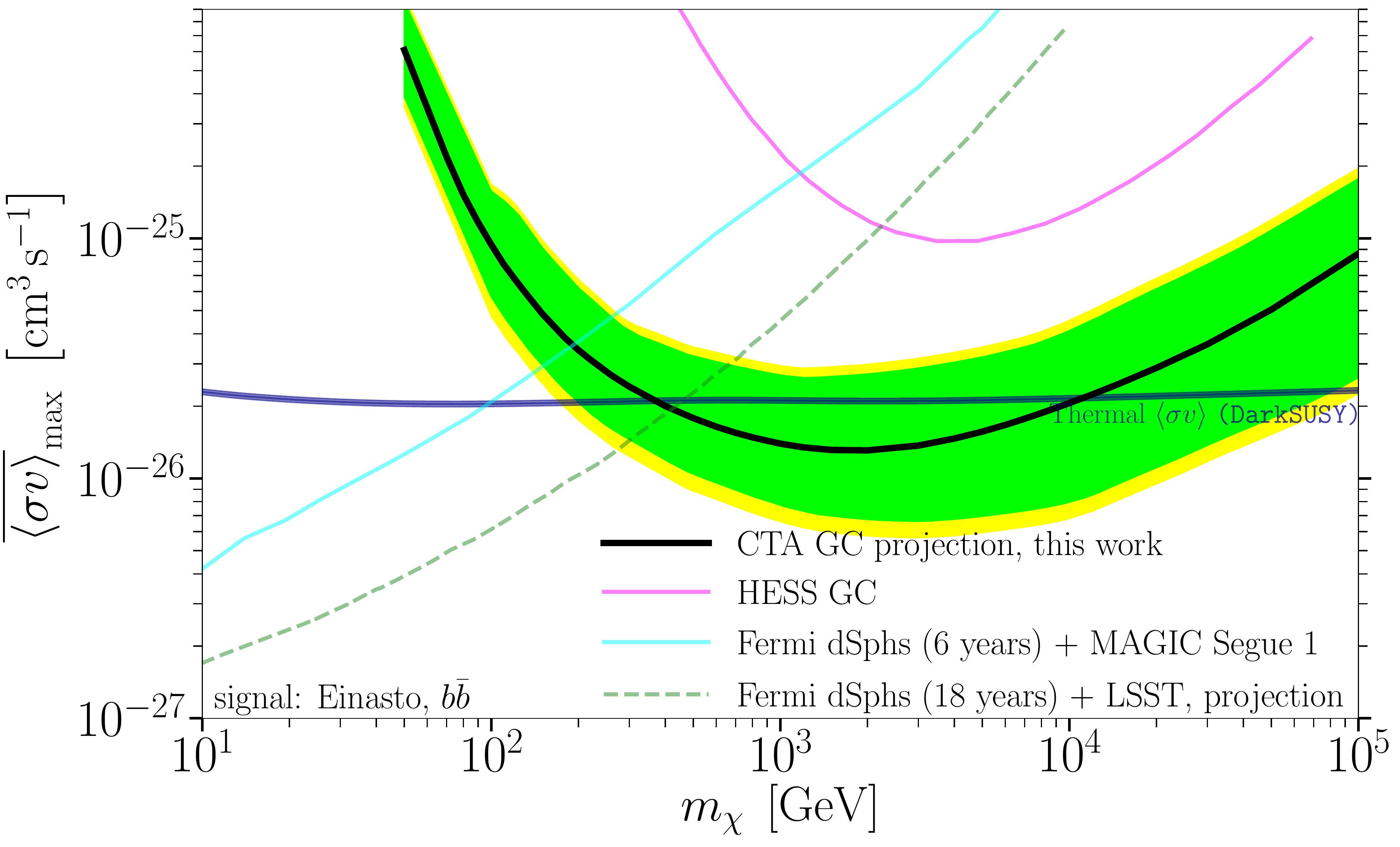}
\centering\includegraphics[width=0.49\linewidth]{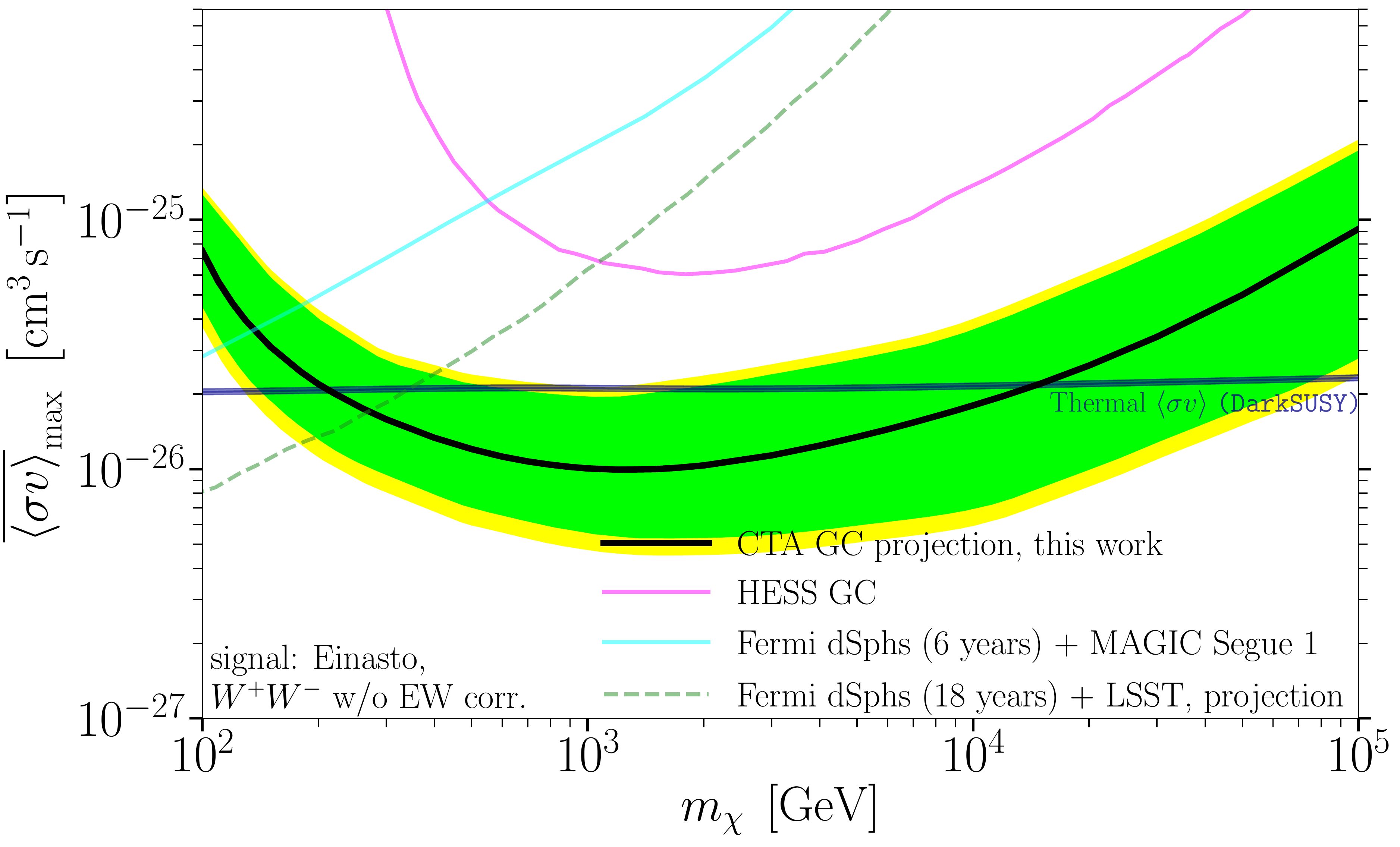}
\caption{The CTA sensitivity curves derived in this work (black line, see Fig.~\ref{fig:bb_MCband}) for the 
$b{\bar b}$ ({\t left}) and $W^+W^-$ ({\it  right}) channels, shown together with the current limits from 
Fermi-LAT observation of dSph  galaxies (cyan) \cite{Ackermann:2015zua} and H.E.S.S. observations of 
the GC (purple) \cite{Abdallah:2016ygi}. In addition we show the projection \cite{Drlica-Wagner:2019xan} 
of the Fermi-LAT sensitivity where future dSphs discoveries with LSST  are taken into account (dashed green). 
Note that the projected sensitivity of CTA shown here includes our estimate of systematic uncertainties
(1\% overall normalisation error and a spatial correlation length of $0.1^\circ$);
for the corresponding results for the initial construction configuration of CTA, see Appendix \ref{app:Phase1}.
\label{fig:summary-sensitivity}
}
\end{figure}

Sub-threshold sources could turn out to be an additional important source of 
systematics, though based on the current study their impact appears to be more limited.
We also included templates for
the Fermi bubbles in our analysis, finding that their 
impact on DM limits is generally subdominant compared to that of the IE (Section \ref{sec:PS_FB}). 
This conclusion rests on the assumption that the 
underlying models are sufficiently reliable, a hypothesis that strictly speaking will only be possible to test with 
sufficient accuracy once actual data are present. While we do not expect major impacts on the expected {\it limit} 
on a DM signal that CTA will be able to put, the impact of modelling uncertainties in these components on 
a potential {\it signal} claim would likely be much larger. Compared to the case of the IE template,
which can be improved with input from complementary observations, it may not be straight-forward 
to reduce the effects of
sub-threshold sources or the Fermi bubbles. In that case, these emission components would potentially present 
an irreducible background that would prevent a robust DM discovery.

To provide a somewhat broader perspective, we conclude by comparing, in Fig.~\ref{fig:summary-sensitivity}, 
the CTA sensitivity derived in this 
work to current (and, for the case of Fermi-LAT, projected) 
DM limits from complementary observations. 
Noting that the CTA sensitivity projections take into account instrumental systematic uncertainties, 
this figure nicely illustrates how CTA will improve on the pioneering
work of the current generation of IACTs 
to test WIMP DM, and to significantly 
extend the range of DM masses where we can robustly probe the theoretically important benchmark that is
provided by the thermal annihilation rate. In that sense CTA will indeed provide a unique opportunity 
to test the WIMP paradigm, in particular when keeping in mind that even annihilation rates a factor of a few 
larger than the `thermal' annihilation rate are not uncommon among proposed models to explain the
particle nature of DM. 
In Appendix \ref{app:Phase1} we demonstrate that much of the discovery space remains available 
even for the reduced 
observational programme associated with the initial  construction configuration of CTA -- though of course 
the baseline array considered in the main text will clearly probe more of the critical 
parameter space and hence have a significantly better leverage to test the WIMP hypothesis.
For either of these scenarios we believe that the chance to obtain unique clues about the nature of DM, newly 
assessed and confirmed here, makes this part of CTA's science programme truly imperative.

\newpage
\input{AppendixA}

\bigskip\bigskip

\newpage
\input{CTA_Acknowledgments_July2020_v2.tex}

\bibliographystyle{JHEP}
\bibliography{CTA_GC.bib}

\end{document}

%% file: author-list-jcap.tex
\author{A.~Acharyya,$^{1}$}
\author{R.~Adam,$^{2}$}
\author{C.~Adams,$^{3}$}
\author{I.~Agudo,$^{4}$}
\author{A.~Aguirre-Santaella,$^{5}$}
\author{R.~Alfaro,$^{6}$}
\author{J.~Alfaro,$^{7}$}
\author{C.~Alispach,$^{8}$}
\author{R.~Aloisio,$^{9}$}
\author{R.~Alves Batista,$^{10}$}
% ...

 \bigskip 
 \bigskip 
\affiliation{$^{1} \ $Dept. of Physics and Centre for Advanced Instrumentation, Durham University, South Road, Durham DH1 3LE, United Kingdom}

\affiliation{$^{2} \ $Laboratoire Leprince-Ringuet, École Polytechnique (UMR 7638, CNRS/IN2P3, Institut Polytechnique de Paris), 91128 Palaiseau, France}

\affiliation{$^{3} \ $Department of Physics, Columbia University, 538 West 120th Street, New York, NY 10027, USA}

\affiliation{$^{4} \ $Instituto de Astrofísica de Andalucía-CSIC, Glorieta de la Astronomía s/n, E-18008, Granada, Spain}

\affiliation{$^{5} \ $Instituto de Física Teórica UAM/CSIC and Departamento de Física Teórica, Campus Cantoblanco, Universidad Autónoma de Madrid, c/ Nicolás Cabrera 13-15, Campus de Cantoblanco UAM, 28049 Madrid, Spain}

\affiliation{$^{6} \ $Universidad Nacional Autónoma de México, Delegación Coyoacán, 04510 Ciudad de México, Mexico}

\affiliation{$^{7} \ $Pontificia Universidad Católica de Chile, Av. Libertador Bernardo O'Higgins 340, Santiago, Chile}

\affiliation{$^{8} \ $University of Geneva - Département de physique nucléaire et corpusculaire, 24 rue du Général-Dufour, 1211 Genève 4, Switzerland}

\affiliation{$^{9} \ $INFN Dipartimento di Scienze Fisiche e Chimiche - Università degli Studi dell'Aquila and Gran Sasso Science Institute, Via Vetoio 1, Viale Crispi 7, 67100 L'Aquila, Italy}

\affiliation{$^{10} \ $Instituto de Astronomia, Geofísico, e Ciências Atmosféricas - Universidade de São Paulo, Cidade Universitária, R. do Matão, 1226, CEP 05508-090, São Paulo, SP, Brazil}

% ...

%% file: AppendixA.tex
\newpage

\appendix

\section{Initial construction configuration}
\label{app:Phase1}

\begin{figure}[t!]
\centering
\includegraphics[width=0.49\linewidth]{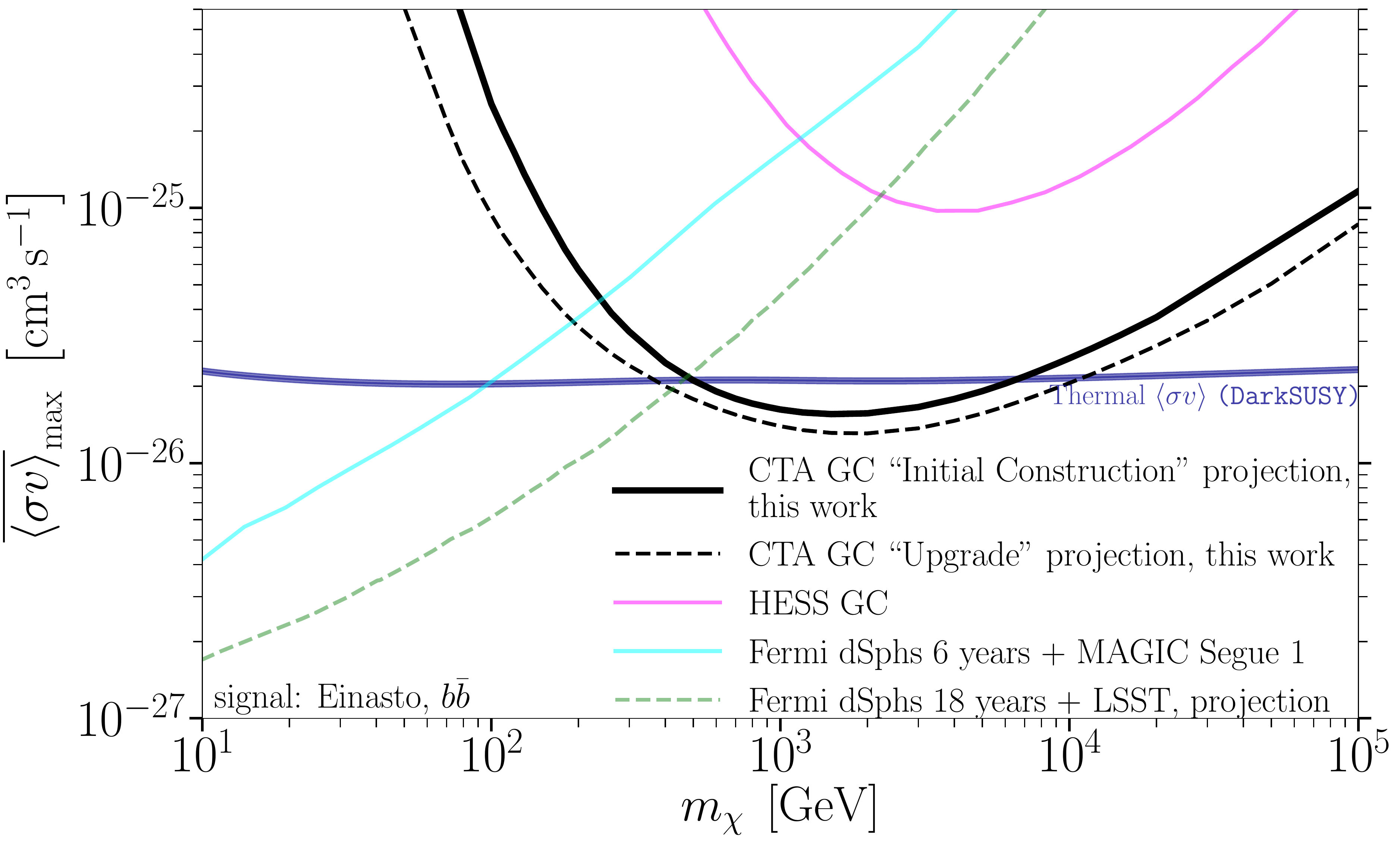}
\includegraphics[width=0.49\linewidth]{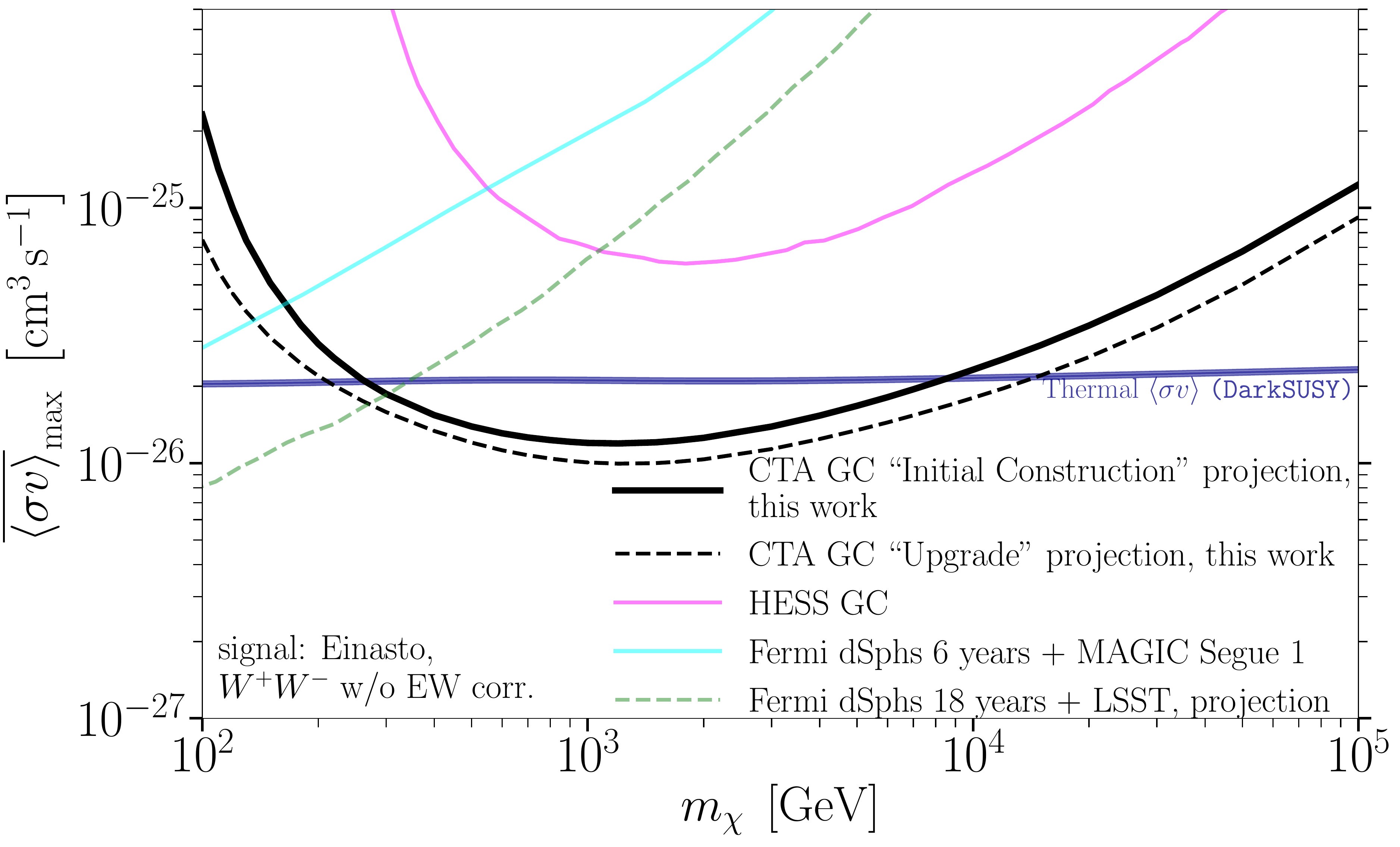}
\caption{Same as Fig.~\ref{fig:summary-sensitivity}, but the solid black line now shows the 
sensitivity projection for the reduced initial construction configuration (while the dashed black line shows the result for 
the benchmark analysis setting presented in the main text).
\label{fig:Sensitivity-phase1}} 
\end{figure}

Given the substantial investment in infrastructure that is required for an instrument with the size of CTA, 
it is not surprising that current planning calls for the telescope arrays to be constructed in phases. 
In the main text we have 
discussed the `baseline array', i.e.~the array configuration corresponding to the original design goal. 
In this Appendix we instead consider a slimmed-down (initial) construction configuration and discuss the
impact of this preliminary configuration on CTA's sensitivity to a DM signal and  its ability to test the
WIMP paradigm. 
It is worth stressing that the construction configuration could be realised with the funding that is currently 
available at the time of this writing.

 The reduced South array we considered is composed of 15 MSTs, 50 SSTs and no LST, which compares to 
 a baseline South array of 4 LSTs, 25 MSTs and 70 SSTs that was considered in the main text.
Here we follow exactly the same analysis steps as described in the main text, in particular concerning the 
treatment of systematic errors, but generate templates and mock data based on IRFs describing this
initial configuration instead.

In Fig.~\ref{fig:Sensitivity-phase1} we illustrate the projected sensitivity for this array configuration (black solid lines)
in analogy to Fig.~\ref{fig:summary-sensitivity} in the main text, including for convenience also the sensitivity
for the full baseline array derived there (black dashed lines). 
The loss in sensitivity of the reduced array is clearly visible and can, for DM masses above 200\,GeV, 
mainly be attributed to the reduction in the number of MSTs; for smaller DM masses the lack of LSTs leads
to a further clearly visible decrease in sensitivity (see also Appendix \ref{TelescopeTypes}). 
When only focussing on this direct comparison between the two array layouts, the difference between 
the two configurations 
may still not appear very dramatic.
However, in comparison to expected results from complementary techniques, in particular the projected 
limits from Fermi LAT, it becomes clear that this impression is misleading. While
there are many WIMP realisations somewhat above the `thermal' line, the number increases substantially 
as one gets close to the line (and slightly below it). Losing the opportunity to robustly exclude annihilation 
cross‐sections within a factor of a few around this 'thermal' value thus results in a significant loss in 
theoretical models that can be probed, correspondingly diminishing the prospects for the detection of 
thermally produced DM. Accordingly, it remains a critical goal to eventually reach the baseline CTA 
configuration that is discussed in the main text.

\begin{figure}[t!]
\centering
\includegraphics[width=0.6\linewidth]{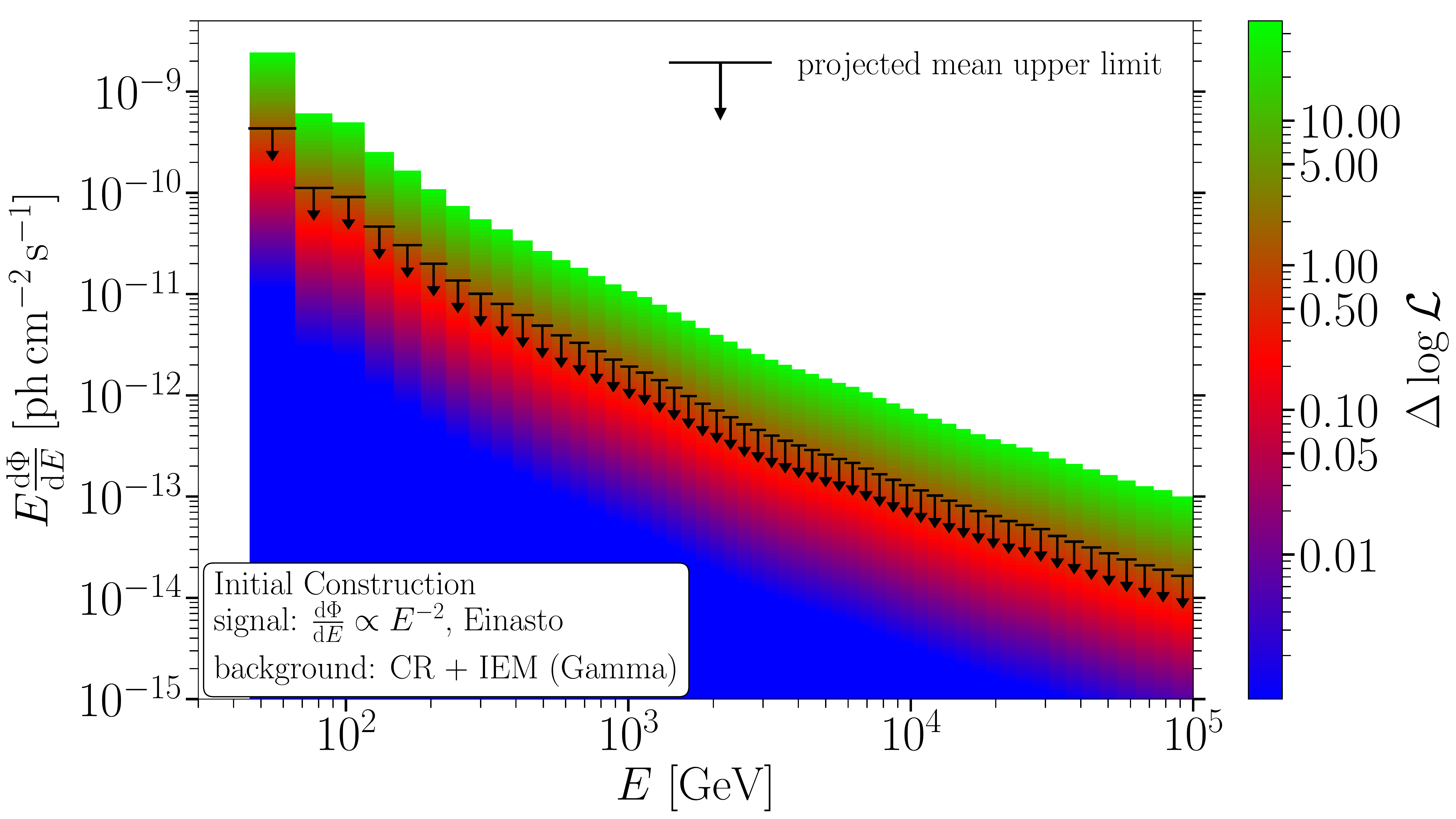}\hspace*{0.5cm}
\includegraphics[width=0.38\linewidth]{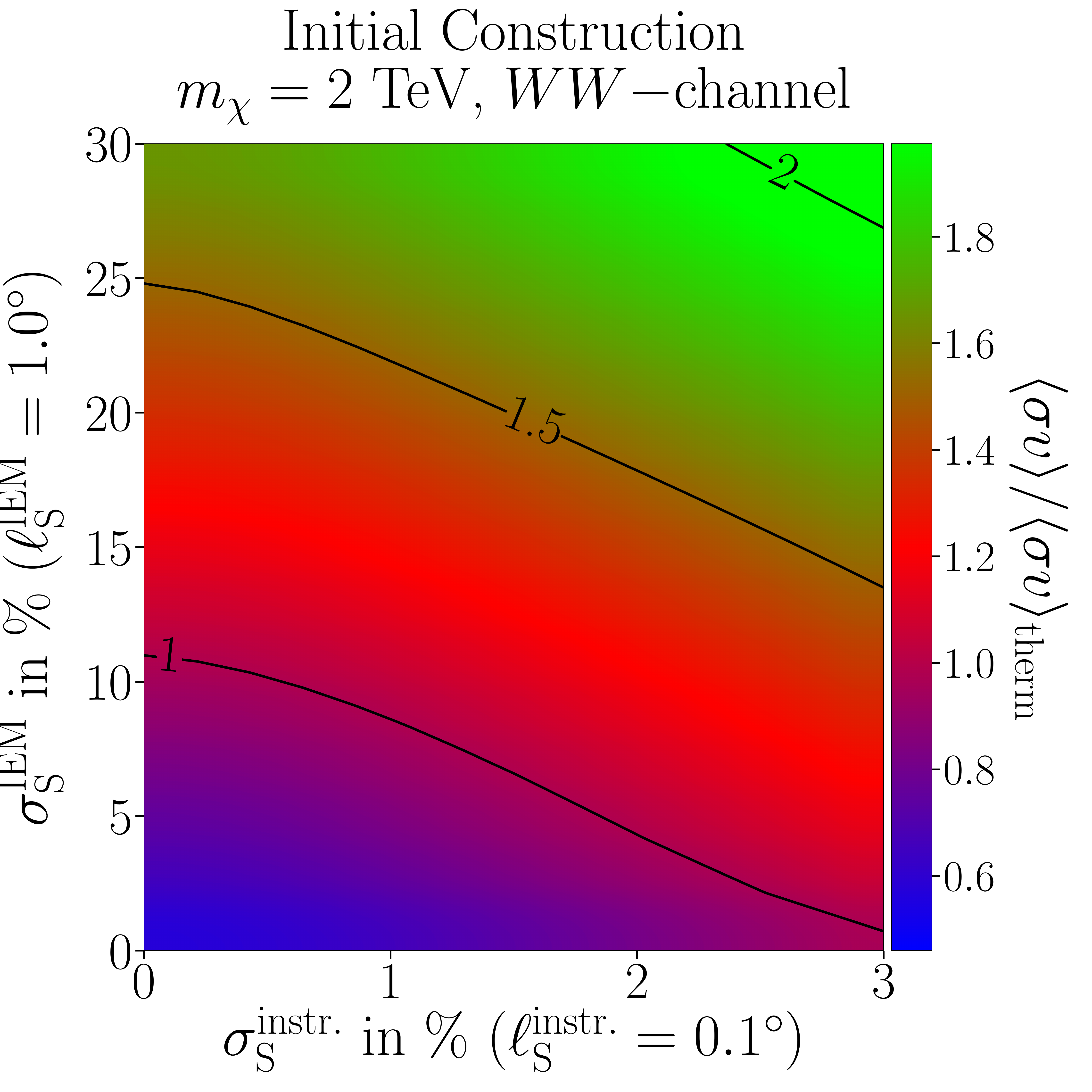}
\caption{{\it Left.} Differential flux (times $E$) as a function of source energy $E$, with the (logarithmic) colour scale indicating the change in the total likelihood due to the contribution from a DM signal in the respective energy bin.
Same as Fig.~\ref{fig:flux_sensitivity_DMGDE}, but with respect to the initial construction configuration of the array.
Also in this case the full likelihood table is available for download at 
zenodo~\cite{bringmann_torsten_2020_4057987}.
{\it Right.}
Mutual impact of instrumental and GDE model systematic uncertainties on the upper limit on the DM 
annihilation cross-section. As in Fig.~\ref{fig:uncertainty_GDE} (right panel), but now with respect to the 
construction configuration.
\label{fig:likelihood_function_systematics_-phase1}} 
\end{figure}

As discussed in section \ref{sec:limit_general}, directly providing the bin-to-bin flux sensitivity to DM signals 
in general allows one to test DM annihilation in the most model-independent way, in particular for models where 
the annihilation spectra deviate from those expected for pure annihilation channels like $\bar bb$ or $W^+W^-$. 
In Fig.~\ref{fig:likelihood_function_systematics_-phase1} (left panel) we show this binned likelihood for
the construction configuration; in particular, this allows us to use Eq.~(\ref{eq:sens_est})
to calculate limits in the same straightforward way as before. Just as for the baseline
configuration, we make this likelihood available in tabulated form in the supplemental material.

We conclude our discussion of the initial CTA configuration by showing, in the right panel of 
Fig.~\ref{fig:likelihood_function_systematics_-phase1}, the required level of instrumental systematic uncertainties and 
modelling uncertainty in the IE component to reach the thermal cross-section. This should be directly compared to 
Fig.~\ref{fig:uncertainty_GDE} in the main text. As expected, a somewhat better control of instrumental
systematic uncertainties is needed to achieve the same performance goals as for the baseline array.

\section{Details of IE models}
\label{app:GDEdetails}

\subsection{Spectral differences in the Galactic Ridge region}
\label{app:Ridge}

\begin{figure}[t!]
\centering
\includegraphics[width=0.71\linewidth]{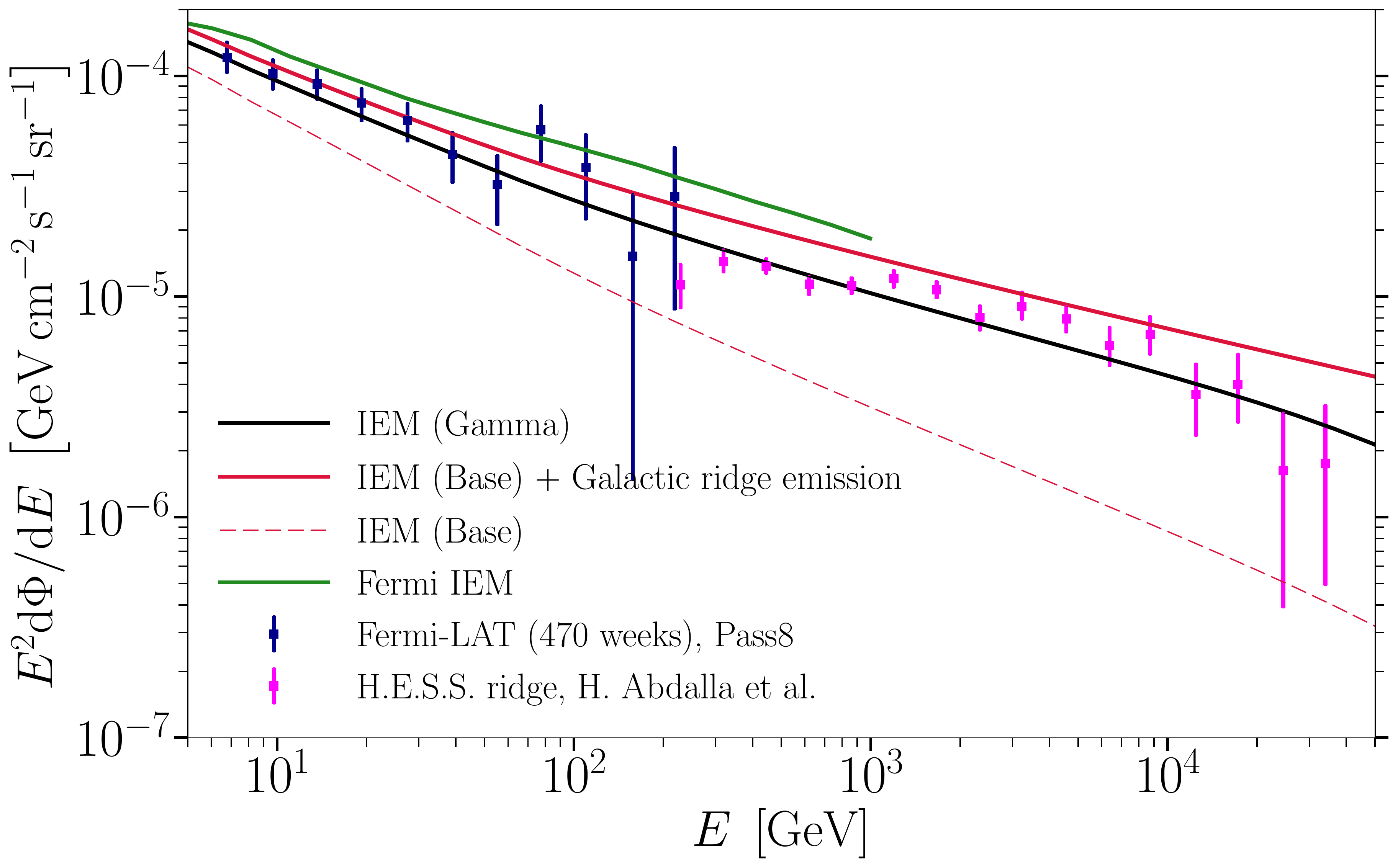}
\caption{Differential $\gamma-$ray flux (times $E^2$) of the Galactic ridge region ($\left|b\right| \leq 0.3^{\circ}$, 
$\left|l\right| \leq 1.0^{\circ}$) as measured by H.E.S.S.~\cite{Abdalla:2017xja} (magenta) and 
\emph{Fermi}-LAT (blue)~\cite{Marinelli:2017csa}. We also show three IE/GDE models as described in Section~\ref{subsec:intro_GDE}: 
two phenomenological models~\cite{Gaggero:2017jts}, {\it Gamma} (black) and {\it Base} + Galactic ridge 
(red), and a  data-driven Pass8 Fermi-LAT diffuse model (green). 
\label{fig:flux-allcomponents-Ridge}} 
\end{figure}

In Fig.~\ref{fig:flux-allcomponents-2} in the main text we compared the expected photon counts resulting 
from our emission components (including all three IE/GDE models that we use) integrated over the full  ROI. 
At TeV energies there are a few existing {\it measurements} of a diffuse component that, e.g.~, have been performed by the 
H.E.S.S.~collaboration \cite{Aharonian:2006au, Abdalla:2017xja} or the VERITAS collaboration \cite{Archer:2016ein, Buchovecky:VERITAS}, albeit mostly restricted to the Galactic Ridge region.
Here we mostly refer to the published results by H.E.S.S.~and in Fig.~\ref{fig:flux-allcomponents-Ridge}, we therefore complement  Fig.~\ref{fig:flux-allcomponents-2} by 
showing how the three IE/GDE models compare to these data in this significantly smaller region. 

Concretely, we plot the flux inside this region as expected from the   {\it Gamma} model (black) and the 
data-driven Pass8 Fermi-LAT diffuse model (green). As already pointed out in the main text, 
the construction of the {\it Base} model (red dashed) does not explicitly 
account for the TeV emission from the Galactic Ridge; we therefore add to this model (red solid)
a spatial template for the Galactic ridge adopted from Ref.~\cite{Aharonian:2006au}, with a spectrum
modelled as a power-law with spectral index $\gamma=-2.28$~\cite{Abdalla:2017xja} and normalised 
to the  H.E.S.S.~measurement.
For comparison we also show in this figure data points from Fermi-LAT (blue) and H.E.S.S.~(magenta).
The former are adopted from  Ref.~\cite{Marinelli:2017csa} and represent 470 weeks of 
Pass8 data after subtracting the emission of known point sources from the 3FGL catalogue.
Neither of these data sets has the isotropic emission component subtracted, which however is negligible
compared to the expected IEM  in this region.

We conclude that all three models are in reasonable agreement with the data in this region, 
and are  therefore also reasonable to use at small scales.

\subsection{Morphological differences}
\label{app:GDEmorphology}

\begin{figure}[t!]
\includegraphics[width=0.48\linewidth]{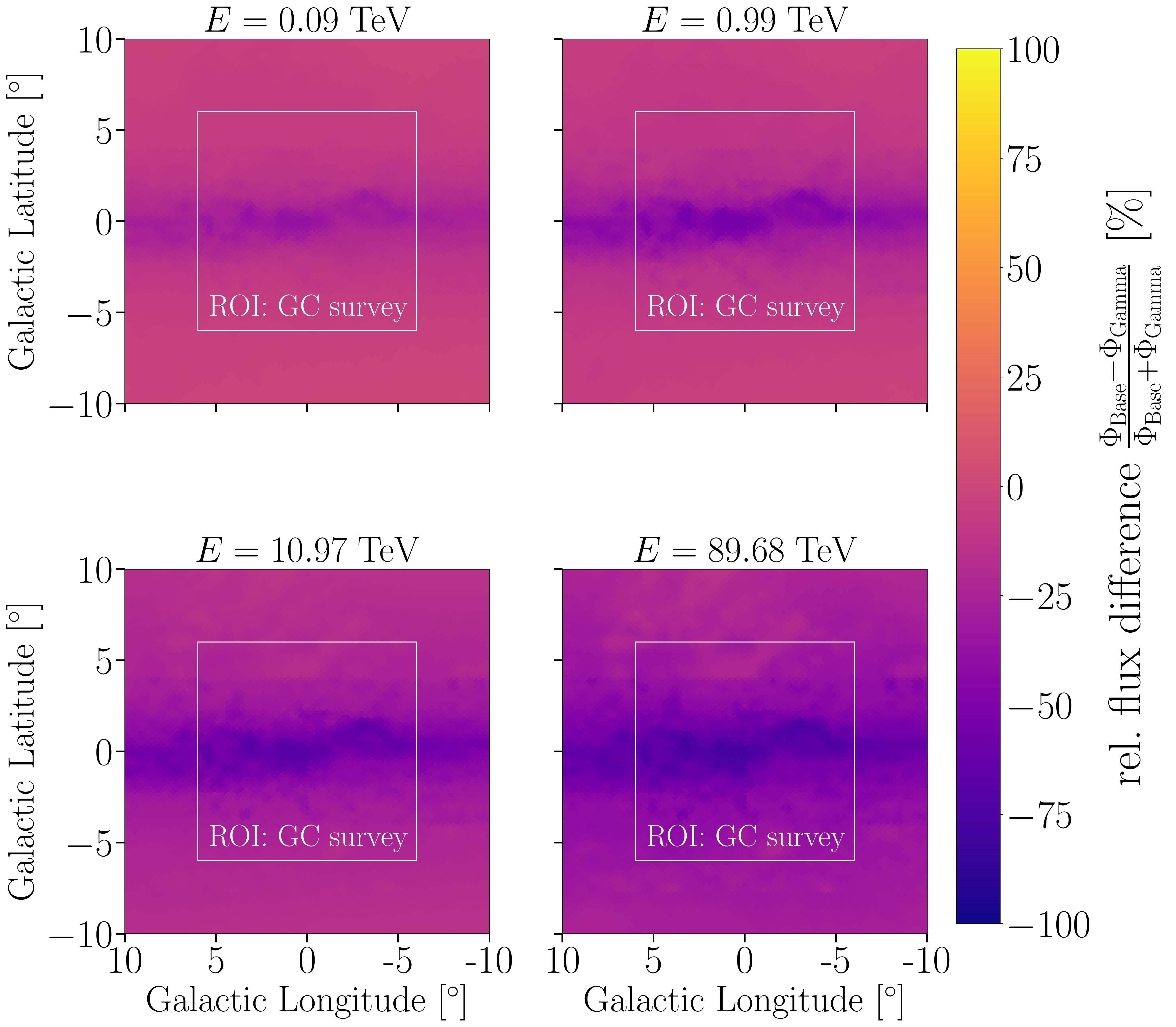}
~~~~\includegraphics[width=0.48\linewidth]{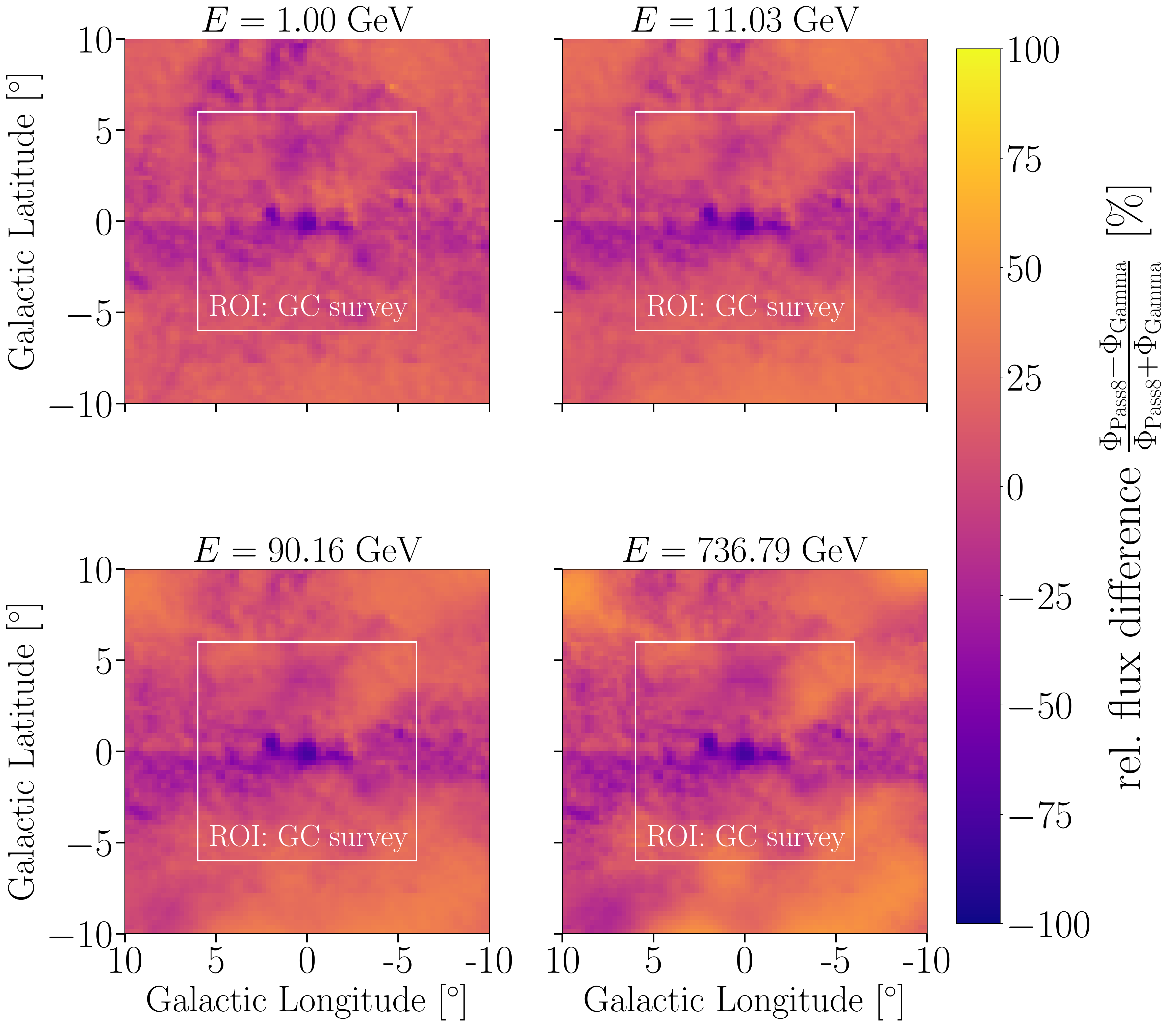}
\caption{{\it Left.} Relative flux difference between the {\it Gamma} and {\it Base} IEMs, 
the square marking our ROI. 
{\it Right.} Same, but for {\it Gamma} and Pass8 Fermi-LAT IEMs.
\label{fig:fluxratio-GammaBase}} 
\end{figure}

As described in the main text, our three benchmark IE models are chosen to represent a fully complementary 
approach to modelling the GC region. Here, we further explore these models by investigating {\it morphological} 
differences between them.
 
The {\it Gamma} and  {\it Base} models, in particular, are produced using the same 
numerical code, {DRAGON~\cite{Evoli:2016xgn}, based on the same target gas and ISRF distribution
(and identical to that given in version v.54 of the GALPROP code~\cite{Vladimirov:2010aq}), so 
the only difference consists in how CR diffusion is treated.  
In the left panel of Fig.~\ref{fig:fluxratio-GammaBase} we show the normalised difference between
the expected flux for these models, spatially resolved in the GC region and for four selected values of the 
gamma-ray energy between about 100\,GeV and 100\,TeV.
This difference is, as expected, largest in the central regions, at a level of up to about $60\%$;
even at the outskirts of our ROI, however, the differences can be larger than $20\%$, especially at higher
energies. We note that these numbers should be roughly compared to the parameter $\sigma_S ^{\rm IEM}$
that we discuss in section \ref{sec:astro_uncertainty} of the main text. 

In the right panel of Fig.~\ref{fig:fluxratio-GammaBase} we perform the same comparison, but for the 
flux ratio between the {\it Gamma} and Fermi-LAT Pass 8 GDE models, where the latter now is based on different 
gas maps. A qualitative difference in this case is that the residuals can be both positive
and negative. The negative residuals displayed here mainly coincide with the location of interstellar clouds.

\subsection{Effect of masking}
\label{app:masking}

\begin{figure}[t!]
\centering\includegraphics[width=0.71\linewidth]{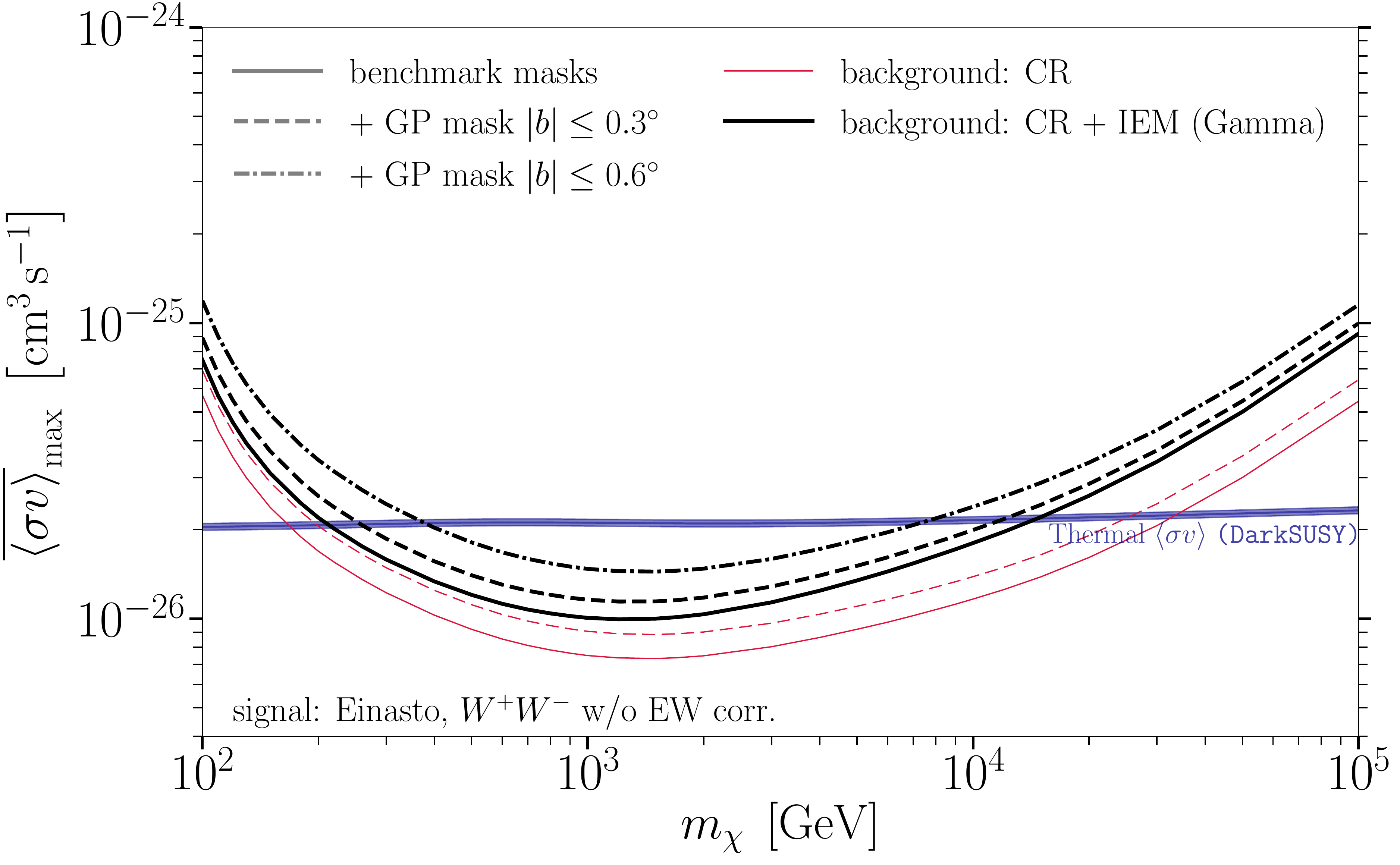}
\caption{CTA sensitivity to a DM signal, and how it is affected by masking the Galactic plane. 
Solid lines indicate the result when not
applying any masks (beyond those for the extended sources shown in the top left panel of Fig.~\ref{fig:template_collection}),
and dashed (dash-dotted) lines indicate the result of implementing additional masks with $\left|b\right|\leq 0.3^\circ$
($\left|b\right|\leq 0.6^\circ$).
Black lines assume the standard CR and IEM background, while purple lines assume instrumental (CR) backgrounds
alone. 
The thick solid line is thus identical to one in the left panel of Fig.~\ref{fig:bb_MCband}, showing the CTA DM
sensitivity as obtained in our benchmark analysis setting.
\label{fig:info-flux-mask}}
\end{figure}

The discussion of the GDE modelling uncertainty is closely related to the choice of the masking, which is traditionally 
used to limit its impact on DM searches. In Fig.~\ref{fig:info-flux-mask} we plot the change in the DM sensitivity 
caused by masking, assuming that we mask the whole galactic plane (GP) out to latitudes of $|b|<0.3^\circ$ and 
$|b|<0.6^\circ$, respectively. We see that masking has a rather limited effect when GDE is 
accounted for (both in the data and in modelling), as shown by the thick lines; in contrast, the decrease in 
sensitivity is more pronounced in the absence of a GDE component (thin lines). 

This is consistent with our Signal-to-Noise Ratio (SNR) studies 
(Appendix \ref{app:InfoFlux}) demonstrating that little information comes from the plane when a diffuse emission is 
present. We note that, while our limits can worsen by up to a factor of 2 (for 2 TeV DM, 30\% IEM uncertainty 
and $0.5^\circ$ correlation length, c.f.~Section \ref{sec:astro_uncertainty}), they worsen by only up to a factor of
1.5 when a $0.6^\circ$ region is masked. If all IE uncertainty was 
localised along the plane, masking would thus indeed be beneficial. However, as just
illustrated in Appendix \ref{app:GDEmorphology}, 
model differences (i.e.~IEM systematic uncertainties) cover most of our ROI, making 
the masking approach challenging to implement.
In practice, this demonstrates that once real data is available, a careful study of the GDE uncertainty will be needed 
and will guide the masking choices and/or interpretation of potential discovery hints.

\section{Further analysis details}
\subsection{Signal-to-Noise ratio and information flux}
\label{app:InfoFlux}

In this section, we study in more detail the SNR in our ROI, which is useful for understanding various
aspects of the general discussion. At a technical level, we use \sword~\cite{Edwards:2017kqw} 
to calculate a generalisation of the 
SNR that also captures the effect of systematic errors, namely the so-called {\it Fisher Information Flux} 
$\mathfrak{F}$~\cite{Edwards:2017mnf}. 
In the statistics-dominated regime, by definition, $\mathfrak{F}$ becomes identical to SNR.

 \begin{figure}[t!]
\centering\includegraphics[width=0.65\linewidth]{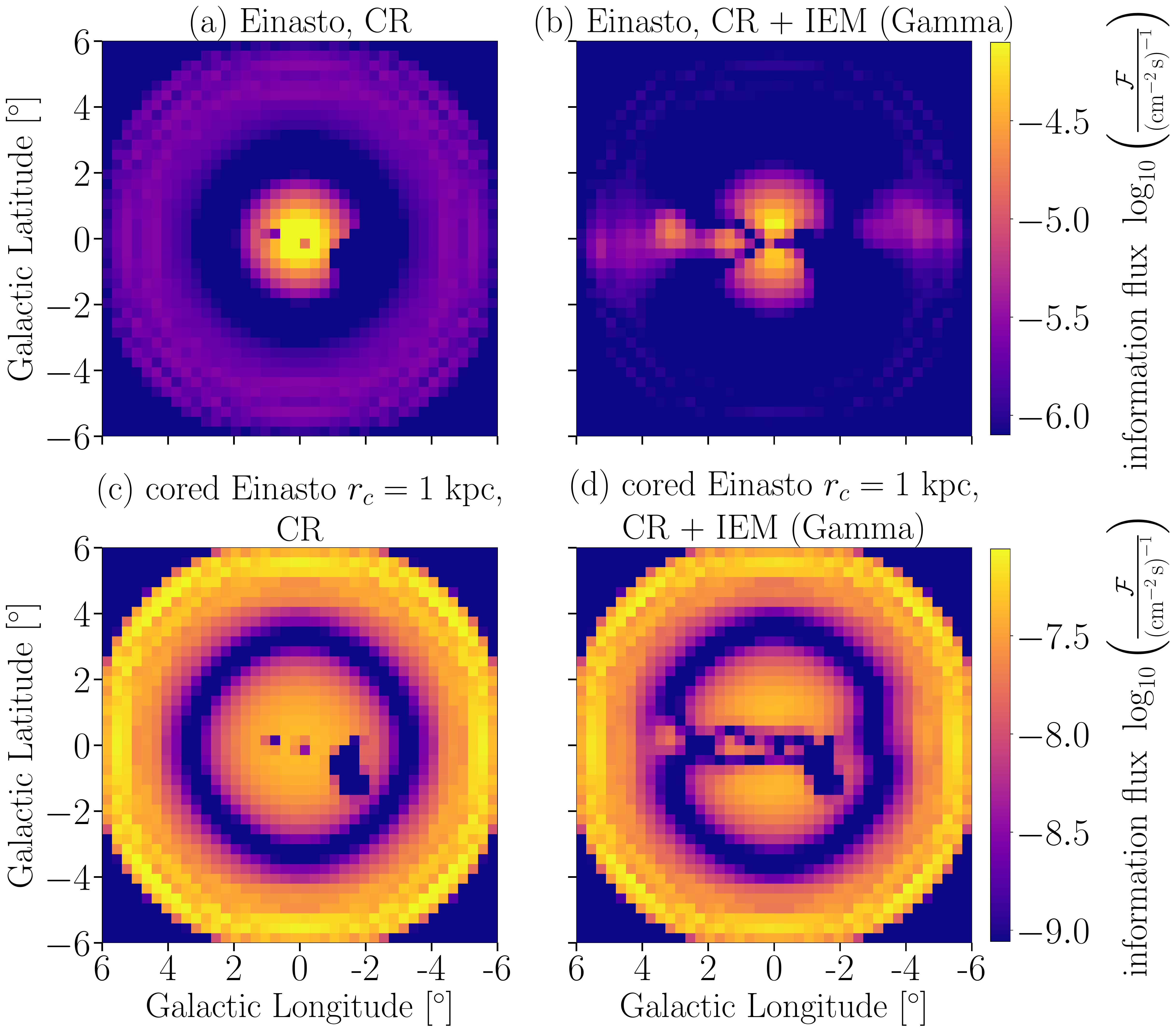}
\caption{{\it Left column:} Information flux $\mathfrak{F}$ (logarithmic colour scale), for Einasto (upper panel) and cored DM profiles (lower panel), 
assuming only the misidentified CR background (and no IE). {\it Right column:} Same, but with the added 
emission based on the {\it Gamma} model. Regions with large $\mathfrak{F}$ yield most of the constraining 
power to a DM signal. These $\mathfrak{F}$ maps are based on the full GC survey, assuming
our benchmark scheme of treating instrumental systematic uncertainties. 
\label{fig:info-flux}} 
\end{figure}

In the left column of Fig.~\ref{fig:info-flux} we show this quantity for the Einasto (upper panel) and a cored Einasto DM 
profile (lower panel), respectively, assuming only the background of misidentified CRs. The resulting 
$\mathfrak{F}$ is spherical, following as expected the signal shape. The blue ring-like feature indicates a sharp drop in 
$\mathfrak{F}$ that separates two regimes: {\it i)} a central region, 
containing information on the signal emission, and {\it ii)} the outer regions that provide constraining information on the 
CR background (which is much more isotropic than any of the other components).  
In this situation, most of the constraining power comes from the inner 0.5$^\circ$ for an Einasto 
profile (which we have used in Section \ref{sec:bm_sys}) and $\sim 1^\circ$  for a cored profile.
Adding an IE component (right column) deforms the spherical region of large $\mathfrak{F}$ into an hourglass shape
for cuspy profiles -- which indeed is the general expectation for the optimal ROI shape in such a 
case~\cite{Bringmann:2012vr}. However, sizeable excesses in $\mathfrak{F}$ also appear along the Galactic plane -- 
though less pronounced than in the centre -- with size and position depending on the chosen spatial 
correlation length $\ell_{\mathrm{S}}$; these excesses are related to regions where more observation time is 
necessary to break the relatively strong degeneracy between IE and DM templates. In comparison, fixing the 
exact normalisation of the CR template is less crucial, which explains the smaller amount of information 
to be gained from the edges of the ROI (as compared to the upper left panel of the figure).
This situation changes somewhat for a cored profile, as shown in the bottom right panel, where the degeneracy
between the IE and DM templates is less severe.  As a consequence, spending observation time on the edges of 
the ROI is still profitable in terms of constraining power. 
The upper right panel of Fig.~\ref{fig:info-flux} provides a solid ground to argue why a Galactic plane mask 
with moderate size does not strongly affect the DM limits, cf.~Fig~\ref{fig:info-flux-mask}. Especially for a 
mask of $\left|b\right| \leq0.3^{\circ}$, the effect is rather mild since the maximal values of $\mathfrak{F}$ are 
concentrated in the upper and lower region of the hourglass outside of the mask.

\begin{figure}[t!]
\centering\includegraphics[width=0.71\linewidth]{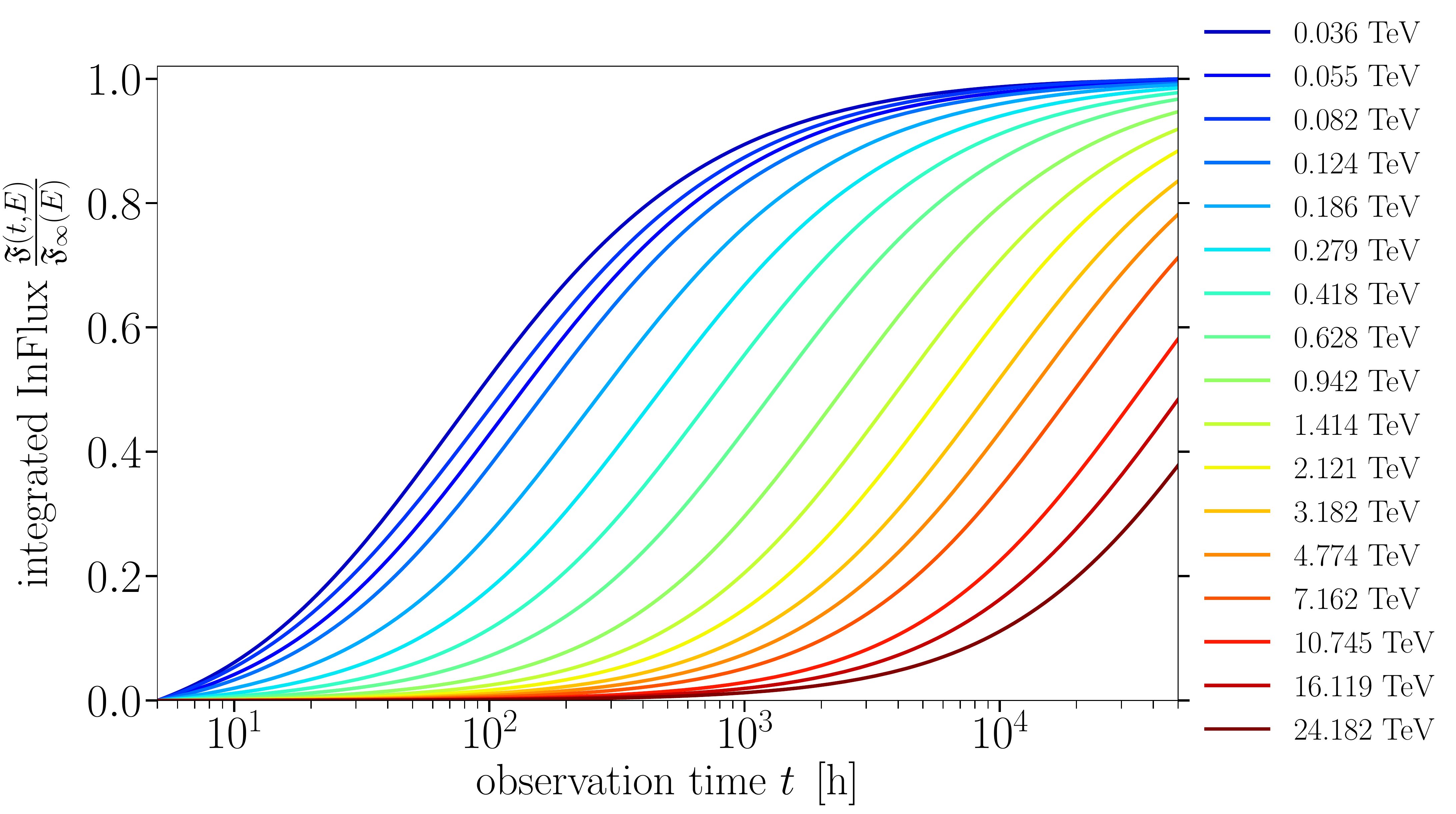}
\caption{Time-integrated information flux $\mathfrak{F}(E,t)$ per energy $E$ as a function of observation time $t$,
normalised to the maximally available integrated information flux $\mathfrak{F}_{\infty}(E)$.
Here we assume a circular ROI of radius $2^{\circ}$ centred on the GC, an Einasto DM density profile, a 
generic DM annihilation spectrum $\Phi\propto E^{-2}$ and the {\it Gamma} IEM, adopting our benchmark 
scheme for treating instrumental systematic uncertainties. In order to avoid excessive use of computational 
resources, the sensitivity 
predictions in this figure are based on only 20 (equally log-spaced) energy bins.
\label{fig:INFL_per_energy_bin}} 
\end{figure}

In Fig. \ref{fig:INFL_per_energy_bin} we show instead how the time-integrated information flux 
evolves with the observation time, for different energy bins (quoted energy values correspond to the central 
energy of each bin). Here, for better comparison, we have (for each energy bin separately) 
normalised $\mathfrak{F}$ to the asymptotic value it would take after an `infinite' observation time.
The figure illustrates that most information at the lowest  energies, for $E\lesssim 200$ GeV, is already extracted 
after around 100\,h of observation time. Gaining a significant share of the in principle available 
information ($\gtrsim0.5$ in this normalisation) 
at TeV energies, on the other hand, clearly requires observation times of at least 1000\,h.

\subsection{Different DM profiles and impact of energy correlations}
\label{app:profiles}

\begin{figure}[t]
\centering\includegraphics[width=0.71\linewidth]{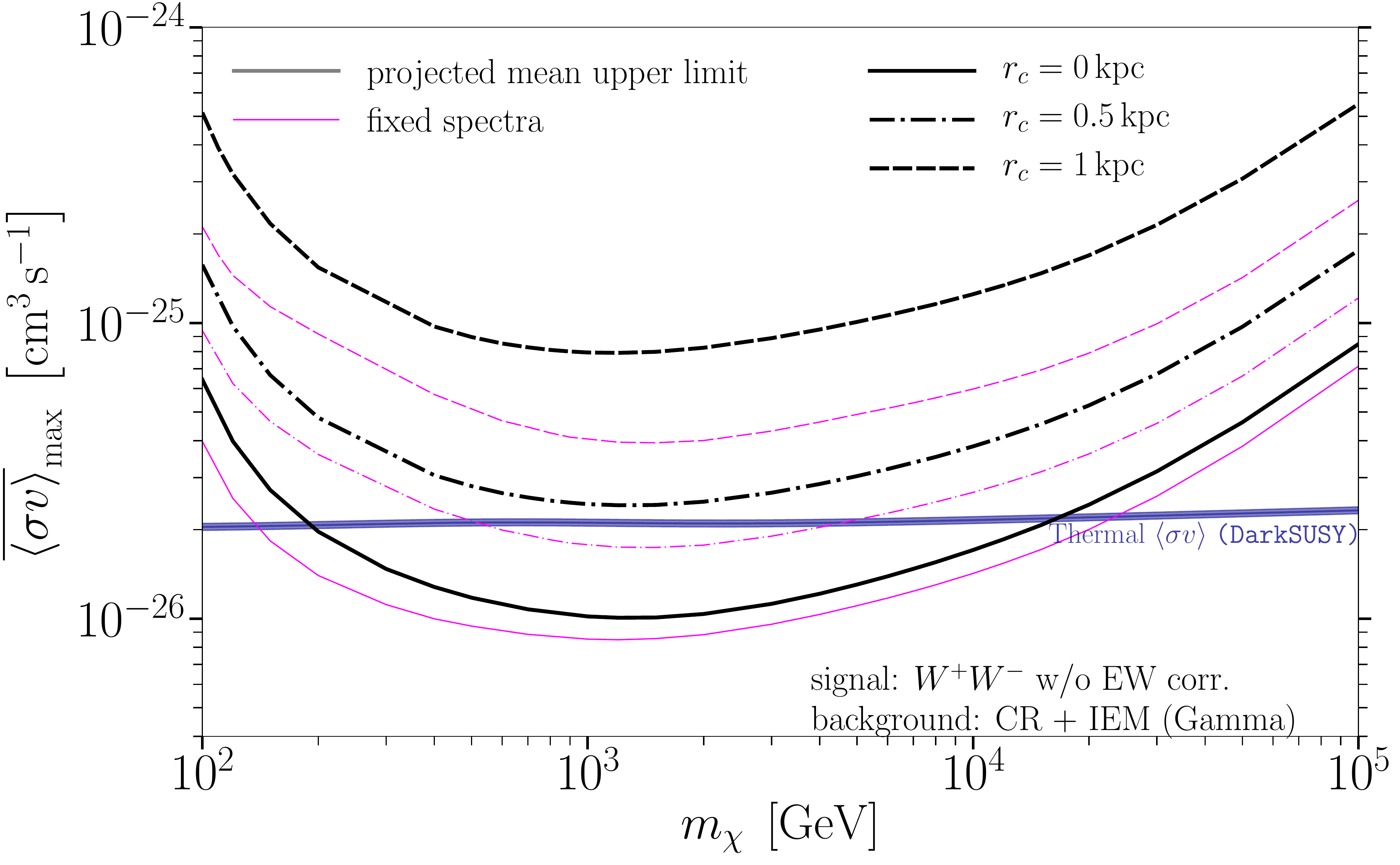}
\caption{
CTA sensitivity to a DM signal, for the $W^+W^-$ channel, for various choices of the DM profile.
Solid lines show the case of an Einasto profile with no central core, while dashed (dash-dotted) lines show the case 
of an Einasto profile with a central core of radius 1\,kpc (0.5\,kpc). Black lines correspond to the result of our
benchmark analysis setting, while magenta lines demonstrate the improvement of these limits when 
adding on top of the (spatial) instrumental systematic uncertainties a long-range correlation in energy, 
which effectively fixes the spectral shapes of the background 
components.  For definiteness, we choose here an energy correlation length of $\ell_{\rm E} = 8$ dex, and adopt 
overall  normalisation uncertainties in the CR (IEM) component of $\sigma^{\rm instr.}_{\rm E} = 1\%$ 
($\sigma^{\rm IEM}_{\rm E} =10\%$). In order to avoid excessive use of computational resources, the 
sensitivity predictions in this figure are based on only 20 (equally log-spaced) energy bins. 
\label{fig:cored_energycorrs}
}
\end{figure}

For the distinct morphology of our benchmark (Einasto) DM profile, as stressed several times,
uncertainties in the {\it spectral} information of the templates only play a sub-dominant role. 
This can, however, change in 
the presence of the cored profiles discussed in section \ref{sec:cores}, where adding spectral information can be 
expected to help in distinguishing the emission components, and thus to improve the DM sensitivity.
A detailed study of how to potentially improve the modelling of background components
is beyond the scope of this work, also because we do not yet have actual data to compare to, but  
in Fig.~\ref{fig:cored_energycorrs} we illustrate the expected effect in terms of systematic uncertainties 
associated with the spectral components. 
The black lines show our benchmark analysis setup, assuming no energy correlations, 
while the magenta lines show the effect of adding such correlations over a range larger than the 
whole energy window adopted in our analysis (thus effectively fixing the spectrum in the 
analysis).\footnote{%
We restrict our discussion here to the effect of such large-scale fluctuations, since computational
limitations anyway restrict us to use relatively large bins in energy (corresponding to the energy
resolution at the 2$\sigma$ level), such that small-scale fluctuations are not relevant for our analysis. 
For smaller energy bins, as well as for an unbinned analysis,
a realistic treatment of the expected energy correlations at (slightly) smaller energy differences would 
be mandatory. 
}
 Solid, dashed and dotted lines refer, as 
in Fig.~\ref{fig:cored_templates_wwo_extended}, to the DM density profile that is adopted.
For the Einasto profile, the sensitivity to a DM signal improves by up to 40\,\% when taking into account the 
additional information contained in the assumed energy correlations. 
For a 1\,kpc core, on the other hand, the improvement is much more significant and can be 
larger than a factor of 2.

\begin{figure}[t]
\centering\includegraphics[width=0.71\linewidth]{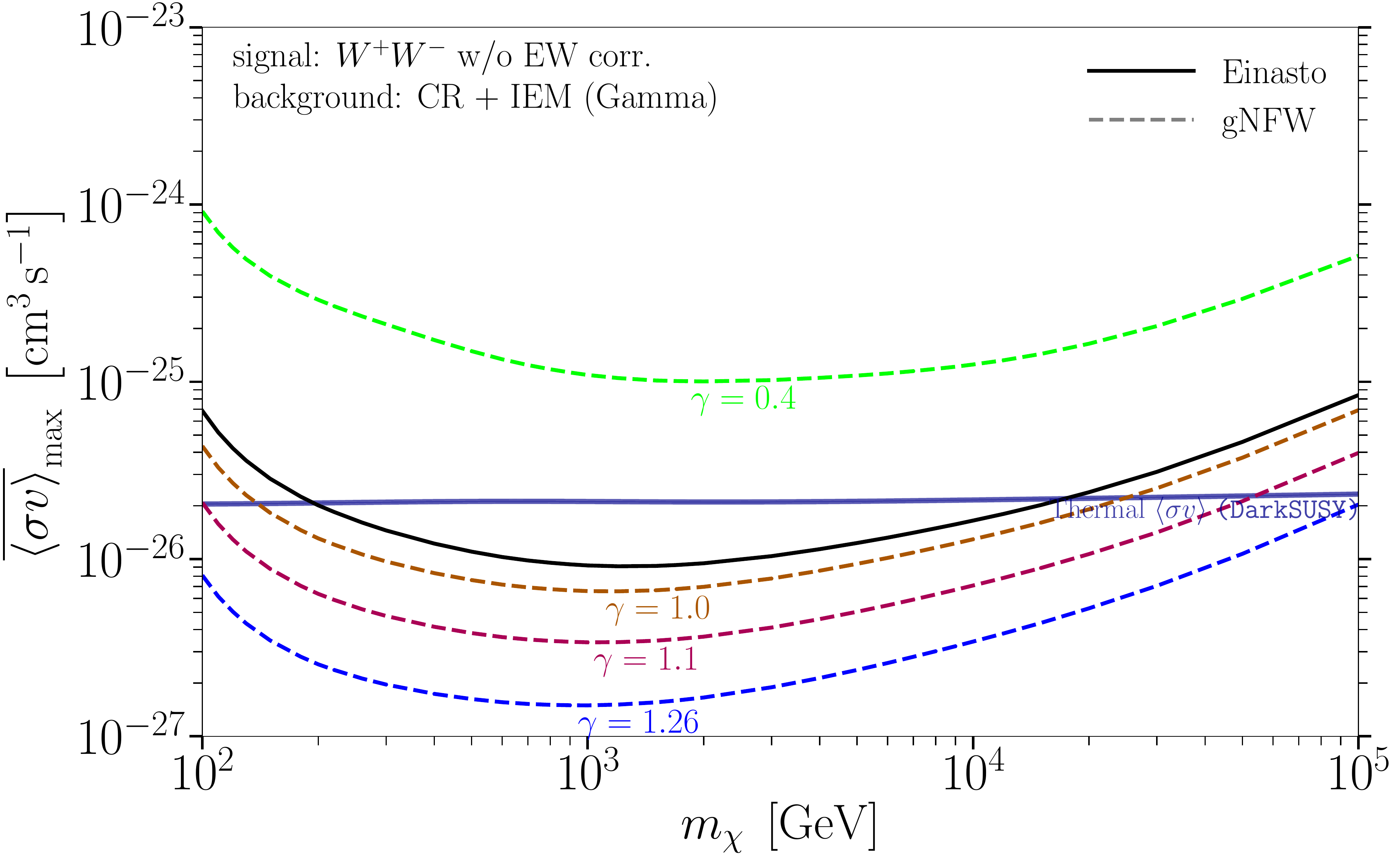}
\caption{CTA sensitivity to a signal from DM annihilating to $W^+W^-$ for a gNFW DM 
density profile (dashed lines), for various choices of the slope parameter $\gamma$, as indicated
(and corresponding best-fit values for scale radius $r_s$ and local DM density $\rho_{\odot}$ as derived
in Ref.~\cite{Karukes:2019jxv}). The solid black line
shows, for comparison, the upper limit for
our standard Einasto profile. All upper limits incorporate our benchmark scheme for the  treatment of 
instrumental systematic 
uncertainties.  {We stress that  the range of DM density profiles shown here  does not  include the 
possible enhancement of the DM density in the central ($r\ll1$\,kpc) regions of the Galaxy, 
which would further contribute to an increase of the CTA sensitivity.  Possible spectral 
correlations (increasing the sensitivity for $\gamma \leq1$ profiles) are
also not considered here.}
\label{fig:profile_scan_gNFW}
}
\end{figure}

In the final part of this section we illustrate the uncertainty of the projected sensitivity 
due to the rather poorly constrained spatial profile of DM in the Galactic centre in a complementary
way, this time assuming no energy correlations (and thus returning to our benchmark settings, for better 
comparison with the results of the main text). In Fig.~\ref{fig:profile_scan_gNFW} we show a summary of 
expected limits for exemplary 
parameter choices with respect to a generalised NFW (gNFW) DM density profile:
\begin{equation}
\label{eq:profile_gNFW}
\rho_{{\rm gNFW}}\!\left(r\right)=\frac{\rho_{s}}{\left(\frac{r\vphantom{x_s^2}}{r_s}\right)^{\gamma}\left(1 + \frac{r\vphantom{x_s^2}}{r_{s}}\right)^{3-\gamma}}{\rm .}
\end{equation}
Our parameter choices are motivated by Fig.~4 of a recent Bayesian analysis of
MW data~\cite{Karukes:2019jxv}. In particular, we pick a range of values for  $\gamma$, 
within $1\,\sigma$ of its best fit value, and then choose the corresponding best-fit values for the remaining 
two parameters, $ \left(r_{s}, \rho_{\odot}\right)$, which we collectively list in Tab.~\ref{tab:parameters_gNFW}. 
Fig.~\ref{fig:profile_scan_gNFW} illustrates that 
NFW limits are somewhat more stringent than those obtained with our benchmark Einasto profile, 
and that the adopted range of $\gamma$ values results in sensitivities symmetric with respect to our 
benchmark case.  Note that  the sensitivities are calculated for the extended survey (which increases 
sensitivity for more cored profiles), but without spectral information (implying that the results shown for 
$\gamma  = 0.4$ are overly  pessimistic).  
We stress that the most recent numerical simulations, when correlated with Gaia data of the baryonic 
content of the MW, tend to produce a DM profile that is {\it steeper} in the inner Galaxy than the Einasto 
case~\cite{Cautun:2019eaf}. Furthermore, 
a possible enhancement of the DM density in the very central ($r\ll1$\,kpc) region of the Galaxy, 
e.g.~due to the presence of the supermassive black hole~\cite{Gondolo:1999ef,Ullio:2001fb,Merritt:2002vj}, 
cannot presently be constrained by kinematic analyses like that of Ref.~\cite{Karukes:2019jxv}. 
In conclusion, even smaller cross-sections than indicated in Fig.~\ref{fig:profile_scan_gNFW} can 
potentially be tested with CTA.

\begin{table}[!h]
\centering
\begin{tabular}{c c c}
\toprule[1.4pt]
$\gamma$ & $r_s\;\left[\mathrm{kpc}\right]$ & $\rho_{\odot}\;\left[\mathrm{GeV}\,\mathrm{cm}^{-3}\right]$ \\
\cmidrule{1-3}
0.4 & 6.7 & 0.44 \\
1.0 & 10.7 & 0.43 \\
1.1 & 11.7 & 0.43 \\
1.26 & 13.2 & 0.43 \\
\bottomrule[1.4pt]
\end{tabular}
\caption{Summary of the gNFW profile parameters extracted from Fig.~4 of \cite{Karukes:2019jxv}. \label{tab:parameters_gNFW}}
\end{table}

\subsection{Individual contribution from different telescope types}
\label{TelescopeTypes}

\begin{figure}[t!]
\centering\includegraphics[width=0.71\textwidth]{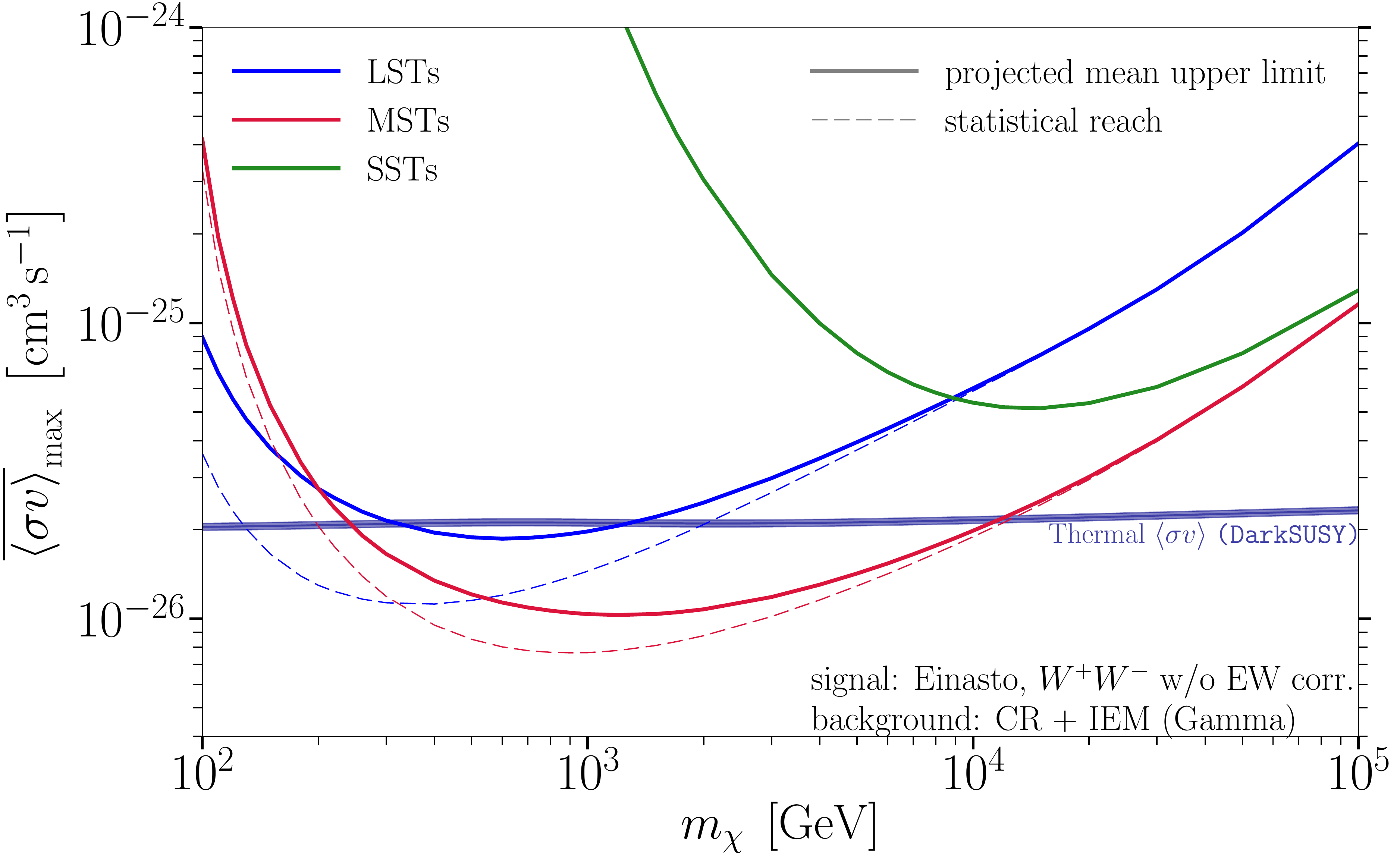}
\caption{CTA sensitivity to a DM signal, independently derived for the three telescope types 
(LSTs -- blue, MSTs -- red, SSTs -- green) according to 
the Southern Array layout, both for our standard analysis pipeline (solid) and when neglecting systematic uncertainties
(dashed). Note that in the case of SSTs the solid and dashed (green) lines are overlapping.
} \label{fig:TelescopeTypes}
\end{figure}

In this section, we discuss how the three main telescope types individually contribute to the DM sensitivity.  
On the one hand,
this potentially helps to assess how CTA's sensitivity will improve during the first years of data 
taking, once the deployment schedule is set. On the other hand, the breakdown of telescope sensitivities
with respect to the telescope types is clearly also relevant in view of potential upgrade steps following
the initial configuration discussed in Appendix \ref{app:Phase1}.

In Fig.~\ref{fig:TelescopeTypes} we illustrate the DM sensitivity that results when only taking into account
one type of telescopes, respectively, in the Southern array Baseline Configuration and adopting out benchmark 
treatment of systematic uncertainties (solid lines). For comparison, we also indicate the resulting limits
in the case when no systematic uncertainties are explicitly added in the analysis (dashed lines).
As expected, 
the LSTs dominate the sensitivity for the lowest DM masses accessible to CTA 
(below $\sim100$\,GeV). In the middle mass range the MSTs dominate, increasing the overall reach due to 
the larger number of telescopes (25 MSTs  vs.~4 LSTs are planned for the Southern site). SSTs are most 
relevant for the highest energies, but their sensitivity only starts to be competitive for DM masses around 
100\,TeV. As clearly visible also when broken down to individual telescope types, limits are dominated
by systematic errors for low DM masses/photon energies, and statistics-dominated for high DM 
masses/photon energies.


\subsection{ON/OFF analysis}
\label{app:onoff}

\begin{figure}[t!]
\centering\includegraphics[width=0.5\linewidth]{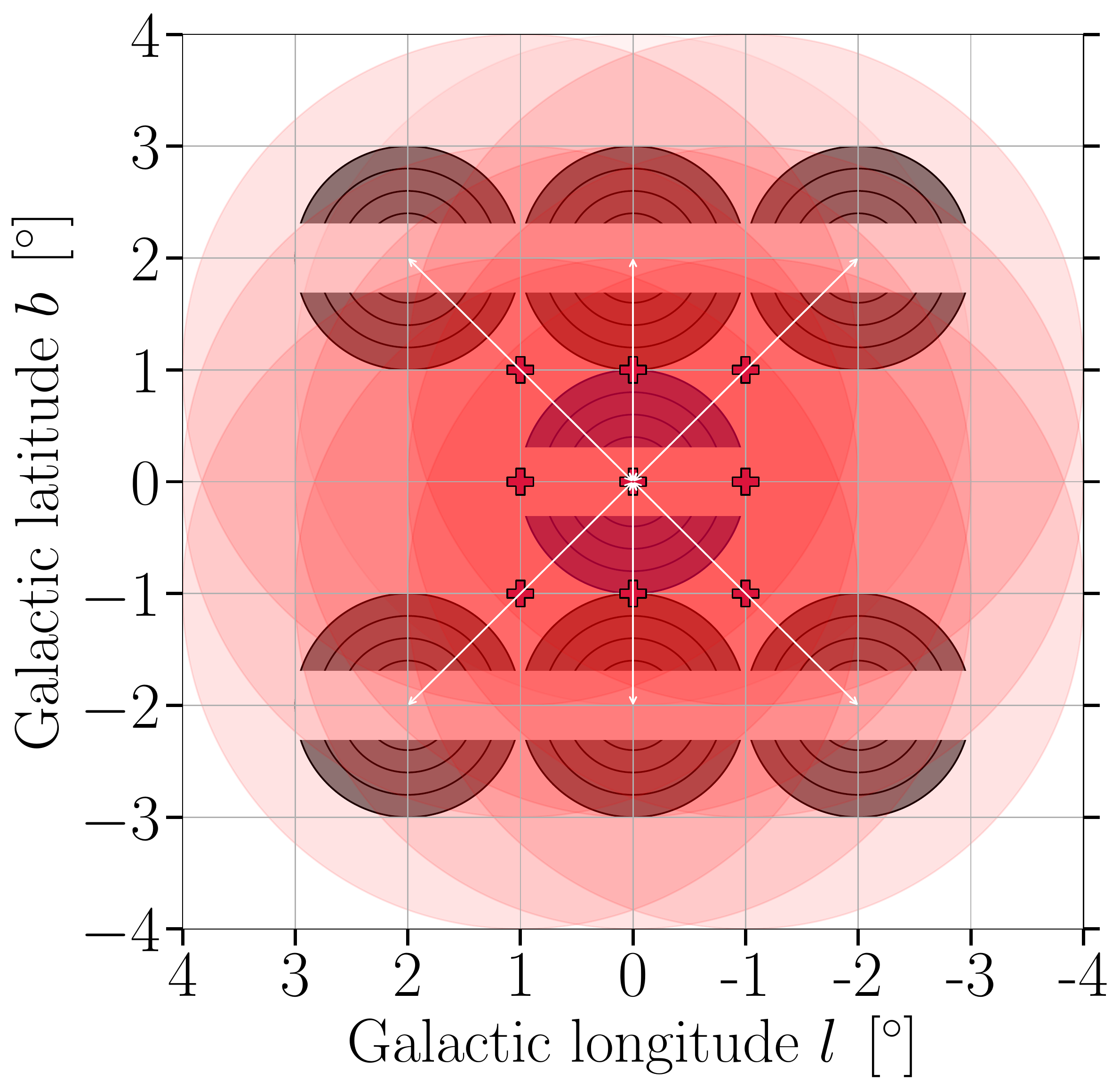}
\caption{Schematic visualisation of the chosen ON (purple) and OFF (black) regions in the context of 
CTA's Galactic centre survey. The choice of the latter is based on the six pointing positions with Galactic 
latitude $|b|=1^{\circ}$ (indicated by crosses), see text for details.
The Galactic plane in the ON region is masked for $\left|b\right| \leq 0.3^{\circ}$, which is reflected in the 
definition of the OFF region(s) as well. Both ON and OFF regions are split into five concentric annuli of width 
$0.2^{\circ}$.
\label{fig:observations_ONOFF}}
\end{figure}

Here we briefly compare the performance of our default template fitting technique
to the  ON/OFF type of analysis discussed in Section \ref{sec:data}. For the latter, we adopt a 
likelihood that is a product of Poisson likelihood functions $\mathcal{L}_{ij}$ over the $i-$th energy, $j-$th 
ON region (ring) and $k-$th pointing position (as used in, e.g., Ref.~\cite{Abdallah:2018qtu}):

\begin{align}
\label{eq:ONOFFlikelihood}
    \mathcal{L}\!\left(M_{\textrm{DM}}, \langle\sigma v\rangle\right) = &\prod_{ijk}\,\mathcal{L}_{ij}\!\left(\mathbf{N^{\textrm{S}}_{k}},\mathbf{N^{\textrm{B}}_{k}}, \mathbf{\kappa_k}\left.\right|\mathbf{N^{\textrm{ON}}_{k}}, \mathbf{N^{\textrm{OFF}}_{k}}\right)\\
    =&\prod_{ijk}\,\,\frac{\left(N^{\textrm{S}}_{ijk}\!\!+\!\!\kappa_{ijk}N^{\textrm{B}}_{ijk}\right)^{N^{\textrm{ON}}_{ijk}}}{N^{\textrm{ON}}_{ijk}!}e^{-(N^{\textrm{S}}_{ijk}+\kappa_{ijk}N^{\textrm{B}}_{ijk})}
    \times\frac{\left(N^{\textrm{B}}_{ijk}\right)^{N^{\textrm{OFF}}_{ijk}}}{N^{\textrm{OFF}}_{ijk}!}e^{-N^{\textrm{B}}_{ijk}}\,,\nonumber
\end{align}
with $N^{\textrm{S}}_{ijk}$ denoting the expected number of signal events in 
the ON (`signal') region. $\mathbf{N_k}^{\textrm{ON}}$ and 
$\mathbf{N_k}^{\textrm{OFF}}$ refer to the measured photon events in ON and OFF region for observation 
$k$ which we prepare as a single Asimov data set from a selection of background source components 
described in Sec.~\ref{sec:conv_astrop}. $\kappa_{ijk}$ is in general a normalisation factor 
to account for the different background acceptance in the ON and OFF regions, but in our case it will by 
construction be equal to one for all bins.

Fixing the value of $m_\chi$, we again choose the likelihood ratio as test statistic to constrain 
$\langle\sigma v\rangle$. To this end, we adopt Eq.~(\ref{eq:TS_stat_reach}) to our purposes here, by 
explicitly profiling over the nuisance parameters $N^{\textrm{B}}_{ijk}$; as a result, we obtain a 
one-dimensional likelihood function depending only on the signal strength $\langle\sigma v\rangle$.
The definition of ON and OFF regions closely follows the scheme outlined in Ref.~\cite{Abdallah:2018qtu}.
We define our ON region as a circular ROI of $1^{\circ}$ 
radius centred at the GC, divided into five concentric annuli with a width of $0.2^{\circ}$, 
and mask the region with $\left|b\right|\leq0.3^{\circ}$ to remove the brightest very high-energy 
point sources in the vicinity of the GC and parts of the IEM.
The position of the OFF regions is chosen as the point-symmetrical image of the ON region with respect to the 
respective observational pointing position (see Fig.~\ref{fig:observations_ONOFF}),
which, under the assumption of an instrument response that depends only on offset from the pointing direction 
but not azimuth, implies that ON and OFF regions share the same solid angle, exposure and acceptance.
Fig.~\ref{fig:DM_benchmarks} in the main text illustrates that for cuspy DM density 
profiles the $J-$factors of ON and OFF region can differ by up to one order of magnitude so that the OFF region 
should indeed only feature minor contaminations by the signal source.
Splitting the ON and OFF ROI into multiple annuli improves the performance of indirect DM searches following 
an ON/OFF analysis~\cite{Abdallah:2016ygi}.

\begin{figure}[t!]
\centering\includegraphics[width=0.71\textwidth]{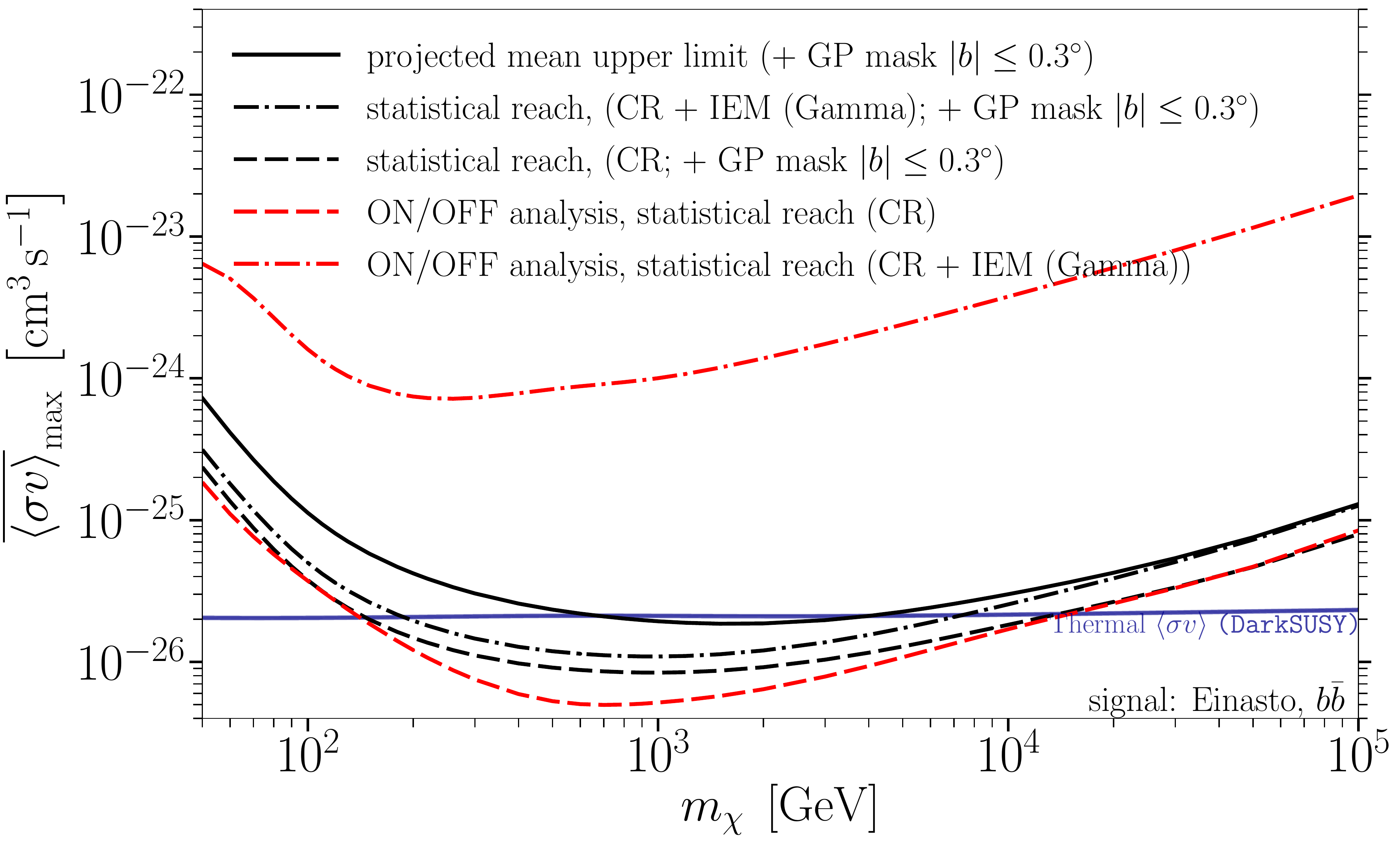}
\caption{Mean expected CTA upper limits on the DM annihilation cross-section for the $\bar{b}b$-channel and an 
Einasto DM density profile, adopting the default morphology analysis (black) and ON/OFF strategy (red), respectively. 
The black solid line represents the benchmark situation (including instrumental systematic errors) where we 
additionally apply a mask to the band $b\leq0.3^{\circ}$ to reduce the impact of the IEM along the Galactic plane. 
All other lines refer to `statistics-only' limits, where the impact of systematic errors is neglected; dashed (dash-dotted)
lines assume that no IE (IE as expected in the {\it Gamma} model) is included in the mock data. 
Note that in each of the cases shown here the Galactic plane is masked. In order to avoid excessive use of 
computational  resources, the sensitivity predictions in this figure are  based on only 20 (equally log-spaced) energy bins.
 \label{fig:ON/OFFmorph}}
\end{figure}

In Fig.~\ref{fig:ON/OFFmorph} we compare expected limits from our benchmark analysis (black lines) with those 
resulting from the ON/OFF analysis as described above (red lines). 
The first crucial observation is that the two approaches result in a comparable sensitivity 
when only including the smoother background of misidentified CRs, and no IE, in the simulated data 
(dashed lines);\footnote{%
Note that our 3D CR-only case assumes a variable CR spectrum (to $10\%$) while the ON/OFF analysis fixes it to 
the measurement in the OFF region. This  is  probably  the main reason why this analysis  set-up ON/OFF 
performs better than  the 3D analysis.
}
this provides a nice consistency check of the two methods in the absence of extended diffuse emission
components -- which indeed is the basic assumption that previous DM searches with IACTs have relied on.
Adding IE (our default {\it Gamma} model) to the simulated data, however, 
the ON/OFF analysis suffers as expected from a significant loss of sensitivity (red dash-dotted), 
while limits obtained from the morphology analysis (black dash-dotted) are much less affected.  
This illustrates, as already argued previously, that the ON/OFF analysis can only be expected to provide 
realistic DM limits if the expected large-scale diffuse emission components are either below the nominal sensitivity
of the instrument, or exhibit a rather low gradient within the ROI. In all other cases, 
at least under the assumption that the spatial IEM templates can be modelled more or less realistically, 
the morphological analysis appears to be more promising.

%% file: CTA_Acknowledgments_July2020_v2.tex
%

\section*{CTA Acknowledgements}

\bigskip

We gratefully acknowledge financial support from the following agencies and organisations:

\bigskip

State Committee of Science of Armenia, Armenia;
The Australian Research Council, Astronomy Australia Ltd, The University of Adelaide, Australian National University, Monash University, The University of New South Wales, The University of Sydney, Western Sydney University, Australia;
Federal Ministry of Education, Science and Research, and Innsbruck University, Austria;
Conselho Nacional de Desenvolvimento Cient\'{\i}fico e Tecnol\'{o}gico (CNPq), Funda\c{c}\~{a}o de Amparo \`{a} Pesquisa do Estado do Rio de Janeiro (FAPERJ), Funda\c{c}\~{a}o de Amparo \`{a} Pesquisa do Estado de S\~{a}o Paulo (FAPESP), Ministry of Science, Technology, Innovations and Communications (MCTIC), and Instituto Serrapilheira, Brasil;
Ministry of Education and Science, National RI Roadmap Project DO1-153/28.08.2018, Bulgaria;
The Natural Sciences and Engineering Research Council of Canada and the Canadian Space Agency, Canada;
CONICYT-Chile grants CATA AFB 170002, ANID PIA/APOYO AFB 180002, ACT 1406, FONDECYT-Chile grants, 1161463, 1170171, 1190886, 1171421, 1170345, 1201582, Gemini-ANID 32180007, Chile;
Croatian Science Foundation, Rudjer Boskovic Institute, University of Osijek, University of Rijeka, University of Split, Faculty of Electrical Engineering, Mechanical Engineering and Naval Architecture, University of Zagreb, Faculty of Electrical Engineering and Computing, Croatia;
Ministry of Education, Youth and Sports, MEYS  LM2015046, LM2018105, LTT17006, EU/MEYS CZ.02.1.01/0.0/0.0/16\_013/0001403, CZ.02.1.01/0.0/0.0/18\_046/0016007 and CZ.02.1.01/0.0/0.0/16\_019/0000754, Czech Republic; 
Academy of Finland (grant nr.317636, 320045, 317383 and 320085), Finland;
Ministry of Higher Education and Research, CNRS-INSU and CNRS-IN2P3, CEA-Irfu, ANR, Regional Council Ile de France, Labex ENIGMASS, OSUG2020, P2IO and OCEVU, France;
Max Planck Society, BMBF, DESY, Helmholtz Association, Germany;
Department of Atomic Energy, Department of Science and Technology, India;
Istituto Nazionale di Astrofisica (INAF), Istituto Nazionale di Fisica Nucleare (INFN), MIUR, Istituto Nazionale di Astrofisica (INAF-OABRERA) Grant Fondazione Cariplo/Regione Lombardia ID 2014-1980/RST\_ERC, Italy;
ICRR, University of Tokyo, JSPS, MEXT, Japan;
Netherlands Research School for Astronomy (NOVA), Netherlands Organization for Scientific Research (NWO), Netherlands;
University of Oslo, Norway;
Ministry of Science and Higher Education, \seqsplit{DIR/WK/2017/12}, the National Centre for Research and Development and the National Science Centre, UMO-2016/22/M/ST9/00583, Poland;
Slovenian Research Agency, grants P1-0031, P1-0385, I0-0033, J1-9146, J1-1700, N1-0111, and the Young Researcher program, Slovenia; 
South African Department of Science and Technology and National Research Foundation through the South African Gamma-Ray Astronomy Programme, South Africa;
The Spanish Ministry of Science and Innovation and the Spanish Research State Agency (AEI) through grants AYA2016-79724-C4-1-P, AYA2016-80889-P, AYA2016-76012-C3-1-P, BES-2016-076342, \seqsplit{ESP2017- 87055-C2-1-P}, FPA2017-82729-C6-1-R, FPA2017-82729-C6-2-R, FPA2017-82729-C6-3-R, FPA2017-82729-C6-4-R, FPA2017-82729-C6-5-R, \seqsplit{FPA2017-82729-C6-6-R}, PGC2018-095161-B-I00, \seqsplit{PGC2018-095512-B-I00}; the “Centro de Excelencia Severo Ochoa” program through grants no. SEV-2015-0548, SEV-2016-0597, SEV-2016-0588, SEV-2017-0709; the “Unidad de Excelencia María de Maeztu” program through grant no. MDM-2015-0509; the “Ram\'{o}n y Cajal” programme through grants RYC-2013-14511, RyC-2013-14660, RYC-2017-22665; and the MultiDark Consolider Network FPA2017-90566-REDC. Atracci\'{o}n de Talento contract no. 2016-T1/TIC-1542 granted by the Comunidad de Madrid; the “Postdoctoral Junior Leader Fellowship” programme from La Caixa Banking Foundation, grants no.~LCF/BQ/LI18/11630014 and LCF/BQ/PI18/11630012; the “Programa Operativo” \seqsplit{FEDER 2014-2020}, Consejer\'{i}a de Econom\'{i}a y Conocimiento de la Junta de Andaluc\'{i}a (Ref. 1257737), PAIDI 2020 (Ref. P18-FR-1580), and Universidad de Ja\'{e}n; the Spanish AEI EQC2018-005094-P FEDER 2014-2020; the European Union’s “Horizon 2020” research and innovation programme under Marie Sk\l{}odowska-Curie grant agreement no. 665919; and the ESCAPE project with grant no. GA:824064, Spain;
Swedish Research Council, Royal Physiographic Society of Lund, Royal Swedish Academy of Sciences, The Swedish National Infrastructure for Computing (SNIC) at Lunarc (Lund), Sweden;
State Secretariat for Education, Research and Innovation (SERI) and Swiss National Science Foundation (SNSF), Switzerland;
Durham University, Leverhulme Trust, Liverpool University, University of Leicester, University of Oxford, Royal Society, Science and Technology Facilities Council, UK;
U.S. National Science Foundation, U.S. Department of Energy, Argonne National Laboratory, Barnard College, University of California, University of Chicago, Columbia University, Georgia Institute of Technology, Institute for Nuclear and Particle Astrophysics (INPAC-MRPI program), Iowa State University, the Smithsonian Institution, Washington University McDonnell Center for the Space Sciences, The University of Wisconsin and the Wisconsin Alumni Research Foundation, USA.

\bigskip

The research leading to these results has received funding from the European Union's Seventh Framework Programme (FP7/2007-2013) under grant agreements No~262053 and No~317446.
This project is receiving funding from the European Union's Horizon 2020 research and innovation programs under agreement No~676134.

%% file: CTA_GC_limits-NewMaster.bbl
\providecommand{\href}[2]{#2}\begingroup\raggedright\begin{thebibliography}{100}

\bibitem{Aghanim:2018eyx}
{\scshape Planck} collaboration, N.~Aghanim et~al., \emph{{Planck 2018 results.
  VI. Cosmological parameters}},
  \href{https://arxiv.org/abs/1807.06209}{{\ttfamily 1807.06209}}.

\bibitem{Bertone:2016nfn}
G.~Bertone and D.~Hooper, \emph{{History of dark matter}},
  \href{https://doi.org/10.1103/RevModPhys.90.045002}{\emph{Rev. Mod. Phys.}
  {\bfseries 90} (2018) 045002},
  [\href{https://arxiv.org/abs/1605.04909}{{\ttfamily 1605.04909}}].

\bibitem{Jungman:1995df}
G.~Jungman, M.~Kamionkowski and K.~Griest, \emph{{Supersymmetric dark matter}},
  \href{https://doi.org/10.1016/0370-1573(95)00058-5}{\emph{Phys. Rept.}
  {\bfseries 267} (1996) 195--373},
  [\href{https://arxiv.org/abs/hep-ph/9506380}{{\ttfamily hep-ph/9506380}}].

\bibitem{Bertone:2004pz}
G.~Bertone, D.~Hooper and J.~Silk, \emph{{Particle dark matter: Evidence,
  candidates and constraints}},
  \href{https://doi.org/10.1016/j.physrep.2004.08.031}{\emph{Phys. Rept.}
  {\bfseries 405} (2005) 279--390},
  [\href{https://arxiv.org/abs/hep-ph/0404175}{{\ttfamily hep-ph/0404175}}].

\bibitem{Feng:2010gw}
J.~L. Feng, \emph{{Dark Matter Candidates from Particle Physics and Methods of
  Detection}},
  \href{https://doi.org/10.1146/annurev-astro-082708-101659}{\emph{Ann. Rev.
  Astron. Astrophys.} {\bfseries 48} (2010) 495--545},
  [\href{https://arxiv.org/abs/1003.0904}{{\ttfamily 1003.0904}}].

\bibitem{Bertone:2018xtm}
G.~Bertone and M.~P. Tait, Tim, \emph{{A new era in the search for dark
  matter}}, \href{https://doi.org/10.1038/s41586-018-0542-z}{\emph{Nature}
  {\bfseries 562} (2018) 51--56},
  [\href{https://arxiv.org/abs/1810.01668}{{\ttfamily 1810.01668}}].

\bibitem{Arcadi:2017kky}
G.~Arcadi, M.~Dutra, P.~Ghosh, M.~Lindner, Y.~Mambrini, M.~Pierre et~al.,
  \emph{{The waning of the WIMP? A review of models, searches, and
  constraints}},
  \href{https://doi.org/10.1140/epjc/s10052-018-5662-y}{\emph{Eur. Phys. J.}
  {\bfseries C78} (2018) 203},
  [\href{https://arxiv.org/abs/1703.07364}{{\ttfamily 1703.07364}}].

\bibitem{Lee:1977ua}
B.~W. Lee and S.~Weinberg, \emph{{Cosmological Lower Bound on Heavy Neutrino
  Masses}}, \href{https://doi.org/10.1103/PhysRevLett.39.165}{\emph{Phys. Rev.
  Lett.} {\bfseries 39} (1977) 165--168}.

\bibitem{Bringmann:2012ez}
T.~Bringmann and C.~Weniger, \emph{{Gamma Ray Signals from Dark Matter:
  Concepts, Status and Prospects}},
  \href{https://doi.org/10.1016/j.dark.2012.10.005}{\emph{Phys. Dark Univ.}
  {\bfseries 1} (2012) 194--217},
  [\href{https://arxiv.org/abs/1208.5481}{{\ttfamily 1208.5481}}].

\bibitem{Bergstrom:1997fj}
L.~Bergstrom, P.~Ullio and J.~H. Buckley, \emph{{Observability of gamma-rays
  from dark matter neutralino annihilations in the Milky Way halo}},
  \href{https://doi.org/10.1016/S0927-6505(98)00015-2}{\emph{Astropart. Phys.}
  {\bfseries 9} (1998) 137--162},
  [\href{https://arxiv.org/abs/astro-ph/9712318}{{\ttfamily
  astro-ph/9712318}}].

\bibitem{CTAweb}
\url{https://www.cta-observatory.org/}.

\bibitem{Acharya:2017ttl}
{\scshape CTA Consortium} collaboration, B.~S. Acharya et~al., \emph{{Science
  with the Cherenkov Telescope Array}}.
\newblock WSP, 2018, \href{https://doi.org/10.1142/10986}{10.1142/10986}.

\bibitem{Doro:2012xx}
{\scshape CTA Consortium} collaboration, M.~Doro et~al., \emph{{Dark Matter and
  Fundamental Physics with the Cherenkov Telescope Array}},
  \href{https://doi.org/10.1016/j.astropartphys.2012.08.002}{\emph{Astropart.
  Phys.} {\bfseries 43} (2013) 189--214},
  [\href{https://arxiv.org/abs/1208.5356}{{\ttfamily 1208.5356}}].

\bibitem{Silverwood:2014yza}
H.~Silverwood, C.~Weniger, P.~Scott and G.~Bertone, \emph{{A realistic
  assessment of the CTA sensitivity to dark matter annihilation}},
  \href{https://doi.org/10.1088/1475-7516/2015/03/055}{\emph{JCAP} {\bfseries
  1503} (2015) 055}, [\href{https://arxiv.org/abs/1408.4131}{{\ttfamily
  1408.4131}}].

\bibitem{Pierre:2014tra}
M.~Pierre, J.~M. Siegal-Gaskins and P.~Scott, \emph{{Sensitivity of CTA to dark
  matter signals from the Galactic Center}},
  \href{https://doi.org/10.1088/1475-7516/2014/10/E01,
  10.1088/1475-7516/2014/06/024}{\emph{JCAP} {\bfseries 1406} (2014) 024},
  [\href{https://arxiv.org/abs/1401.7330}{{\ttfamily 1401.7330}}].

\bibitem{Carr:2015hta}
{\scshape CTA} collaboration, J.~Carr et~al., \emph{{Prospects for Indirect
  Dark Matter Searches with the Cherenkov Telescope Array (CTA)}},
  \href{https://doi.org/10.22323/1.236.1203}{\emph{PoS} {\bfseries ICRC2015}
  (2016) 1203}, [\href{https://arxiv.org/abs/1508.06128}{{\ttfamily
  1508.06128}}].

\bibitem{Lefranc:2015pza}
V.~Lefranc, E.~Moulin, P.~Panci and J.~Silk, \emph{{Prospects for Annihilating
  Dark Matter in the inner Galactic halo by the Cherenkov Telescope Array}},
  \href{https://doi.org/10.1103/PhysRevD.91.122003}{\emph{Phys. Rev.}
  {\bfseries D91} (2015) 122003},
  [\href{https://arxiv.org/abs/1502.05064}{{\ttfamily 1502.05064}}].

\bibitem{Rinchiuso:2020skh}
L.~Rinchiuso, O.~Macias, E.~Moulin, N.~L. Rodd and T.~R. Slatyer,
  \emph{{Prospects for Heavy WIMP Dark Matter with CTA: the Wino and
  Higgsino}},  \href{https://arxiv.org/abs/2008.00692}{{\ttfamily 2008.00692}}.

\bibitem{AMSweb}
\url{https://ams02.space/}.

\bibitem{Bringmann:2006im}
T.~Bringmann and P.~Salati, \emph{{The galactic antiproton spectrum at high
  energies: Background expectation vs. exotic contributions}},
  \href{https://doi.org/10.1103/PhysRevD.75.083006}{\emph{Phys. Rev.}
  {\bfseries D75} (2007) 083006},
  [\href{https://arxiv.org/abs/astro-ph/0612514}{{\ttfamily
  astro-ph/0612514}}].

\bibitem{Cuoco:2017iax}
A.~Cuoco, J.~Heisig, M.~Korsmeier and M.~Krämer, \emph{{Constraining heavy
  dark matter with cosmic-ray antiprotons}},
  \href{https://doi.org/10.1088/1475-7516/2018/04/004}{\emph{JCAP} {\bfseries
  1804} (2018) 004}, [\href{https://arxiv.org/abs/1711.05274}{{\ttfamily
  1711.05274}}].

\bibitem{Bergstrom:2010gh}
L.~Bergstrom, T.~Bringmann and J.~Edsjo, \emph{{Complementarity of direct dark
  matter detection and indirect detection through gamma-rays}},
  \href{https://doi.org/10.1103/PhysRevD.83.045024}{\emph{Phys. Rev.}
  {\bfseries D83} (2011) 045024},
  [\href{https://arxiv.org/abs/1011.4514}{{\ttfamily 1011.4514}}].

\bibitem{Cahill-Rowley:2013dpa}
M.~Cahill-Rowley, R.~Cotta, A.~Drlica-Wagner, S.~Funk, J.~Hewett, A.~Ismail
  et~al., \emph{{Complementarity and Searches for Dark Matter in the pMSSM}},
  in \emph{{Proceedings, 2013 Community Summer Study on the Future of U.S.
  Particle Physics: Snowmass on the Mississippi (CSS2013): Minneapolis, MN,
  USA, July 29-August 6, 2013}}, 2013,
  \href{https://arxiv.org/abs/1305.6921}{{\ttfamily 1305.6921}},
  \href{http://www.slac.stanford.edu/econf/C1307292/docs/submittedArxivFiles/1305.6921.pdf}{http://www.slac.stanford.edu/econf/C1307292/docs/submittedArxivFiles/1305.6921.pdf}.

\bibitem{Arrenberg:2013rzp}
S.~Arrenberg et~al., \emph{{Working Group Report: Dark Matter
  Complementarity}},  in \emph{{Proceedings, 2013 Community Summer Study on the
  Future of U.S. Particle Physics: Snowmass on the Mississippi (CSS2013):
  Minneapolis, MN, USA, July 29-August 6, 2013}}, 2013,
  \href{https://arxiv.org/abs/1310.8621}{{\ttfamily 1310.8621}},
  \href{http://lss.fnal.gov/archive/preprint/fermilab-conf-13-574-ae.shtml}{http://lss.fnal.gov/archive/preprint/fermilab-conf-13-574-ae.shtml}.

\bibitem{Balazs:2017hxh}
C.~Balázs, J.~Conrad, B.~Farmer, T.~Jacques, T.~Li, M.~Meyer et~al.,
  \emph{{Sensitivity of the Cherenkov Telescope Array to the detection of a
  dark matter signal in comparison to direct detection and collider
  experiments}}, \href{https://doi.org/10.1103/PhysRevD.96.083002}{\emph{Phys.
  Rev.} {\bfseries D96} (2017) 083002},
  [\href{https://arxiv.org/abs/1706.01505}{{\ttfamily 1706.01505}}].

\bibitem{4323497}
D.~B. {Hicks}, L.~{Ried} and L.~E. {Peterson}, \emph{X-ray telescope for an
  orbiting solar observatory},
  \href{https://doi.org/10.1109/TNS.1965.4323497}{\emph{IEEE Transactions on
  Nuclear Science} {\bfseries 12} (Feb, 1965) 54--65}.

\bibitem{1968ApJ...153L.203C}
G.~W. {Clark}, G.~P. {Garmire} and W.~L. {Kraushaar}, \emph{{Observation of
  High-Energy Cosmic Gamma Rays}},
  \href{https://doi.org/10.1086/180252}{\emph{The Astrophysical Journal}
  {\bfseries 153} (Sep, 1968) L203}.

\bibitem{1972ApJ...177..341K}
W.~L. {Kraushaar}, G.~W. {Clark}, G.~P. {Garmire}, R.~{Borken}, P.~{Higbie},
  V.~{Leong} et~al., \emph{{High-Energy Cosmic Gamma-Ray Observations from the
  OSO-3 Satellite}}, \href{https://doi.org/10.1086/151713}{\emph{The
  Astrophysical Journal} {\bfseries 177} (Nov, 1972) 341}.

\bibitem{1975ApJ...198..163F}
C.~E. {Fichtel}, R.~C. {Hartman}, D.~A. {Kniffen}, D.~J. {Thompson}, G.~F.
  {Bignami}, H.~{{\"O}gelman} et~al., \emph{{High-energy gamma-ray results from
  the second Small Astronomy Satellite.}},
  \href{https://doi.org/10.1086/153590}{\emph{\apj} {\bfseries 198} (May, 1975)
  163--182}.

\bibitem{1975SSI.....1..245B}
G.~F. {Bignami}, G.~{Boella}, J.~J. {Burger}, P.~{Keirle}, H.~A.
  {Mayer-Hasselwander}, J.~A. {Paul} et~al., \emph{{The COS-B experiment for
  gamma-ray astronomy.}}, {\emph{Space Science Instrumentation} {\bfseries 1}
  (Aug., 1975) 245--268}.

\bibitem{Hartman:1999fc}
{\scshape EGRET} collaboration, R.~C. Hartman et~al., \emph{{The Third EGRET
  catalog of high-energy gamma-ray sources}},
  \href{https://doi.org/10.1086/313231}{\emph{Astrophys. J. Suppl.} {\bfseries
  123} (1999) 79}.

\bibitem{Tavani:2008sp}
{\scshape AGILE} collaboration, M.~Tavani et~al., \emph{{The AGILE Mission}},
  \href{https://doi.org/10.1051/0004-6361/200810527}{\emph{Astron. Astrophys.}
  {\bfseries 502} (2009) 995--1013},
  [\href{https://arxiv.org/abs/0807.4254}{{\ttfamily 0807.4254}}].

\bibitem{Atwood:2009ez}
{\scshape Fermi-LAT} collaboration, W.~B. Atwood et~al., \emph{{The Large Area
  Telescope on the Fermi Gamma-ray Space Telescope Mission}},
  \href{https://doi.org/10.1088/0004-637X/697/2/1071}{\emph{Astrophys. J.}
  {\bfseries 697} (2009) 1071--1102},
  [\href{https://arxiv.org/abs/0902.1089}{{\ttfamily 0902.1089}}].

\bibitem{1989ApJ...342..379W}
T.~C. {Weekes}, M.~F. {Cawley}, D.~J. {Fegan}, K.~G. {Gibbs}, A.~M. {Hillas},
  P.~W. {Kowk} et~al., \emph{{Observation of TeV gamma rays from the Crab
  nebula using the atmospheric Cerenkov imaging technique}},
  \href{https://doi.org/10.1086/167599}{\emph{Astrophys. J.} {\bfseries 342}
  (July, 1989) 379--395}.

\bibitem{DiSciascio:2019lse}
G.~Di~Sciascio, \emph{{Ground-based Gamma-Ray Astronomy: an Introduction}},  in
  \emph{{18th International Baikal Summer School on Physics of Elementary
  Particles and Astrophysics: Exploring the Universe through multiple
  messengers (ISAPP-Baikal 2018) Bolshie Koty, Lake Baikal, Russia, July 12-21,
  2018}}, 2019, \href{https://arxiv.org/abs/1904.06218}{{\ttfamily
  1904.06218}}.

\bibitem{Atkins_2004}
R.~Atkins, W.~Benbow, D.~Berley, E.~Blaufuss, J.~Bussons, D.~G. Coyne et~al.,
  \emph{{{TeV} Gamma-Ray Survey of the Northern Hemisphere Sky Using the
  Milagro Observatory}}, \href{https://doi.org/10.1086/420880}{\emph{The
  Astrophysical Journal} {\bfseries 608} (jun, 2004) 680--685}.

\bibitem{Abdo_2008}
A.~A. Abdo, B.~Allen, T.~Aune, D.~Berley, E.~Blaufuss, S.~Casanova et~al.,
  \emph{{A Measurement of the Spatial Distribution of Diffuse {TeV} Gamma-Ray
  Emission from the Galactic Plane with Milagro}},
  \href{https://doi.org/10.1086/592213}{\emph{The Astrophysical Journal}
  {\bfseries 688} (dec, 2008) 1078--1083}.

\bibitem{Abdo_2009}
A.~A. Abdo, B.~T. Allen, T.~Aune, D.~Berley, S.~Casanova, C.~Chen et~al.,
  \emph{{{THE} {LARGE}-{SCALE} {COSMIC}-{RAY} {ANISOTROPY} {AS} {OBSERVED}
  {WITH} {MILAGRO}}},
  \href{https://doi.org/10.1088/0004-637x/698/2/2121}{\emph{The Astrophysical
  Journal} {\bfseries 698} (jun, 2009) 2121--2130}.

\bibitem{2013ApJ...779...27B}
B.~{Bartoli}, P.~{Bernardini}, X.~J. {Bi}, I.~{Bolognino}, P.~{Branchini},
  A.~{Budano} et~al., \emph{{TeV Gamma-Ray Survey of the Northern Sky Using the
  ARGO-YBJ Detector}},
  \href{https://doi.org/10.1088/0004-637X/779/1/27}{\emph{\apj} {\bfseries 779}
  (Dec., 2013) 27}, [\href{https://arxiv.org/abs/1311.3376}{{\ttfamily
  1311.3376}}].

\bibitem{Abeysekara:2013tza}
A.~Abeysekara et~al., \emph{{Sensitivity of the High Altitude Water Cherenkov
  Detector to Sources of Multi-TeV Gamma Rays}},
  \href{https://doi.org/10.1016/j.astropartphys.2013.08.002}{\emph{Astropart.
  Phys.} {\bfseries 50-52} (2013) 26--32},
  [\href{https://arxiv.org/abs/1306.5800}{{\ttfamily 1306.5800}}].

\bibitem{Bai:2019khm}
X.~Bai et~al., \emph{{The Large High Altitude Air Shower Observatory (LHAASO)
  Science White Paper}},  \href{https://arxiv.org/abs/1905.02773}{{\ttfamily
  1905.02773}}.

\bibitem{Guo:2020ooo}
Y.~Guo, X.~Chang, H.~Hu and Z.~Yao, \emph{{Prospects for a Multi-TeV Gamma-ray
  Sky Survey with the LHAASO Water Cherenkov Detector Array}},
  \href{https://arxiv.org/abs/2002.04819}{{\ttfamily 2002.04819}}.

\bibitem{Bernlohr:2008kv}
K.~Bernlohr, \emph{{Simulation of Imaging Atmospheric Cherenkov Telescopes with
  CORSIKA and simtelarray}},
  \href{https://doi.org/10.1016/j.astropartphys.2008.07.009}{\emph{Astropart.
  Phys.} {\bfseries 30} (2008) 149--158},
  [\href{https://arxiv.org/abs/0808.2253}{{\ttfamily 0808.2253}}].

\bibitem{Becherini:2011pb}
Y.~Becherini, A.~Djannati-Atai, V.~Marandon, M.~Punch and S.~Pita, \emph{{A new
  analysis strategy for detection of faint gamma-ray sources with Imaging
  Atmospheric Cherenkov Telescopes}},
  \href{https://doi.org/10.1016/j.astropartphys.2011.03.005}{\emph{Astropart.
  Phys.} {\bfseries 34} (2011) 858--870},
  [\href{https://arxiv.org/abs/1104.5359}{{\ttfamily 1104.5359}}].

\bibitem{hess}
{H.E.S.S.}, ``{High Energy Stereoscopic System}.''
  \url{https://www.mpi-hd.mpg.de/hfm/HESS/}.

\bibitem{MAGIC}
{MAGIC}, ``{Major Atmospheric Gamma Imaging Cherenkov Telescopes}.''
  \url{https://magic.mpp.mpg.de/}.

\bibitem{VERITAS}
{VERITAS}, ``{Very Energetic Radiation Imaging Telescope Array System}.''
  \url{https://veritas.sao.arizona.edu/}.

\bibitem{tevcat2}
\url{http://tevcat2.uchicago.edu/}.

\bibitem{Rinchiuso:2019etv}
{\scshape H.E.S.S.} collaboration, L.~Rinchiuso, E.~Moulin, C.~Armand and
  V.~Poireau, \emph{{Dark Matter search with H.E.S.S. towards ultra-faint dwarf
  nearby DES satellites of the Milky Way}},  in \emph{{36th International
  Cosmic Ray Conference (ICRC 2019) Madison, Wisconsin, USA, July 24-August 1,
  2019}}, 2019, \href{https://arxiv.org/abs/1908.04311}{{\ttfamily
  1908.04311}}.

\bibitem{Rinchiuso:2019rrh}
{\scshape H.E.S.S.} collaboration, L.~Rinchiuso, \emph{{Latest results on dark
  matter searches with H.E.S.S}},
  \href{https://doi.org/10.1051/epjconf/201920901023}{\emph{EPJ Web Conf.}
  {\bfseries 209} (2019) 01023},
  [\href{https://arxiv.org/abs/1901.05299}{{\ttfamily 1901.05299}}].

\bibitem{Zitzer:2017xlo}
{\scshape VERITAS} collaboration, B.~Zitzer, \emph{{The VERITAS Dark Matter
  Program}}, \href{https://doi.org/10.22323/1.301.0904}{\emph{PoS} {\bfseries
  ICRC2017} (2018) 904}, [\href{https://arxiv.org/abs/1708.07447}{{\ttfamily
  1708.07447}}].

\bibitem{Doro:2017dqn}
{\scshape MAGIC} collaboration, M.~Doro, \emph{{A review of the past and
  present MAGIC dark matter search program and a glimpse at the future}},  in
  \emph{{25th European Cosmic Ray Symposium (ECRS 2016) Turin, Italy, September
  04-09, 2016}}, 2017, \href{https://arxiv.org/abs/1701.05702}{{\ttfamily
  1701.05702}}.

\bibitem{CTA_performance}
\url{https://www.cta-observatory.org/science/cta-performance/}.

\bibitem{Knodlseder:2016nnv}
J.~Knödlseder et~al., \emph{{GammaLib and ctools: A software framework for the
  analysis of astronomical gamma-ray data}},
  \href{https://doi.org/10.1051/0004-6361/201628822}{\emph{Astron. Astrophys.}
  {\bfseries 593} (2016) A1},
  [\href{https://arxiv.org/abs/1606.00393}{{\ttfamily 1606.00393}}].

\bibitem{Servant:2002aq}
G.~Servant and T.~M.~P. Tait, \emph{{Is the lightest Kaluza-Klein particle a
  viable dark matter candidate?}},
  \href{https://doi.org/10.1016/S0550-3213(02)01012-X}{\emph{Nucl. Phys.}
  {\bfseries B650} (2003) 391--419},
  [\href{https://arxiv.org/abs/hep-ph/0206071}{{\ttfamily hep-ph/0206071}}].

\bibitem{Cembranos:2003mr}
J.~Cembranos, A.~Dobado and A.~L. Maroto, \emph{{Brane world dark matter}},
  \href{https://doi.org/10.1103/PhysRevLett.90.241301}{\emph{Phys. Rev. Lett.}
  {\bfseries 90} (2003) 241301},
  [\href{https://arxiv.org/abs/hep-ph/0302041}{{\ttfamily hep-ph/0302041}}].

\bibitem{Batell:2017rol}
B.~Batell, T.~Han and B.~Shams Es~Haghi, \emph{{Indirect Detection of Neutrino
  Portal Dark Matter}},
  \href{https://doi.org/10.1103/PhysRevD.97.095020}{\emph{Phys. Rev.}
  {\bfseries D97} (2018) 095020},
  [\href{https://arxiv.org/abs/1704.08708}{{\ttfamily 1704.08708}}].

\bibitem{Bandyopadhyay:2018qcv}
P.~Bandyopadhyay, E.~J. Chun, R.~Mandal and F.~S. Queiroz, \emph{{Scrutinizing
  Right-Handed Neutrino Portal Dark Matter With Yukawa Effect}},
  \href{https://doi.org/10.1016/j.physletb.2018.12.003}{\emph{Phys. Lett.}
  {\bfseries B788} (2019) 530--534},
  [\href{https://arxiv.org/abs/1807.05122}{{\ttfamily 1807.05122}}].

\bibitem{Athron:2018hpc}
{\scshape GAMBIT} collaboration, P.~Athron et~al., \emph{{Global analyses of
  Higgs portal singlet dark matter models using GAMBIT}},
  \href{https://doi.org/10.1140/epjc/s10052-018-6513-6}{\emph{Eur. Phys. J.}
  {\bfseries C79} (2019) 38},
  [\href{https://arxiv.org/abs/1808.10465}{{\ttfamily 1808.10465}}].

\bibitem{Siqueira:2019wdg}
C.~Siqueira, \emph{{Secluded Dark Matter in light of the Cherenkov Telescope
  Array (CTA)}},
  \href{https://doi.org/10.1016/j.physletb.2019.134840}{\emph{Phys. Lett.}
  {\bfseries B797} (2019) 134840},
  [\href{https://arxiv.org/abs/1901.11055}{{\ttfamily 1901.11055}}].

\bibitem{Bergstrom:2004cy}
L.~Bergstrom, T.~Bringmann, M.~Eriksson and M.~Gustafsson, \emph{{Gamma rays
  from Kaluza-Klein dark matter}},
  \href{https://doi.org/10.1103/PhysRevLett.94.131301}{\emph{Phys. Rev. Lett.}
  {\bfseries 94} (2005) 131301},
  [\href{https://arxiv.org/abs/astro-ph/0410359}{{\ttfamily
  astro-ph/0410359}}].

\bibitem{Bringmann:2007nk}
T.~Bringmann, L.~Bergstrom and J.~Edsjo, \emph{{New Gamma-Ray Contributions to
  Supersymmetric Dark Matter Annihilation}},
  \href{https://doi.org/10.1088/1126-6708/2008/01/049}{\emph{JHEP} {\bfseries
  01} (2008) 049}, [\href{https://arxiv.org/abs/0710.3169}{{\ttfamily
  0710.3169}}].

\bibitem{Rinchiuso:2018ajn}
L.~Rinchiuso, N.~L. Rodd, I.~Moult, E.~Moulin, M.~Baumgart, T.~Cohen et~al.,
  \emph{{Hunting for Heavy Winos in the Galactic Center}},
  \href{https://doi.org/10.1103/PhysRevD.98.123014}{\emph{Phys. Rev.}
  {\bfseries D98} (2018) 123014},
  [\href{https://arxiv.org/abs/1808.04388}{{\ttfamily 1808.04388}}].

\bibitem{Baumgart:2018yed}
M.~Baumgart, T.~Cohen, E.~Moulin, I.~Moult, L.~Rinchiuso, N.~L. Rodd et~al.,
  \emph{{Precision Photon Spectra for Wino Annihilation}},
  \href{https://doi.org/10.1007/JHEP01(2019)036}{\emph{JHEP} {\bfseries 01}
  (2019) 036}, [\href{https://arxiv.org/abs/1808.08956}{{\ttfamily
  1808.08956}}].

\bibitem{Beneke:2018ssm}
M.~Beneke, A.~Broggio, C.~Hasner and M.~Vollmann, \emph{{Energetic
  $\gamma$-rays from TeV scale dark matter annihilation resummed}},
  \href{https://doi.org/10.1016/j.physletb.2018.10.008}{\emph{Phys. Lett.}
  {\bfseries B786} (2018) 347--354},
  [\href{https://arxiv.org/abs/1805.07367}{{\ttfamily 1805.07367}}].

\bibitem{Beneke:2019vhz}
M.~Beneke, A.~Broggio, C.~Hasner, K.~Urban and M.~Vollmann, \emph{{Resummed
  photon spectrum from dark matter annihilation for intermediate and narrow
  energy resolution}},
  \href{https://doi.org/10.1007/JHEP08(2019)103}{\emph{JHEP} {\bfseries 08}
  (2019) 103}, [\href{https://arxiv.org/abs/1903.08702}{{\ttfamily
  1903.08702}}].

\bibitem{Aparicio:2016qqb}
L.~Aparicio, M.~Cicoli, B.~Dutta, F.~Muia and F.~Quevedo, \emph{{Light Higgsino
  Dark Matter from Non-thermal Cosmology}},
  \href{https://doi.org/10.1007/JHEP11(2016)038}{\emph{JHEP} {\bfseries 11}
  (2016) 038}, [\href{https://arxiv.org/abs/1607.00004}{{\ttfamily
  1607.00004}}].

\bibitem{Kowalska:2018toh}
K.~Kowalska and E.~M. Sessolo, \emph{{The discreet charm of higgsino dark
  matter - a pocket review}},
  \href{https://doi.org/10.1155/2018/6828560}{\emph{Adv. High Energy Phys.}
  {\bfseries 2018} (2018) 6828560},
  [\href{https://arxiv.org/abs/1802.04097}{{\ttfamily 1802.04097}}].

\bibitem{Beneke:2019gtg}
M.~Beneke, C.~Hasner, K.~Urban and M.~Vollmann, \emph{{Precise yield of
  high-energy photons from Higgsino dark matter annihilation}},
  \href{https://arxiv.org/abs/1912.02034}{{\ttfamily 1912.02034}}.

\bibitem{Hisano:2004ds}
J.~Hisano, S.~Matsumoto, M.~M. Nojiri and O.~Saito, \emph{{Non-perturbative
  effect on dark matter annihilation and gamma ray signature from galactic
  center}}, \href{https://doi.org/10.1103/PhysRevD.71.063528}{\emph{Phys. Rev.}
  {\bfseries D71} (2005) 063528},
  [\href{https://arxiv.org/abs/hep-ph/0412403}{{\ttfamily hep-ph/0412403}}].

\bibitem{Griest:1989wd}
K.~Griest and M.~Kamionkowski, \emph{{Unitarity Limits on the Mass and Radius
  of Dark Matter Particles}},
  \href{https://doi.org/10.1103/PhysRevLett.64.615}{\emph{Phys. Rev. Lett.}
  {\bfseries 64} (1990) 615}.

\bibitem{Smirnov:2019ngs}
J.~Smirnov and J.~F. Beacom, \emph{{TeV-Scale Thermal WIMPs: Unitarity and its
  Consequences}},
  \href{https://doi.org/10.1103/PhysRevD.100.043029}{\emph{Phys. Rev.}
  {\bfseries D100} (2019) 043029},
  [\href{https://arxiv.org/abs/1904.11503}{{\ttfamily 1904.11503}}].

\bibitem{Gondolo:1990dk}
P.~Gondolo and G.~Gelmini, \emph{{Cosmic abundances of stable particles:
  Improved analysis}},
  \href{https://doi.org/10.1016/0550-3213(91)90438-4}{\emph{Nucl. Phys.}
  {\bfseries B360} (1991) 145--179}.

\bibitem{Griest:1990kh}
K.~Griest and D.~Seckel, \emph{{Three exceptions in the calculation of relic
  abundances}}, \href{https://doi.org/10.1103/PhysRevD.43.3191}{\emph{Phys.
  Rev. D} {\bfseries 43} (1991) 3191--3203}.

\bibitem{Baer:2003bp}
H.~Baer and J.~O'Farrill, \emph{{Probing neutralino resonance annihilation via
  indirect detection of dark matter}},
  \href{https://doi.org/10.1088/1475-7516/2004/04/005}{\emph{JCAP} {\bfseries
  04} (2004) 005}, [\href{https://arxiv.org/abs/hep-ph/0312350}{{\ttfamily
  hep-ph/0312350}}].

\bibitem{Kakizaki:2005en}
M.~Kakizaki, S.~Matsumoto, Y.~Sato and M.~Senami, \emph{{Significant effects of
  second KK particles on LKP dark matter physics}},
  \href{https://doi.org/10.1103/PhysRevD.71.123522}{\emph{Phys. Rev. D}
  {\bfseries 71} (2005) 123522},
  [\href{https://arxiv.org/abs/hep-ph/0502059}{{\ttfamily hep-ph/0502059}}].

\bibitem{Ibe:2008ye}
M.~Ibe, H.~Murayama and T.~Yanagida, \emph{{Breit-Wigner Enhancement of Dark
  Matter Annihilation}},
  \href{https://doi.org/10.1103/PhysRevD.79.095009}{\emph{Phys. Rev. D}
  {\bfseries 79} (2009) 095009},
  [\href{https://arxiv.org/abs/0812.0072}{{\ttfamily 0812.0072}}].

\bibitem{Arina:2014fna}
C.~Arina, T.~Bringmann, J.~Silk and M.~Vollmann, \emph{{Enhanced Line Signals
  from Annihilating Kaluza-Klein Dark Matter}},
  \href{https://doi.org/10.1103/PhysRevD.90.083506}{\emph{Phys. Rev. D}
  {\bfseries 90} (2014) 083506},
  [\href{https://arxiv.org/abs/1409.0007}{{\ttfamily 1409.0007}}].

\bibitem{2019A&A...625L..10G}
{Gravity Collaboration}, R.~{Abuter}, A.~{Amorim}, M.~{Baub{\"o}ck}, J.~P.
  {Berger}, H.~{Bonnet} et~al., \emph{{A geometric distance measurement to the
  Galactic center black hole with 0.3\% uncertainty}},
  \href{https://doi.org/10.1051/0004-6361/201935656}{\emph{\aap} {\bfseries
  625} (May, 2019) L10}, [\href{https://arxiv.org/abs/1904.05721}{{\ttfamily
  1904.05721}}].

\bibitem{Abuter:2020dou}
{\scshape GRAVITY} collaboration, R.~Abuter et~al., \emph{{Detection of the
  Schwarzschild precession in the orbit of the star S2 near the Galactic centre
  massive black hole}},  \href{https://arxiv.org/abs/2004.07187}{{\ttfamily
  2004.07187}}.

\bibitem{Catena:2009mf}
R.~Catena and P.~Ullio, \emph{{A novel determination of the local dark matter
  density}}, \href{https://doi.org/10.1088/1475-7516/2010/08/004}{\emph{JCAP}
  {\bfseries 1008} (2010) 004},
  [\href{https://arxiv.org/abs/0907.0018}{{\ttfamily 0907.0018}}].

\bibitem{McMillan:2011wd}
P.~J. McMillan, \emph{{Mass models of the Milky Way}},
  \href{https://doi.org/10.1111/j.1365-2966.2011.18564.x}{\emph{Mon. Not. Roy.
  Astron. Soc.} {\bfseries 414} (2011) 2446--2457},
  [\href{https://arxiv.org/abs/1102.4340}{{\ttfamily 1102.4340}}].

\bibitem{Benito:2019ngh}
M.~Benito, A.~Cuoco and F.~Iocco, \emph{{Handling the Uncertainties in the
  Galactic Dark Matter Distribution for Particle Dark Matter Searches}},
  \href{https://arxiv.org/abs/1901.02460}{{\ttfamily 1901.02460}}.

\bibitem{Karukes:2019jxv}
E.~V. Karukes, M.~Benito, F.~Iocco, R.~Trotta and A.~Geringer-Sameth,
  \emph{{Bayesian reconstruction of the Milky Way dark matter distribution}},
  \href{https://arxiv.org/abs/1901.02463}{{\ttfamily 1901.02463}}.

\bibitem{Iocco:2011jz}
F.~Iocco, M.~Pato, G.~Bertone and P.~Jetzer, \emph{{Dark Matter distribution in
  the Milky Way: microlensing and dynamical constraints}},
  \href{https://doi.org/10.1088/1475-7516/2011/11/029}{\emph{JCAP} {\bfseries
  1111} (2011) 029}, [\href{https://arxiv.org/abs/1107.5810}{{\ttfamily
  1107.5810}}].

\bibitem{Gammaldi:2016uhg}
V.~Gammaldi, V.~Avila-Reese, O.~Valenzuela and A.~X. Gonzales-Morales,
  \emph{{Analysis of the very inner Milky Way dark matter distribution and
  gamma-ray signals}},
  \href{https://doi.org/10.1103/PhysRevD.94.121301}{\emph{Phys. Rev.}
  {\bfseries D94} (2016) 121301},
  [\href{https://arxiv.org/abs/1607.02012}{{\ttfamily 1607.02012}}].

\bibitem{Evans:2003sc}
N.~Evans, F.~Ferrer and S.~Sarkar, \emph{{A 'Baedecker' for the dark matter
  annihilation signal}},
  \href{https://doi.org/10.1103/PhysRevD.69.123501}{\emph{Phys. Rev. D}
  {\bfseries 69} (2004) 123501},
  [\href{https://arxiv.org/abs/astro-ph/0311145}{{\ttfamily
  astro-ph/0311145}}].

\bibitem{Navarro:1995iw}
J.~F. Navarro, C.~S. Frenk and S.~D.~M. White, \emph{{The Structure of cold
  dark matter halos}}, \href{https://doi.org/10.1086/177173}{\emph{Astrophys.
  J.} {\bfseries 462} (1996) 563--575},
  [\href{https://arxiv.org/abs/astro-ph/9508025}{{\ttfamily
  astro-ph/9508025}}].

\bibitem{Navarro:1996gj}
J.~F. Navarro, C.~S. Frenk and S.~D.~M. White, \emph{{A Universal density
  profile from hierarchical clustering}},
  \href{https://doi.org/10.1086/304888}{\emph{Astrophys. J.} {\bfseries 490}
  (1997) 493--508}, [\href{https://arxiv.org/abs/astro-ph/9611107}{{\ttfamily
  astro-ph/9611107}}].

\bibitem{Zavala:2019gpq}
J.~Zavala and C.~S. Frenk, \emph{{Dark matter haloes and subhaloes}},
  \href{https://doi.org/10.3390/galaxies7040081}{\emph{Galaxies} {\bfseries 7}
  (2019) 81}, [\href{https://arxiv.org/abs/1907.11775}{{\ttfamily
  1907.11775}}].

\bibitem{1965TrAlm...5...87E}
J.~{Einasto}, \emph{{On the Construction of a Composite Model for the Galaxy
  and on the Determination of the System of Galactic Parameters}}, {\emph{Trudy
  Astrofizicheskogo Instituta Alma-Ata} {\bfseries 5} (1965) 87--100}.

\bibitem{Karukes:2019jwa}
E.~V. Karukes, M.~Benito, F.~Iocco, R.~Trotta and A.~Geringer-Sameth, \emph{{A
  robust estimate of the Milky Way mass from rotation curve data}},
  \href{https://arxiv.org/abs/1912.04296}{{\ttfamily 1912.04296}}.

\bibitem{Diemand:2008in}
J.~Diemand, M.~Kuhlen, P.~Madau, M.~Zemp, B.~Moore, D.~Potter et~al.,
  \emph{{Clumps and streams in the local dark matter distribution}},
  \href{https://doi.org/10.1038/nature07153}{\emph{Nature} {\bfseries 454}
  (2008) 735--738}, [\href{https://arxiv.org/abs/0805.1244}{{\ttfamily
  0805.1244}}].

\bibitem{Blumenthal:1985qy}
G.~R. Blumenthal, S.~M. Faber, R.~Flores and J.~R. Primack, \emph{{Contraction
  of Dark Matter Galactic Halos Due to Baryonic Infall}},
  \href{https://doi.org/10.1086/163867}{\emph{Astrophys. J.} {\bfseries 301}
  (1986) 27}.

\bibitem{Gnedin:2004cx}
O.~Y. Gnedin, A.~V. Kravtsov, A.~A. Klypin and D.~Nagai, \emph{{Response of
  dark matter halos to condensation of baryons: Cosmological simulations and
  improved adiabatic contraction model}},
  \href{https://doi.org/10.1086/424914}{\emph{Astrophys. J.} {\bfseries 616}
  (2004) 16--26}, [\href{https://arxiv.org/abs/astro-ph/0406247}{{\ttfamily
  astro-ph/0406247}}].

\bibitem{Gustafsson:2006gr}
M.~Gustafsson, M.~Fairbairn and J.~Sommer-Larsen, \emph{{Baryonic Pinching of
  Galactic Dark Matter Haloes}},
  \href{https://doi.org/10.1103/PhysRevD.74.123522}{\emph{Phys. Rev.}
  {\bfseries D74} (2006) 123522},
  [\href{https://arxiv.org/abs/astro-ph/0608634}{{\ttfamily
  astro-ph/0608634}}].

\bibitem{DiCintio:2013qxa}
A.~Di~Cintio, C.~B. Brook, A.~V. Macciò, G.~S. Stinson, A.~Knebe, A.~A. Dutton
  et~al., \emph{{The dependence of dark matter profiles on the stellar-to-halo
  mass ratio: a prediction for cusps versus cores}},
  \href{https://doi.org/10.1093/mnras/stt1891}{\emph{Mon. Not. Roy. Astron.
  Soc.} {\bfseries 437} (2014) 415--423},
  [\href{https://arxiv.org/abs/1306.0898}{{\ttfamily 1306.0898}}].

\bibitem{Cautun:2019eaf}
M.~Cautun, A.~Benitez-Llambay, A.~J. Deason, C.~S. Frenk, A.~Fattahi, F.~A.
  Gómez et~al., \emph{{The Milky Way total mass profile as inferred from Gaia
  DR2}},  \href{https://arxiv.org/abs/1911.04557}{{\ttfamily 1911.04557}}.

\bibitem{Vogelsberger:2019ynw}
M.~Vogelsberger, F.~Marinacci, P.~Torrey and E.~Puchwein, \emph{{Cosmological
  Simulations of Galaxy Formation}},
  \href{https://arxiv.org/abs/1909.07976}{{\ttfamily 1909.07976}}.

\bibitem{Sawala:2015cdf}
T.~Sawala et~al., \emph{{The APOSTLE simulations: solutions to the Local
  Group's cosmic puzzles}},
  \href{https://doi.org/10.1093/mnras/stw145}{\emph{Mon. Not. Roy. Astron.
  Soc.} {\bfseries 457} (2016) 1931--1943},
  [\href{https://arxiv.org/abs/1511.01098}{{\ttfamily 1511.01098}}].

\bibitem{Tollet:2015gqa}
E.~Tollet et~al., \emph{{NIHAO – IV: core creation and destruction in dark
  matter density profiles across cosmic time}},
  \href{https://doi.org/10.1093/mnras/stz1376,
  10.1093/mnras/stv2856}{\emph{Mon. Not. Roy. Astron. Soc.} {\bfseries 456}
  (2016) 3542--3552}, [\href{https://arxiv.org/abs/1507.03590}{{\ttfamily
  1507.03590}}].

\bibitem{Pillepich:2017jle}
A.~Pillepich et~al., \emph{{Simulating Galaxy Formation with the IllustrisTNG
  Model}}, \href{https://doi.org/10.1093/mnras/stx2656}{\emph{Mon. Not. Roy.
  Astron. Soc.} {\bfseries 473} (2018) 4077--4106},
  [\href{https://arxiv.org/abs/1703.02970}{{\ttfamily 1703.02970}}].

\bibitem{Springel:2017tpz}
V.~Springel et~al., \emph{{First results from the IllustrisTNG simulations:
  matter and galaxy clustering}},
  \href{https://doi.org/10.1093/mnras/stx3304}{\emph{Mon. Not. Roy. Astron.
  Soc.} {\bfseries 475} (2018) 676--698},
  [\href{https://arxiv.org/abs/1707.03397}{{\ttfamily 1707.03397}}].

\bibitem{Bovy:2013raa}
J.~Bovy and H.-W. Rix, \emph{{A Direct Dynamical Measurement of the Milky Way's
  Disk Surface Density Profile, Disk Scale Length, and Dark Matter Profile at 4
  kpc $\stackrel{<}{\sim}$ R $\stackrel{<}{\sim}$ 9 kpc}},
  \href{https://doi.org/10.1088/0004-637X/779/2/115}{\emph{Astrophys. J.}
  {\bfseries 779} (2013) 115},
  [\href{https://arxiv.org/abs/1309.0809}{{\ttfamily 1309.0809}}].

\bibitem{Bringmann:2018lay}
T.~Bringmann, J.~Edsjö, P.~Gondolo, P.~Ullio and L.~Bergström,
  \emph{{DarkSUSY 6 : An Advanced Tool to Compute Dark Matter Properties
  Numerically}},
  \href{https://doi.org/10.1088/1475-7516/2018/07/033}{\emph{JCAP} {\bfseries
  1807} (2018) 033}, [\href{https://arxiv.org/abs/1802.03399}{{\ttfamily
  1802.03399}}].

\bibitem{2020SciPy-NMeth}
P.~{Virtanen}, R.~{Gommers}, T.~E. {Oliphant}, M.~{Haberland}, T.~{Reddy},
  D.~{Cournapeau} et~al., \emph{{SciPy 1.0: Fundamental Algorithms for
  Scientific Computing in Python}},
  \href{https://doi.org/https://doi.org/10.1038/s41592-019-0686-2}{\emph{Nature
  Methods} (2020) }.

\bibitem{bonnivard2016clumpy}
V.~Bonnivard et~al., \emph{{CLUMPY: Jeans analysis, $\gamma$-ray and $\nu$
  fluxes from dark matter (sub-) structures}}, {\emph{Computer physics
  communications} {\bfseries 200} (2016) 336--349}.

\bibitem{charbonnier2012clumpy}
A.~Charbonnier, C.~Combet and D.~Maurin, \emph{{CLUMPY: A code for $\gamma$-ray
  signals from dark matter structures}}, {\emph{Computer Physics
  Communications} {\bfseries 183} (2012) 656--668}.

\bibitem{Hutten:2018aix}
M.~H\"utten, C.~Combet and D.~Maurin, \emph{{{CLUMPY v3: $\gamma$-ray and $\nu$
  signals from dark matter at all scales}}},
  \href{https://arxiv.org/abs/1806.08639}{{\ttfamily 1806.08639}}.

\bibitem{Gorski_2005}
K.~M. Gorski, E.~Hivon, A.~J. Banday, B.~D. Wandelt, F.~K. Hansen, M.~Reinecke
  et~al., \emph{{HEALPix}: A framework for high-resolution discretization and
  fast analysis of data distributed on the sphere},
  \href{https://doi.org/10.1086/427976}{\emph{The Astrophysical Journal}
  {\bfseries 622} (apr, 2005) 759--771}.

\bibitem{Sjostrand:2007gs}
T.~Sjostrand, S.~Mrenna and P.~Z. Skands, \emph{{A Brief Introduction to PYTHIA
  8.1}}, \href{https://doi.org/10.1016/j.cpc.2008.01.036}{\emph{Comput. Phys.
  Commun.} {\bfseries 178} (2008) 852--867},
  [\href{https://arxiv.org/abs/0710.3820}{{\ttfamily 0710.3820}}].

\bibitem{Corcella:2000bw}
G.~Corcella, I.~G. Knowles, G.~Marchesini, S.~Moretti, K.~Odagiri,
  P.~Richardson et~al., \emph{{HERWIG 6: An Event generator for hadron emission
  reactions with interfering gluons (including supersymmetric processes)}},
  \href{https://doi.org/10.1088/1126-6708/2001/01/010}{\emph{JHEP} {\bfseries
  01} (2001) 010}, [\href{https://arxiv.org/abs/hep-ph/0011363}{{\ttfamily
  hep-ph/0011363}}].

\bibitem{Cembranos:2013cfa}
J.~A.~R. Cembranos, A.~de~la Cruz-Dombriz, V.~Gammaldi, R.~A. Lineros and A.~L.
  Maroto, \emph{{Reliability of Monte Carlo event generators for gamma ray dark
  matter searches}}, \href{https://doi.org/10.1007/JHEP09(2013)077}{\emph{JHEP}
  {\bfseries 09} (2013) 077},
  [\href{https://arxiv.org/abs/1305.2124}{{\ttfamily 1305.2124}}].

\bibitem{Bringmann:2017sko}
T.~Bringmann, F.~Calore, A.~Galea and M.~Garny, \emph{{Electroweak and Higgs
  Boson Internal Bremsstrahlung: General considerations for Majorana dark
  matter annihilation and application to MSSM neutralinos}},
  \href{https://doi.org/10.1007/JHEP09(2017)041}{\emph{JHEP} {\bfseries 09}
  (2017) 041}, [\href{https://arxiv.org/abs/1705.03466}{{\ttfamily
  1705.03466}}].

\bibitem{Bergstrom:1989jr}
L.~Bergstrom, \emph{{Radiative Processes in Dark Matter Photino Annihilation}},
  \href{https://doi.org/10.1016/0370-2693(89)90585-6}{\emph{Phys. Lett.}
  {\bfseries B225} (1989) 372--380}.

\bibitem{Flores:1989ru}
R.~Flores, K.~A. Olive and S.~Rudaz, \emph{{Radiative Processes in Lsp
  Annihilation}},
  \href{https://doi.org/10.1016/0370-2693(89)90760-0}{\emph{Phys. Lett.}
  {\bfseries B232} (1989) 377--382}.

\bibitem{Bringmann:2015cpa}
T.~Bringmann, A.~J. Galea and P.~Walia, \emph{{Leading QCD Corrections for
  Indirect Dark Matter Searches: a Fresh Look}},
  \href{https://doi.org/10.1103/PhysRevD.93.043529}{\emph{Phys. Rev.}
  {\bfseries D93} (2016) 043529},
  [\href{https://arxiv.org/abs/1510.02473}{{\ttfamily 1510.02473}}].

\bibitem{Kachelriess:2009zy}
M.~Kachelriess, P.~D. Serpico and M.~A. Solberg, \emph{{On the role of
  electroweak bremsstrahlung for indirect dark matter signatures}},
  \href{https://doi.org/10.1103/PhysRevD.80.123533}{\emph{Phys. Rev.}
  {\bfseries D80} (2009) 123533},
  [\href{https://arxiv.org/abs/0911.0001}{{\ttfamily 0911.0001}}].

\bibitem{Ciafaloni:2011sa}
P.~Ciafaloni, M.~Cirelli, D.~Comelli, A.~De~Simone, A.~Riotto and A.~Urbano,
  \emph{{On the Importance of Electroweak Corrections for Majorana Dark Matter
  Indirect Detection}},
  \href{https://doi.org/10.1088/1475-7516/2011/06/018}{\emph{JCAP} {\bfseries
  1106} (2011) 018}, [\href{https://arxiv.org/abs/1104.2996}{{\ttfamily
  1104.2996}}].

\bibitem{Bell:2011if}
N.~F. Bell, J.~B. Dent, A.~J. Galea, T.~D. Jacques, L.~M. Krauss and T.~J.
  Weiler, \emph{{W/Z Bremsstrahlung as the Dominant Annihilation Channel for
  Dark Matter, Revisited}},
  \href{https://doi.org/10.1016/j.physletb.2011.10.057}{\emph{Phys. Lett.}
  {\bfseries B706} (2011) 6--12},
  [\href{https://arxiv.org/abs/1104.3823}{{\ttfamily 1104.3823}}].

\bibitem{Garny:2011cj}
M.~Garny, A.~Ibarra and S.~Vogl, \emph{{Antiproton constraints on dark matter
  annihilations from internal electroweak bremsstrahlung}},
  \href{https://doi.org/10.1088/1475-7516/2011/07/028}{\emph{JCAP} {\bfseries
  1107} (2011) 028}, [\href{https://arxiv.org/abs/1105.5367}{{\ttfamily
  1105.5367}}].

\bibitem{Garny:2011ii}
M.~Garny, A.~Ibarra and S.~Vogl, \emph{{Dark matter annihilations into two
  light fermions and one gauge boson: General analysis and antiproton
  constraints}},
  \href{https://doi.org/10.1088/1475-7516/2012/04/033}{\emph{JCAP} {\bfseries
  1204} (2012) 033}, [\href{https://arxiv.org/abs/1112.5155}{{\ttfamily
  1112.5155}}].

\bibitem{Ciafaloni:2010ti}
P.~Ciafaloni, D.~Comelli, A.~Riotto, F.~Sala, A.~Strumia and A.~Urbano,
  \emph{{Weak Corrections are Relevant for Dark Matter Indirect Detection}},
  \href{https://doi.org/10.1088/1475-7516/2011/03/019}{\emph{JCAP} {\bfseries
  1103} (2011) 019}, [\href{https://arxiv.org/abs/1009.0224}{{\ttfamily
  1009.0224}}].

\bibitem{Cirelli:2010xx}
M.~Cirelli, G.~Corcella, A.~Hektor, G.~Hutsi, M.~Kadastik, P.~Panci et~al.,
  \emph{{PPPC 4 DM ID: A Poor Particle Physicist Cookbook for Dark Matter
  Indirect Detection}}, \href{https://doi.org/10.1088/1475-7516/2012/10/E01,
  10.1088/1475-7516/2011/03/051}{\emph{JCAP} {\bfseries 1103} (2011) 051},
  [\href{https://arxiv.org/abs/1012.4515}{{\ttfamily 1012.4515}}].

\bibitem{Bergstrom:2005ss}
L.~Bergstrom, T.~Bringmann, M.~Eriksson and M.~Gustafsson, \emph{{Gamma rays
  from heavy neutralino dark matter}},
  \href{https://doi.org/10.1103/PhysRevLett.95.241301}{\emph{Phys. Rev. Lett.}
  {\bfseries 95} (2005) 241301},
  [\href{https://arxiv.org/abs/hep-ph/0507229}{{\ttfamily hep-ph/0507229}}].

\bibitem{Regis:2008ij}
M.~Regis and P.~Ullio, \emph{{Multi-wavelength signals of dark matter
  annihilations at the Galactic center}},
  \href{https://doi.org/10.1103/PhysRevD.78.043505}{\emph{Phys. Rev.}
  {\bfseries D78} (2008) 043505},
  [\href{https://arxiv.org/abs/0802.0234}{{\ttfamily 0802.0234}}].

\bibitem{Cirelli:2009vg}
M.~Cirelli and P.~Panci, \emph{{Inverse Compton constraints on the Dark Matter
  e+e- excesses}},
  \href{https://doi.org/10.1016/j.nuclphysb.2009.06.034}{\emph{Nucl. Phys.}
  {\bfseries B821} (2009) 399--416},
  [\href{https://arxiv.org/abs/0904.3830}{{\ttfamily 0904.3830}}].

\bibitem{Belikov:2009cx}
A.~V. Belikov and D.~Hooper, \emph{{The Contribution Of Inverse Compton
  Scattering To The Diffuse Extragalactic Gamma-Ray Background From
  Annihilating Dark Matter}},
  \href{https://doi.org/10.1103/PhysRevD.81.043505}{\emph{Phys. Rev.}
  {\bfseries D81} (2010) 043505},
  [\href{https://arxiv.org/abs/0906.2251}{{\ttfamily 0906.2251}}].

\bibitem{Abazajian:2010zb}
K.~N. Abazajian, S.~Blanchet and J.~P. Harding, \emph{{Current and Future
  Constraints on Dark Matter from Prompt and Inverse-Compton Photon Emission in
  the Isotropic Diffuse Gamma-ray Background}},
  \href{https://doi.org/10.1103/PhysRevD.85.043509}{\emph{Phys. Rev.}
  {\bfseries D85} (2012) 043509},
  [\href{https://arxiv.org/abs/1011.5090}{{\ttfamily 1011.5090}}].

\bibitem{Delahaye:2010ji}
T.~Delahaye, J.~Lavalle, R.~Lineros, F.~Donato and N.~Fornengo, \emph{{Galactic
  electrons and positrons at the Earth:new estimate of the primary and
  secondary fluxes}},
  \href{https://doi.org/10.1051/0004-6361/201014225}{\emph{Astron. Astrophys.}
  {\bfseries 524} (2010) A51},
  [\href{https://arxiv.org/abs/1002.1910}{{\ttfamily 1002.1910}}].

\bibitem{Genzel:2010zy}
R.~Genzel, F.~Eisenhauer and S.~Gillessen, \emph{{The Galactic Center Massive
  Black Hole and Nuclear Star Cluster}},
  \href{https://doi.org/10.1103/RevModPhys.82.3121}{\emph{Rev. Mod. Phys.}
  {\bfseries 82} (2010) 3121--3195},
  [\href{https://arxiv.org/abs/1006.0064}{{\ttfamily 1006.0064}}].

\bibitem{Morris:1996th}
M.~Morris and E.~Serabyn, \emph{{The galactic center environment}},
  \href{https://doi.org/10.1146/annurev.astro.34.1.645}{\emph{Ann. Rev. Astron.
  Astrophys.} {\bfseries 34} (1996) 645--701}.

\bibitem{Ackermann:2012pya}
{\scshape Fermi-LAT} collaboration, M.~Ackermann et~al., \emph{{Fermi-LAT
  Observations of the Diffuse Gamma-Ray Emission: Implications for Cosmic Rays
  and the Interstellar Medium}},
  \href{https://doi.org/10.1088/0004-637X/750/1/3}{\emph{Astrophys. J.}
  {\bfseries 750} (2012) 3}, [\href{https://arxiv.org/abs/1202.4039}{{\ttfamily
  1202.4039}}].

\bibitem{2017arXiv170505332M}
E.~A.~C. {Mills}, \emph{{The Milky Way's Central Molecular Zone}}, {\emph{arXiv
  e-prints} (May, 2017) arXiv:1705.05332},
  [\href{https://arxiv.org/abs/1705.05332}{{\ttfamily 1705.05332}}].

\bibitem{2017IAUS..322..164B}
R.~{Blackwell}, M.~{Burton} and G.~{Rowell}, \emph{{Mopra Central Molecular
  Zone Carbon Monoxide Survey Status}},  in \emph{The Multi-Messenger
  Astrophysics of the Galactic Centre} (R.~M. {Crocker}, S.~N. {Longmore} and
  G.~V. {Bicknell}, eds.), vol.~322 of \emph{IAU Symposium}, pp.~164--165,
  Jan., 2017, \href{https://doi.org/10.1017/S1743921316012035}{DOI}.

\bibitem{Aharonian:2006au}
{\scshape H.E.S.S.} collaboration, F.~Aharonian et~al., \emph{{Discovery of
  very-high-energy gamma-rays from the galactic centre ridge}},
  \href{https://doi.org/10.1038/nature04467}{\emph{Nature} {\bfseries 439}
  (2006) 695--698}, [\href{https://arxiv.org/abs/astro-ph/0603021}{{\ttfamily
  astro-ph/0603021}}].

\bibitem{Abdalla:2017xja}
{\scshape HESS} collaboration, H.~Abdalla et~al., \emph{{{Characterising the
  VHE diffuse emission in the central 200 parsecs of our Galaxy with
  H.E.S.S}}}, \href{https://doi.org/10.1051/0004-6361/201730824}{\emph{Astron.
  Astrophys.} {\bfseries 612} (2018) A9},
  [\href{https://arxiv.org/abs/1706.04535}{{\ttfamily 1706.04535}}].

\bibitem{Archer:2016ein}
A.~Archer et~al., \emph{{TeV Gamma-ray Observations of The Galactic Center
  Ridge By VERITAS}},
  \href{https://doi.org/10.3847/0004-637X/821/2/129}{\emph{Astrophys. J.}
  {\bfseries 821} (2016) 129},
  [\href{https://arxiv.org/abs/1602.08522}{{\ttfamily 1602.08522}}].

\bibitem{Abramowski:2016mir}
{\scshape H.E.S.S.} collaboration, A.~Abramowski et~al., \emph{{Acceleration of
  petaelectronvolt protons in the Galactic Centre}},
  \href{https://doi.org/10.1038/nature17147}{\emph{Nature} {\bfseries 531}
  (2016) 476}, [\href{https://arxiv.org/abs/1603.07730}{{\ttfamily
  1603.07730}}].

\bibitem{Neronov:2019ncc}
A.~Neronov and D.~Semikoz, \emph{{Galactic diffuse gamma-ray emission at TeV
  energy}}, \href{https://doi.org/10.1051/0004-6361/201936368}{\emph{Astron.
  Astrophys.} {\bfseries 633} (2020) A94},
  [\href{https://arxiv.org/abs/1907.06061}{{\ttfamily 1907.06061}}].

\bibitem{Abramowski:2014vox}
{\scshape H.E.S.S.} collaboration, A.~Abramowski et~al., \emph{{Diffuse
  Galactic gamma-ray emission with H.E.S.S}},
  \href{https://doi.org/10.1103/PhysRevD.90.122007}{\emph{Phys. Rev.}
  {\bfseries D90} (2014) 122007},
  [\href{https://arxiv.org/abs/1411.7568}{{\ttfamily 1411.7568}}].

\bibitem{Abdo:2008if}
A.~A. Abdo et~al., \emph{{A Measurement of the Spatial Distribution of Diffuse
  TeV Gamma Ray Emission from the Galactic Plane with Milagro}},
  \href{https://doi.org/10.1086/592213}{\emph{Astrophys. J.} {\bfseries 688}
  (2008) 1078--1083}, [\href{https://arxiv.org/abs/0805.0417}{{\ttfamily
  0805.0417}}].

\bibitem{Zhou:2017lgv}
{\scshape HAWC} collaboration, H.~Zhou, C.~D. Rho and G.~Vianello,
  \emph{{Probing Galactic Diffuse TeV Gamma-Ray Emission with the HAWC
  Observatory}}, \href{https://doi.org/10.22323/1.301.0689}{\emph{PoS}
  {\bfseries ICRC2017} (2018) 689},
  [\href{https://arxiv.org/abs/1709.03619}{{\ttfamily 1709.03619}}].

\bibitem{Gaggero:2017jts}
D.~Gaggero, D.~Grasso, A.~Marinelli, M.~Taoso and A.~Urbano, \emph{{Diffuse
  cosmic rays shining in the Galactic center: A novel interpretation of
  H.E.S.S. and Fermi-LAT $\gamma$-ray data}},
  \href{https://doi.org/10.1103/PhysRevLett.119.031101}{\emph{Phys. Rev. Lett.}
  {\bfseries 119} (2017) 031101},
  [\href{https://arxiv.org/abs/1702.01124}{{\ttfamily 1702.01124}}].

\bibitem{H.E.S.S.:2018zkf}
{\scshape HESS} collaboration, H.~Abdalla et~al., \emph{{The H.E.S.S. Galactic
  plane survey}},
  \href{https://doi.org/10.1051/0004-6361/201732098}{\emph{Astron. Astrophys.}
  {\bfseries 612} (2018) A1},
  [\href{https://arxiv.org/abs/1804.02432}{{\ttfamily 1804.02432}}].

\bibitem{Ahnen:2016crz}
M.~L. Ahnen et~al., \emph{{Observations of Sagittarius A* during the pericenter
  passage of the G2 object with MAGIC}},
  \href{https://doi.org/10.1051/0004-6361/201629355}{\emph{Astron. Astrophys.}
  {\bfseries 601} (2017) A33},
  [\href{https://arxiv.org/abs/1611.07095}{{\ttfamily 1611.07095}}].

\bibitem{Ackermann:2015uya}
{\scshape Fermi-LAT} collaboration, M.~Ackermann et~al., \emph{{2FHL: The
  Second Catalog of Hard Fermi-LAT Sources}},
  \href{https://doi.org/10.3847/0067-0049/222/1/5}{\emph{Astrophys. J. Suppl.}
  {\bfseries 222} (2016) 5},
  [\href{https://arxiv.org/abs/1508.04449}{{\ttfamily 1508.04449}}].

\bibitem{Fermi-LAT:2019yla}
{\scshape Fermi-LAT} collaboration, \emph{{Fermi Large Area Telescope Fourth
  Source Catalog}},  \href{https://arxiv.org/abs/1902.10045}{{\ttfamily
  1902.10045}}.

\bibitem{GPSPreliminary}
{\scshape CTA} collaboration, \emph{{Survey of the Galactic Plane with the
  Cherenkov Telescope Array}},  \href{https://arxiv.org/abs/in preparation
  within the CTA}{{\ttfamily in preparation within the CTA}}.

\bibitem{2018A&A...612E...1F}
T.~{Forveille}, S.~{Campana} and S.~{Shore}, \emph{{H.E.S.S. phase-I
  observations of the plane of the Milky Way}},
  \href{https://doi.org/10.1051/0004-6361/201833049}{\emph{\aap} {\bfseries
  612} (Apr., 2018) E1}.

\bibitem{Su:2010qj}
M.~Su, T.~R. Slatyer and D.~P. Finkbeiner, \emph{{Giant Gamma-ray Bubbles from
  Fermi-LAT: AGN Activity or Bipolar Galactic Wind?}},
  \href{https://doi.org/10.1088/0004-637X/724/2/1044}{\emph{Astrophys. J.}
  {\bfseries 724} (2010) 1044--1082},
  [\href{https://arxiv.org/abs/1005.5480}{{\ttfamily 1005.5480}}].

\bibitem{Fermi-LAT:2014sfa}
{\scshape Fermi-LAT} collaboration, M.~Ackermann et~al., \emph{{The Spectrum
  and Morphology of the $Fermi$ Bubbles}},
  \href{https://doi.org/10.1088/0004-637X/793/1/64}{\emph{Astrophys. J.}
  {\bfseries 793} (2014) 64},
  [\href{https://arxiv.org/abs/1407.7905}{{\ttfamily 1407.7905}}].

\bibitem{TheFermi-LAT:2017vmf}
{\scshape Fermi-LAT} collaboration, M.~Ackermann et~al., \emph{{The Fermi
  Galactic Center GeV Excess and Implications for Dark Matter}},
  \href{https://doi.org/10.3847/1538-4357/aa6cab}{\emph{Astrophys. J.}
  {\bfseries 840} (2017) 43},
  [\href{https://arxiv.org/abs/1704.03910}{{\ttfamily 1704.03910}}].

\bibitem{Yang:2018bfb}
L.~Yang and S.~Razzaque, \emph{{Constraints on very high energy gamma-ray
  emission from the Fermi Bubbles with future ground-based experiments}},
  \href{https://doi.org/10.1103/PhysRevD.99.083007}{\emph{Phys. Rev. D}
  {\bfseries 99} (2019) 083007},
  [\href{https://arxiv.org/abs/1811.10970}{{\ttfamily 1811.10970}}].

\bibitem{Herold:2019pei}
L.~Herold and D.~Malyshev, \emph{{Hard and bright gamma-ray emission at the
  base of the Fermi bubbles}},
  \href{https://arxiv.org/abs/1904.01454}{{\ttfamily 1904.01454}}.

\bibitem{electrons1}
T.~Edwards,
  \emph{\href{https://www.imprs-hd.mpg.de/143531/thesis_edwards.pdf}{Separation
  of $\gamma$-Ray, Electron andProton induced Air Showers applied to Diffuse
  Emission Studies with H.E.S.S.}}, Ph.D. thesis, Combined Faculties of the
  Natural Sciences and Mathematics of the Ruperto-Carola-University of
  Heidelberg, 2018.

\bibitem{Aguilar:2014fea}
{\scshape AMS} collaboration, M.~Aguilar et~al., \emph{{Precision Measurement
  of the ($e^+ + e^?$) Flux in Primary Cosmic Rays from 0.5 GeV to 1 TeV with
  the Alpha Magnetic Spectrometer on the International Space Station}},
  \href{https://doi.org/10.1103/PhysRevLett.113.221102}{\emph{Phys. Rev. Lett.}
  {\bfseries 113} (2014) 221102}.

\bibitem{Aguilar:2014mma}
{\scshape AMS} collaboration, M.~Aguilar et~al., \emph{{Electron and Positron
  Fluxes in Primary Cosmic Rays Measured with the Alpha Magnetic Spectrometer
  on the International Space Station}},
  \href{https://doi.org/10.1103/PhysRevLett.113.121102}{\emph{Phys. Rev. Lett.}
  {\bfseries 113} (2014) 121102}.

\bibitem{Adriani:2018ktz}
O.~Adriani et~al., \emph{{Extended Measurement of the Cosmic-Ray Electron and
  Positron Spectrum from 11 GeV to 4.8 TeV with the Calorimetric Electron
  Telescope on the International Space Station}},
  \href{https://doi.org/10.1103/PhysRevLett.120.261102}{\emph{Phys. Rev. Lett.}
  {\bfseries 120} (2018) 261102},
  [\href{https://arxiv.org/abs/1806.09728}{{\ttfamily 1806.09728}}].

\bibitem{Ambrosi:2017wek}
{\scshape DAMPE} collaboration, G.~Ambrosi et~al., \emph{{Direct detection of a
  break in the teraelectronvolt cosmic-ray spectrum of electrons and
  positrons}}, \href{https://doi.org/10.1038/nature24475}{\emph{Nature}
  {\bfseries 552} (2017) 63--66},
  [\href{https://arxiv.org/abs/1711.10981}{{\ttfamily 1711.10981}}].

\bibitem{1994APh.....2..137F}
V.~P. {Fomin}, A.~A. {Stepanian}, R.~C. {Lamb}, D.~A. {Lewis}, M.~{Punch} and
  T.~C. {Weekes}, \emph{{New methods of atmospheric Cherenkov imaging for
  gamma-ray astronomy. I. The false source method}},
  \href{https://doi.org/10.1016/0927-6505(94)90036-1}{\emph{Astroparticle
  Physics} {\bfseries 2} (May, 1994) 137--150}.

\bibitem{Knodlseder:2019coy}
J.~Knödlseder et~al., \emph{{Analysis of the H.E.S.S. public data release with
  ctools}}, \href{https://doi.org/10.1051/0004-6361/201936010}{\emph{Astron.
  Astrophys.} {\bfseries 632} (2019) A102},
  [\href{https://arxiv.org/abs/1910.09456}{{\ttfamily 1910.09456}}].

\bibitem{Mohrmann:2019hfq}
L.~Mohrmann, A.~Specovius, D.~Tiziani, S.~Funk, D.~Malyshev, K.~Nakashima
  et~al., \emph{{Validation of open-source science tools and background model
  construction in $\gamma$-ray astronomy}},
  \href{https://doi.org/10.1051/0004-6361/201936452}{\emph{Astron. Astrophys.}
  {\bfseries 632} (2019) A72},
  [\href{https://arxiv.org/abs/1910.08088}{{\ttfamily 1910.08088}}].

\bibitem{Porter:2019wih}
T.~A. {Porter}, G.~{J{\'o}hannesson} and I.~V. {Moskalenko}, \emph{{Deciphering
  Residual Emissions: Time-dependent Models for the Nonthermal Interstellar
  Radiation from the Milky Way}},
  \href{https://doi.org/10.3847/1538-4357/ab5961}{\emph{\apj} {\bfseries 887}
  (Dec., 2019) 250}, [\href{https://arxiv.org/abs/1909.02223}{{\ttfamily
  1909.02223}}].

\bibitem{Abdallah:2016ygi}
{\scshape H.E.S.S.} collaboration, H.~Abdallah et~al., \emph{{Search for dark
  matter annihilations towards the inner Galactic halo from 10 years of
  observations with H.E.S.S}},
  \href{https://doi.org/10.1103/PhysRevLett.117.111301}{\emph{Phys. Rev. Lett.}
  {\bfseries 117} (2016) 111301},
  [\href{https://arxiv.org/abs/1607.08142}{{\ttfamily 1607.08142}}].

\bibitem{Charles:2016pgz}
{\scshape Fermi-LAT} collaboration, E.~Charles et~al., \emph{{Sensitivity
  Projections for Dark Matter Searches with the Fermi Large Area Telescope}},
  \href{https://doi.org/10.1016/j.physrep.2016.05.001}{\emph{Phys. Rept.}
  {\bfseries 636} (2016) 1--46},
  [\href{https://arxiv.org/abs/1605.02016}{{\ttfamily 1605.02016}}].

\bibitem{Storm:2017arh}
E.~Storm, C.~Weniger and F.~Calore, \emph{{SkyFACT: High-dimensional modeling
  of gamma-ray emission with adaptive templates and penalized likelihoods}},
  \href{https://doi.org/10.1088/1475-7516/2017/08/022}{\emph{JCAP} {\bfseries
  1708} (2017) 022}, [\href{https://arxiv.org/abs/1705.04065}{{\ttfamily
  1705.04065}}].

\bibitem{Wilks:1938dza}
S.~S. Wilks, \emph{{The Large-Sample Distribution of the Likelihood Ratio for
  Testing Composite Hypotheses}},
  \href{https://doi.org/10.1214/aoms/1177732360}{\emph{Annals Math. Statist.}
  {\bfseries 9} (1938) 60--62}.

\bibitem{2011EPJC...71.1554C}
G.~{Cowan}, K.~{Cranmer}, E.~{Gross} and O.~{Vitells}, \emph{{Asymptotic
  formulae for likelihood-based tests of new physics}},
  \href{https://doi.org/10.1140/epjc/s10052-011-1554-0}{\emph{European Physical
  Journal C} {\bfseries 71} (Feb., 2011) 1554},
  [\href{https://arxiv.org/abs/1007.1727}{{\ttfamily 1007.1727}}].

\bibitem{Edwards:2017mnf}
T.~D.~P. Edwards and C.~Weniger, \emph{{A Fresh Approach to Forecasting in
  Astroparticle Physics and Dark Matter Searches}},
  \href{https://doi.org/10.1088/1475-7516/2018/02/021}{\emph{JCAP} {\bfseries
  1802} (2018) 021}, [\href{https://arxiv.org/abs/1704.05458}{{\ttfamily
  1704.05458}}].

\bibitem{Edwards:2017kqw}
T.~D.~P. Edwards and C.~Weniger, \emph{{swordfish: Efficient Forecasting of New
  Physics Searches without Monte Carlo}},
  \href{https://arxiv.org/abs/1712.05401}{{\ttfamily 1712.05401}}.

\bibitem{www_swordfish}
\url{https://github.com/cweniger/swordfish/}.

\bibitem{PhysRevLett.120.201101}
{\scshape H.E.S.S. Collaboration} collaboration, H.~Abdallah et~al.,
  \emph{{{Search for $\gamma$-Ray Line Signals from Dark Matter Annihilations
  in the Inner Galactic Halo from 10 Years of Observations with H.E.S.S.}}},
  \href{https://doi.org/10.1103/PhysRevLett.120.201101}{\emph{Phys. Rev. Lett.}
  {\bfseries 120} (May, 2018) 201101}.

\bibitem{PLA}
{\scshape Planck} collaboration, \emph{{Planck Legacy Archive}},
  \href{https://arxiv.org/abs/http://pla.esac.esa.int/pla/}{{\ttfamily
  http://pla.esac.esa.int/pla/}}.

\bibitem{Steigman:2012nb}
G.~Steigman, B.~Dasgupta and J.~F. Beacom, \emph{{Precise Relic WIMP Abundance
  and its Impact on Searches for Dark Matter Annihilation}},
  \href{https://doi.org/10.1103/PhysRevD.86.023506}{\emph{Phys. Rev.}
  {\bfseries D86} (2012) 023506},
  [\href{https://arxiv.org/abs/1204.3622}{{\ttfamily 1204.3622}}].

\bibitem{Drees:2015exa}
M.~Drees, F.~Hajkarim and E.~R. Schmitz, \emph{{The Effects of QCD Equation of
  State on the Relic Density of WIMP Dark Matter}},
  \href{https://doi.org/10.1088/1475-7516/2015/06/025}{\emph{JCAP} {\bfseries
  1506} (2015) 025}, [\href{https://arxiv.org/abs/1503.03513}{{\ttfamily
  1503.03513}}].

\bibitem{Ackermann:2015zua}
{\scshape Fermi-LAT} collaboration, M.~Ackermann et~al., \emph{{Searching for
  Dark Matter Annihilation from Milky Way Dwarf Spheroidal Galaxies with Six
  Years of Fermi Large Area Telescope Data}},
  \href{https://doi.org/10.1103/PhysRevLett.115.231301}{\emph{Phys. Rev. Lett.}
  {\bfseries 115} (2015) 231301},
  [\href{https://arxiv.org/abs/1503.02641}{{\ttfamily 1503.02641}}].

\bibitem{bringmann_torsten_2020_4057987}
T.~Bringmann, C.~Eckner, A.~Sokolenko, L.~Yang and G.~Zaharijas,
  \emph{{Likelihoods for the CTA sensitivity to a dark matter signal from the
  Galactic centre (A. Acharyya et al., [arXiv:2007.16129])}},  Sept., 2020.
\newblock 10.5281/zenodo.4057987.

\bibitem{Ackermann:2015lka}
{\scshape Fermi-LAT} collaboration, M.~Ackermann et~al., \emph{{Updated search
  for spectral lines from Galactic dark matter interactions with pass 8 data
  from the Fermi Large Area Telescope}},
  \href{https://doi.org/10.1103/PhysRevD.91.122002}{\emph{Phys. Rev. D}
  {\bfseries 91} (2015) 122002},
  [\href{https://arxiv.org/abs/1506.00013}{{\ttfamily 1506.00013}}].

\bibitem{Bonnivard:2015xpq}
V.~Bonnivard et~al., \emph{{Dark matter annihilation and decay in dwarf
  spheroidal galaxies: The classical and ultrafaint dSphs}},
  \href{https://doi.org/10.1093/mnras/stv1601}{\emph{Mon. Not. Roy. Astron.
  Soc.} {\bfseries 453} (2015) 849--867},
  [\href{https://arxiv.org/abs/1504.02048}{{\ttfamily 1504.02048}}].

\bibitem{GeringerSameth:2011iw}
A.~Geringer-Sameth and S.~M. Koushiappas, \emph{{Exclusion of canonical WIMPs
  by the joint analysis of Milky Way dwarfs with Fermi}},
  \href{https://doi.org/10.1103/PhysRevLett.107.241303}{\emph{Phys. Rev. Lett.}
  {\bfseries 107} (2011) 241303},
  [\href{https://arxiv.org/abs/1108.2914}{{\ttfamily 1108.2914}}].

\bibitem{Drlica-Wagner:2015xua}
{\scshape Fermi-LAT, DES} collaboration, A.~Drlica-Wagner et~al., \emph{{Search
  for Gamma-Ray Emission from DES Dwarf Spheroidal Galaxy Candidates with
  Fermi-LAT Data}},
  \href{https://doi.org/10.1088/2041-8205/809/1/L4}{\emph{Astrophys. J.}
  {\bfseries 809} (2015) L4},
  [\href{https://arxiv.org/abs/1503.02632}{{\ttfamily 1503.02632}}].

\bibitem{Ahnen:2016qkx}
{\scshape MAGIC, Fermi-LAT} collaboration, M.~L. Ahnen et~al., \emph{{Limits to
  Dark Matter Annihilation Cross-Section from a Combined Analysis of MAGIC and
  Fermi-LAT Observations of Dwarf Satellite Galaxies}},
  \href{https://doi.org/10.1088/1475-7516/2016/02/039}{\emph{JCAP} {\bfseries
  1602} (2016) 039}, [\href{https://arxiv.org/abs/1601.06590}{{\ttfamily
  1601.06590}}].

\bibitem{Ackermann:2011wa}
{\scshape Fermi-LAT} collaboration, M.~Ackermann et~al., \emph{{Constraining
  Dark Matter Models from a Combined Analysis of Milky Way Satellites with the
  Fermi Large Area Telescope}},
  \href{https://doi.org/10.1103/PhysRevLett.107.241302}{\emph{Phys. Rev. Lett.}
  {\bfseries 107} (2011) 241302},
  [\href{https://arxiv.org/abs/1108.3546}{{\ttfamily 1108.3546}}].

\bibitem{Abramowski:2011hc}
{\scshape H.E.S.S.} collaboration, A.~Abramowski et~al., \emph{{Search for a
  Dark Matter annihilation signal from the Galactic Center halo with H.E.S.S}},
  \href{https://doi.org/10.1103/PhysRevLett.106.161301}{\emph{Phys. Rev. Lett.}
  {\bfseries 106} (2011) 161301},
  [\href{https://arxiv.org/abs/1103.3266}{{\ttfamily 1103.3266}}].

\bibitem{Abdallah:2018qtu}
{\scshape HESS} collaboration, H.~Abdallah et~al., \emph{{Search for
  $\gamma$-Ray Line Signals from Dark Matter Annihilations in the Inner
  Galactic Halo from 10 Years of Observations with H.E.S.S.}},
  \href{https://doi.org/10.1103/PhysRevLett.120.201101}{\emph{Phys. Rev. Lett.}
  {\bfseries 120} (2018) 201101},
  [\href{https://arxiv.org/abs/1805.05741}{{\ttfamily 1805.05741}}].

\bibitem{Athron:2017qdc}
{\scshape GAMBIT} collaboration, P.~Athron et~al., \emph{{Global fits of
  GUT-scale SUSY models with GAMBIT}},
  \href{https://doi.org/10.1140/epjc/s10052-017-5167-0}{\emph{Eur. Phys. J.}
  {\bfseries C77} (2017) 824},
  [\href{https://arxiv.org/abs/1705.07935}{{\ttfamily 1705.07935}}].

\bibitem{Athron:2017yua}
{\scshape GAMBIT} collaboration, P.~Athron et~al., \emph{{A global fit of the
  MSSM with GAMBIT}},
  \href{https://doi.org/10.1140/epjc/s10052-017-5196-8}{\emph{Eur. Phys. J.}
  {\bfseries C77} (2017) 879},
  [\href{https://arxiv.org/abs/1705.07917}{{\ttfamily 1705.07917}}].

\bibitem{Shilon:2018xlp}
I.~Shilon, M.~Kraus, M.~Büchele, K.~Egberts, T.~Fischer, T.~L. Holch et~al.,
  \emph{{Application of Deep Learning methods to analysis of Imaging
  Atmospheric Cherenkov Telescopes data}},
  \href{https://doi.org/10.1016/j.astropartphys.2018.10.003}{\emph{Astropart.
  Phys.} {\bfseries 105} (2019) 44--53},
  [\href{https://arxiv.org/abs/1803.10698}{{\ttfamily 1803.10698}}].

\bibitem{Drlica-Wagner:2019xan}
{\scshape LSST Dark Matter Group} collaboration, A.~Drlica-Wagner et~al.,
  \emph{{Probing the Fundamental Nature of Dark Matter with the Large Synoptic
  Survey Telescope}},  \href{https://arxiv.org/abs/1902.01055}{{\ttfamily
  1902.01055}}.

\bibitem{Marinelli:2017csa}
A.~Marinelli, D.~Gaggero, D.~Grasso, M.~Taoso, A.~Urbano and S.~Ventura,
  \emph{{High Energy Neutrino expectations from the Central Molecular Zone}},
  \href{https://doi.org/10.22323/1.301.0939}{\emph{PoS} {\bfseries ICRC2017}
  (2018) 939}.

\bibitem{Buchovecky:VERITAS}
M.~Buchovecky, \emph{{Very High Energy Emission from the Galactic Center with
  VERITAS}}, Ph.D. thesis, University of California Los Angeles, 2019.

\bibitem{Evoli:2016xgn}
C.~Evoli, D.~Gaggero, A.~Vittino, G.~Di~Bernardo, M.~Di~Mauro, A.~Ligorini
  et~al., \emph{{Cosmic-ray propagation with $\small{DRAGON2}$: I. numerical
  solver and astrophysical ingredients}},
  \href{https://doi.org/10.1088/1475-7516/2017/02/015}{\emph{JCAP} {\bfseries
  1702} (2017) 015}, [\href{https://arxiv.org/abs/1607.07886}{{\ttfamily
  1607.07886}}].

\bibitem{Vladimirov:2010aq}
A.~E. Vladimirov, S.~W. Digel, G.~Johannesson, P.~F. Michelson, I.~V.
  Moskalenko, P.~L. Nolan et~al., \emph{{GALPROP WebRun: an internet-based
  service for calculating galactic cosmic ray propagation and associated photon
  emissions}}, \href{https://doi.org/10.1016/j.cpc.2011.01.017}{\emph{Comput.
  Phys. Commun.} {\bfseries 182} (2011) 1156--1161},
  [\href{https://arxiv.org/abs/1008.3642}{{\ttfamily 1008.3642}}].

\bibitem{Bringmann:2012vr}
T.~Bringmann, X.~Huang, A.~Ibarra, S.~Vogl and C.~Weniger, \emph{{Fermi LAT
  Search for Internal Bremsstrahlung Signatures from Dark Matter
  Annihilation}},
  \href{https://doi.org/10.1088/1475-7516/2012/07/054}{\emph{JCAP} {\bfseries
  1207} (2012) 054}, [\href{https://arxiv.org/abs/1203.1312}{{\ttfamily
  1203.1312}}].

\bibitem{Gondolo:1999ef}
P.~Gondolo and J.~Silk, \emph{{Dark matter annihilation at the galactic
  center}}, \href{https://doi.org/10.1103/PhysRevLett.83.1719}{\emph{Phys. Rev.
  Lett.} {\bfseries 83} (1999) 1719--1722},
  [\href{https://arxiv.org/abs/astro-ph/9906391}{{\ttfamily
  astro-ph/9906391}}].

\bibitem{Ullio:2001fb}
P.~Ullio, H.~Zhao and M.~Kamionkowski, \emph{{A Dark matter spike at the
  galactic center?}},
  \href{https://doi.org/10.1103/PhysRevD.64.043504}{\emph{Phys. Rev.}
  {\bfseries D64} (2001) 043504},
  [\href{https://arxiv.org/abs/astro-ph/0101481}{{\ttfamily
  astro-ph/0101481}}].

\bibitem{Merritt:2002vj}
D.~Merritt, M.~Milosavljevic, L.~Verde and R.~Jimenez, \emph{{Dark matter
  spikes and annihilation radiation from the galactic center}},
  \href{https://doi.org/10.1103/PhysRevLett.88.191301}{\emph{Phys. Rev. Lett.}
  {\bfseries 88} (2002) 191301},
  [\href{https://arxiv.org/abs/astro-ph/0201376}{{\ttfamily
  astro-ph/0201376}}].

\end{thebibliography}\endgroup
